\documentclass[11pt,reqno,letter]{amsart}
\usepackage{amsmath}

\usepackage{amsfonts}
\usepackage{amssymb}
\usepackage{amsthm}

\usepackage{graphicx}
\usepackage{hyperref}
\usepackage[left=1.25in,right=1.25in,top=1.5in,bottom=1.25in]{geometry}
\usepackage{multirow}
\usepackage{verbatim}
\usepackage{float}
\usepackage{rotating}
\usepackage[dvipsnames]{xcolor}
\usepackage[normalem]{ulem}
\usepackage{booktabs}

\usepackage{tabu}

\usepackage{float}
\usepackage{booktabs}

\usepackage{subcaption}

\theoremstyle{definition}

\newtheorem*{example*}{Example}

\usepackage{calc,tikz,nicefrac}

\usepackage[section]{placeins}

\usetikzlibrary{positioning}
\usepackage{mathtools}

\usepackage{bm}

\usepackage[authoryear]{natbib}
\usepackage{pdflscape}
\usepackage{afterpage}
\usepackage{capt-of}

\renewcommand{\to}{{\rightarrow}}
\providecommand{\Infected}{{\mathcal{I}}}
\providecommand{\Recovered}{{R}}

\providecommand{\Var}{{\mathrm{Var}}}

\newcommand\indep{\protect\mathpalette{\protect\independenT}{\perp}}
\def\independenT#1#2{\mathrel{\setbox0\hbox{$#1#2$}%
    \copy0\kern-\wd0\mkern4mu\box0}}

\newcommand{\Ep}{{\mathrm{E}}}

\usepackage{neuralnetwork}

\def\bcolor{\color{ForestGreen}}
\def\pcolor{\color{blue}}
\def\icolor{\color{magenta}}
\def\wcolor{\color{gray}}
\def\ycolor{\color{red}}

\begin{document}

\title[Causal Impact of Masks, Policies, Behavior]
{Causal Impact of Masks, Policies, Behavior on Early Covid-19 Pandemic in the U.S.}\thanks{We are grateful to Daron Acemoglu, V.V. Chari, Raj Chetty,  Christian Hansen, Glenn Ellison, Ivan Fernandez-Val, David Green,  Ido Rosen, Konstantin Sonin, James Stock, Elie Tamer, and Ivan Werning for helpful comments. We also thank Chiyoung Ahn,
Joshua Catalano,  Jason Chau, Samuel Gyetvay, Sev Chenyu Hou,
Jordan Hutchings, and Dongxiao Zhang  for excellent research assistance.  We are especially grateful
to   C. Jessica E. Metcalf and an anonymous referee for detailed and insightful comments,
which considerable improved the paper. All mistakes are our own. }

\author{Victor Chernozhukov}
\address{Department of Economics and Center for Statistics and Data Science, MIT,  MA 02139}
\email{vchern@mit.edu}
\author{Hiroyuki Kasahara}
\address{ Vancouver School of Economics, UBC, 6000 Iona Drive, Vancouver, BC.}
\email{hkasahar@mail.ubc.ca}

\author{Paul Schrimpf}
\address{ Vancouver School of Economics, UBC, 6000 Iona Drive, Vancouver, BC.}
\email{schrimpf@mail.ubc.ca}

\date{\today; \textit{First public} version posted to ArXiv:  May 28, 2020}

\begin{abstract}
The paper evaluates the dynamic impact of various policies adopted by US states on the growth rates of confirmed Covid-19  cases and deaths as well as social distancing behavior measured by Google Mobility Reports, where we take into consideration people's voluntarily behavioral response to new information of transmission risks in a causal structural model framework. Our analysis finds that both policies and information on transmission risks are important determinants of Covid-19  cases and deaths and shows that a change in policies explains a large fraction of observed changes in social distancing behavior. Our main counterfactual experiments suggest that nationally mandating face masks for employees early in the pandemic could have reduced the weekly growth rate of cases and deaths  by more than 10 percentage points in late April and could have led to as much as 19 to 47 percent less deaths nationally by the end of May, which roughly translates into 19 to 47 thousand saved lives.   We also find that, without stay-at-home orders, cases would have been larger by  6 to 63 percent and without business closures, cases would have been larger by 17 to 78 percent.  We find considerable uncertainty over the effects of school closures due to lack of cross-sectional variation;  we could not robustly rule out either large or small effects.   Overall, substantial declines in growth rates are attributable to private behavioral response, but policies played an important role as well.   We also carry out sensitivity analyses to find neighborhoods of the models under which the results hold robustly:  the results on mask policies appear to be  much more robust than   the results on business closures and stay-at-home orders. Finally, we stress that our study is observational and therefore should be interpreted with great caution. From a completely agnostic point of view,
our findings uncover predictive effects (association) of observed policies and behavioral changes  on future health outcomes,
 controlling for informational and other confounding variables.

\end{abstract}



\maketitle


 \section{Introduction}

Accumulating evidence suggests that various policies in the US have reduced social interactions and slowed down the growth of Covid-19  infections.\footnote{
See
\citet{courtemanche2020}, \citet{hsiang2020}, \citet{pei2020},  \citet{abouk2020}, and \cite{wright2020}.} An important outstanding issue, however, is how much of the observed slow down in the spread is attributable to the effect of policies as opposed to a voluntarily change in people's behavior out of fear of being infected. This question is critical for evaluating the effectiveness of restrictive policies  in the US relative to an alternative policy of just providing recommendations and information such as the one adopted by Sweden.  More generally, understanding people's dynamic  behavioral response to policies and information  is indispensable for properly evaluating the effect of policies on the spread of Covid-19.

This paper  quantitatively assesses the impact of various policies adopted by US states on the spread of Covid-19, such as non-essential business closure and mandatory  face masks, paying particular attention to how people adjust their behavior in response to policies as well as new information on cases and deaths.

We present a conceptual framework that spells out the causal structure on how the Covid-19 spread is dynamically determined by policies and human behavior. Our approach explicitly recognizes that policies not only directly affect the spread of Covid-19 (e.g., mask requirement) but also indirectly affect its spread by changing people's behavior (e.g., stay-at-home order). It also recognizes that people react to new information on Covid-19 cases and deaths and voluntarily adjust their behavior (e.g., voluntary social distancing and hand washing) even without any policy in place. Our casual model provides a framework to quantitatively decompose the growth of Covid-19 cases and deaths into three components: (1) direct policy effect, (2) policy effect through behavior, and (3) direct behavior effect in response to new information.

Guided by the causal model, our empirical analysis examines how the weekly growth rates of confirmed Covid-19 cases and deaths are determined by (the lags of) policies and behavior using US state-level data. To examine how policies and information affect people's behavior, we also regress social distancing measures  on policy and information variables. Our regression specification for case and death growths is explicitly guided by a SIR model although  our causal approach does not hinge on the validity of a SIR model.

As policy variables, we consider  mandatory face masks for employees in public businesses, stay-at-home orders (or shelter-in-place orders), closure of K-12 schools, closure of restaurants except take out, closure of movie theaters, and closure of non-essential businesses. Our behavior variables are four mobility measures that capture the intensity of visits to ``transit,'' ``grocery,'' ``retail,'' and ``workplaces''  from Google Mobility Reports. We take the lagged growth rate of cases and deaths and the log of lagged cases and deaths at both the state-level and the national-level as our measures of information on infection risks that affects people's  behavior. We also consider the growth rate of tests, month dummies, and state-level characteristics (e.g., population size and total area) as confounders that have to be controlled for in order to identify the causal relationship between policy/behavior and the growth rate of cases and deaths.

Our key findings from regression analysis are as follows.  We find
that both policies and information on past cases and deaths are important
determinants of people's social distancing behavior, where policy
effects explain more than $50\%$ of the observed decline in the four
behavior variables.\footnote{The behavior accounts for the other half. This is in line with theoretical study by \cite{gitmez2020} that investigates the role of private behavior and
negative external effects for individual decisions over policy compliance as well as information acquisition during pandemics.} Our estimates suggest that there are both large policy effects and large behavioral effects on the growth of cases and deaths. Except for mandatory masks, the effect of policies on cases and deaths is indirectly materialized through their impact on behavior; the effect of mandatory mask policy is direct without affecting behavior.

Using the estimated model, we evaluate the dynamic impact of the following three counterfactual policies on Covid-19 cases and deaths: (1) mandating face masks, (2)  allowing all businesses to open, and (3) not implementing a stay-at-home order.  The counterfactual experiments show a large impact of  those policies on  the number of cases and deaths.  They also highlight the importance of voluntary behavioral response to infection risks when evaluating the dynamic policy effects.

\begin{figure}[ht]
  \caption{Relative cumulative effect  on confirmed cases and fatalities 
  of nationally mandating masks for employees on March
    14th in the US \label{fig:US-mask-dgrowth}}
  \begin{minipage}{\linewidth}
    \centering
    \medskip
    \begin{tabular}{cc}
      \includegraphics[width=0.49\textwidth]{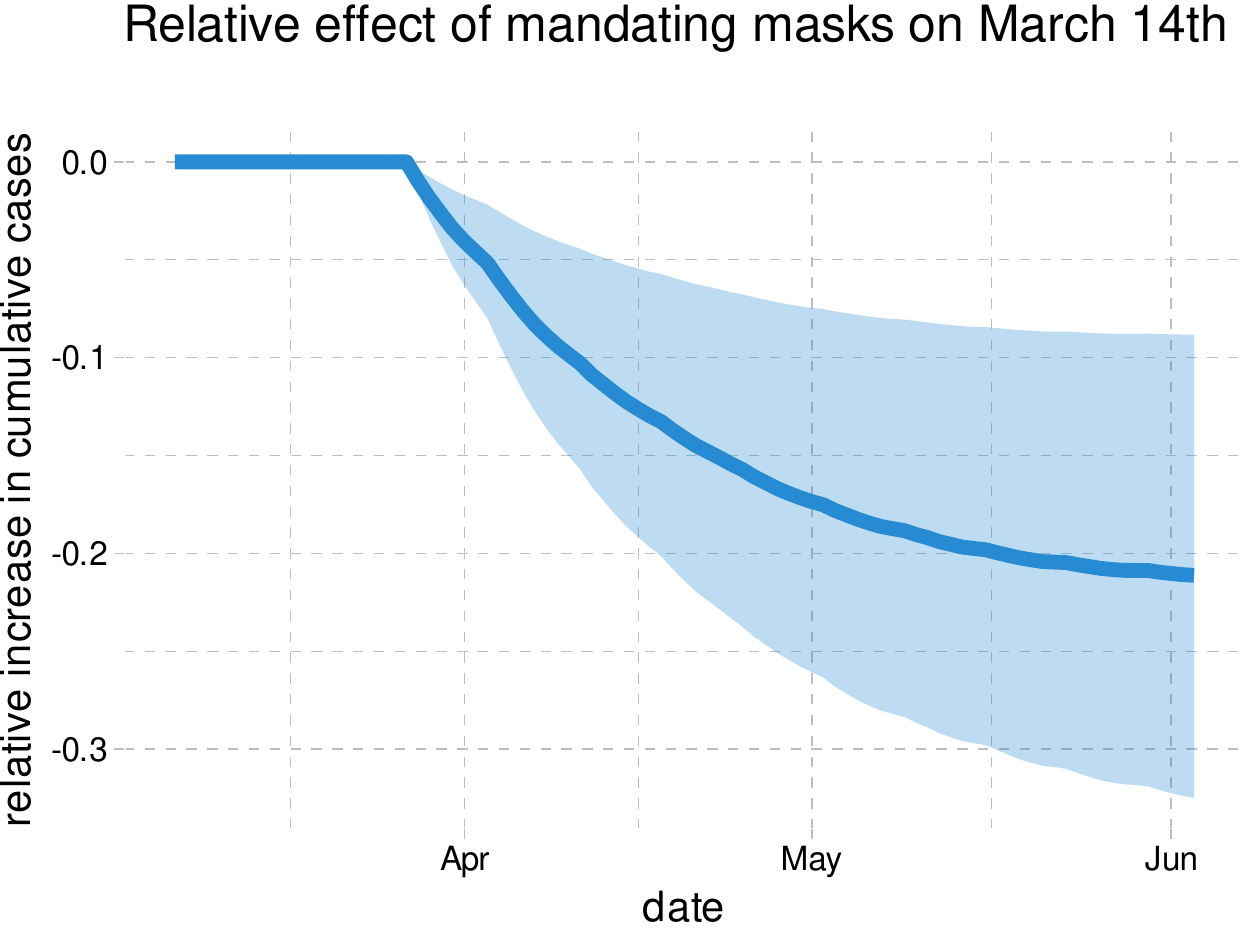}
      & \includegraphics[width=0.49\textwidth]{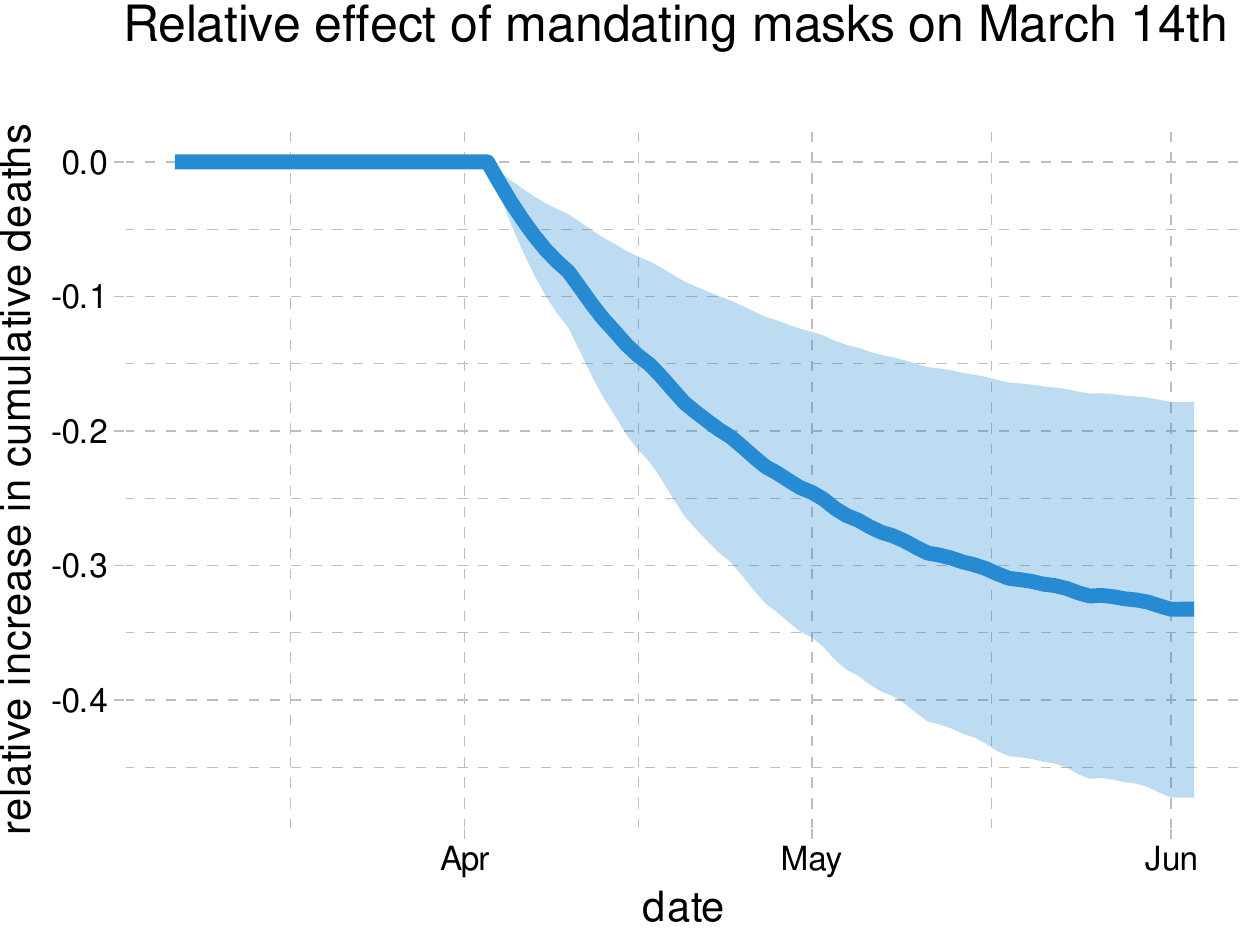}
    \end{tabular}
    \begin{flushleft}
      \footnotesize This figure shows the estimated relative change in
      cumulative cases and deaths if all states had mandated masks on
      March 14th. The thick blue line is the estimated change in
      cumulative cases or deaths relative to the observed number of
      cases or deaths. The shaded region is a pointwise 90\%
      confidence band.
    \end{flushleft}
  \end{minipage}
\end{figure}

Our estimates imply that nationally implementing mandatory face masks
for employees in public businesses on March 14th would have reduced the
growth rate of cases and that of deaths by approximately 10 percentage
points in late April. As shown in Figure \ref{fig:US-mask-dgrowth},
this leads to reductions of $21$\% and $34$\% in cumulative reported
cases and deaths, respectively, by the end of May with 90 percent
confidence intervals of $[9,32]$\% and $[19,47]$\%, which roughly
implies that 34 thousand lives could have been saved. 
This finding is
significant: given this potentially large benefit of reducing the
spread of Covid-19, mandating masks is an attractive policy instrument
especially because it involves relatively little economic
disruption. These estimates contribute to the ongoing efforts towards
designing approaches to minimize risks from reopening
\citep{stock2020b}.

\begin{figure}[ht]
  \caption{Relative cumulative effect  of no business closure policies on cases and fatalities in the US  \label{fig:US-nb-dgrowth}}
  \begin{minipage}{\linewidth}
    \centering
    \begin{tabular}{cc}
      \includegraphics[width=0.45\textwidth]{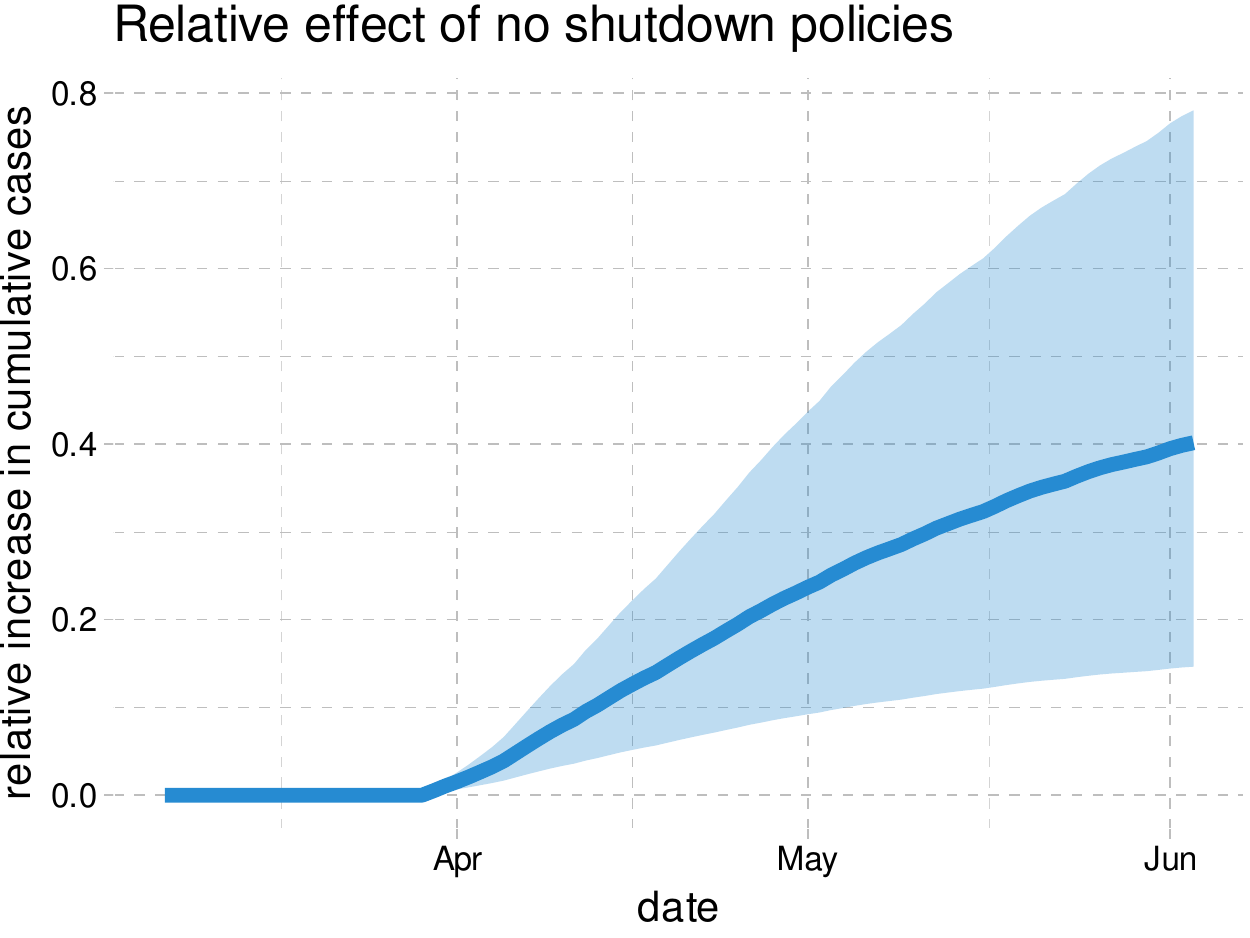}
      & \includegraphics[width=0.45\textwidth]{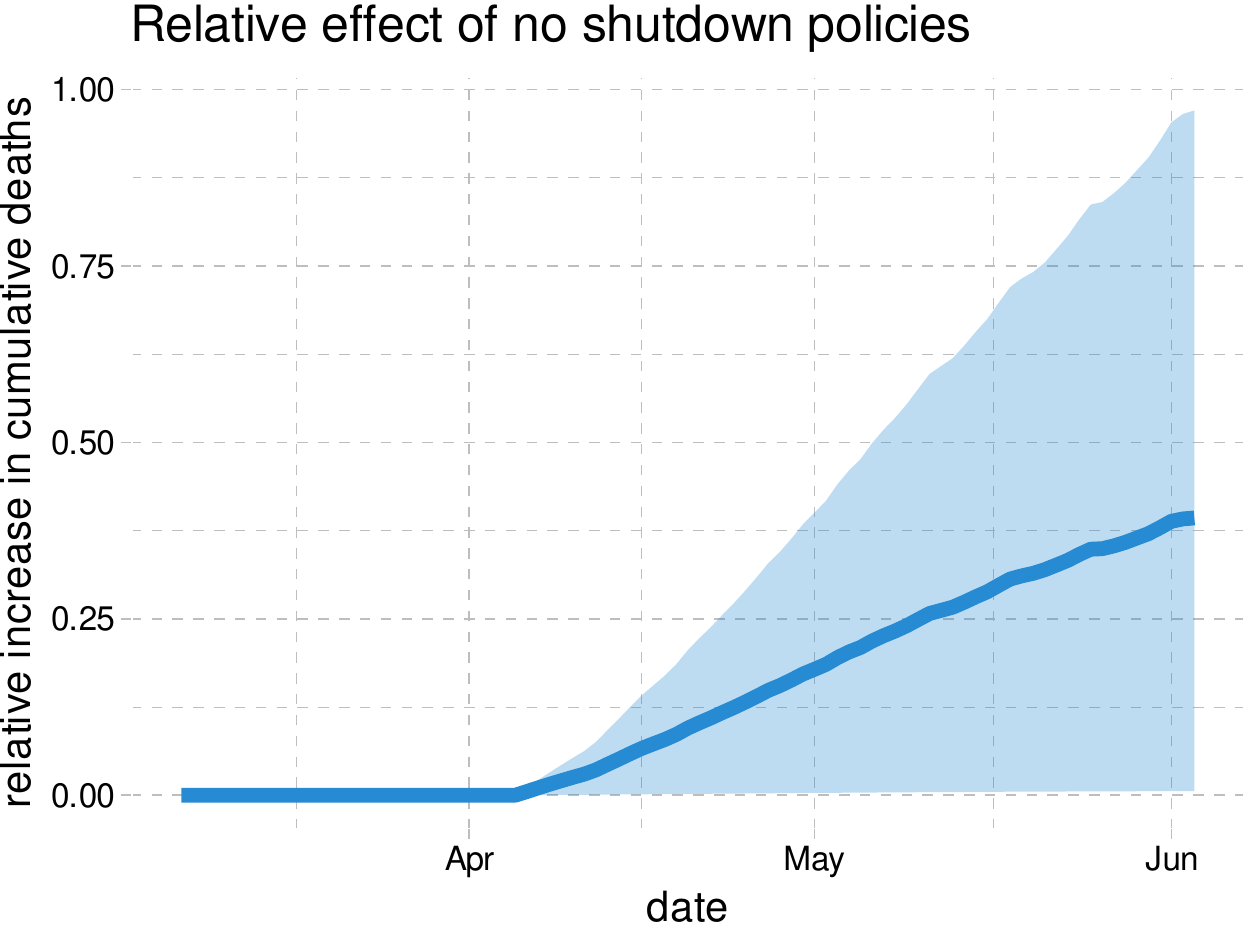}
    \end{tabular}
    \begin{flushleft}
      \footnotesize This figure shows the estimated relative change in
      cases and deaths if no states had ever implemented any business closure
      policies. The thick blue line is the estimated change in cumulative
      cases or deaths relative to the observed number of cumulative cases
      or deaths. The shaded region is a pointwise 90\% confidence
      band.
    \end{flushleft}
  \end{minipage}
\end{figure}

Figure \ref{fig:US-nb-dgrowth} illustrates how never closing any
businesses (no movie theater closure, no non-essential business
closure, and no closure of restaurants except take-out) could have
affected cases and deaths.  We estimate that business shutdowns have 
roughly the same impact on growth rates as mask mandates, albeit with
more uncertainty. The point estimates indicate that keeping all
businesses open could have increased cumulative cases and deaths by
$40\%$ at the end of May (with 90 percent confidence intervals of
$[17,78]$\% for cases and $[1,97]$\% for deaths).

\begin{figure}[ht]
  \caption{Relative cumulative effect  of  not implementing stay-at-home order on cases and fatalities in the US  \label{fig:US-shelter-dgrowth}}
  \begin{minipage}{\linewidth}
    \centering
    \begin{tabular}{cc}
         \includegraphics[width=0.45\textwidth]{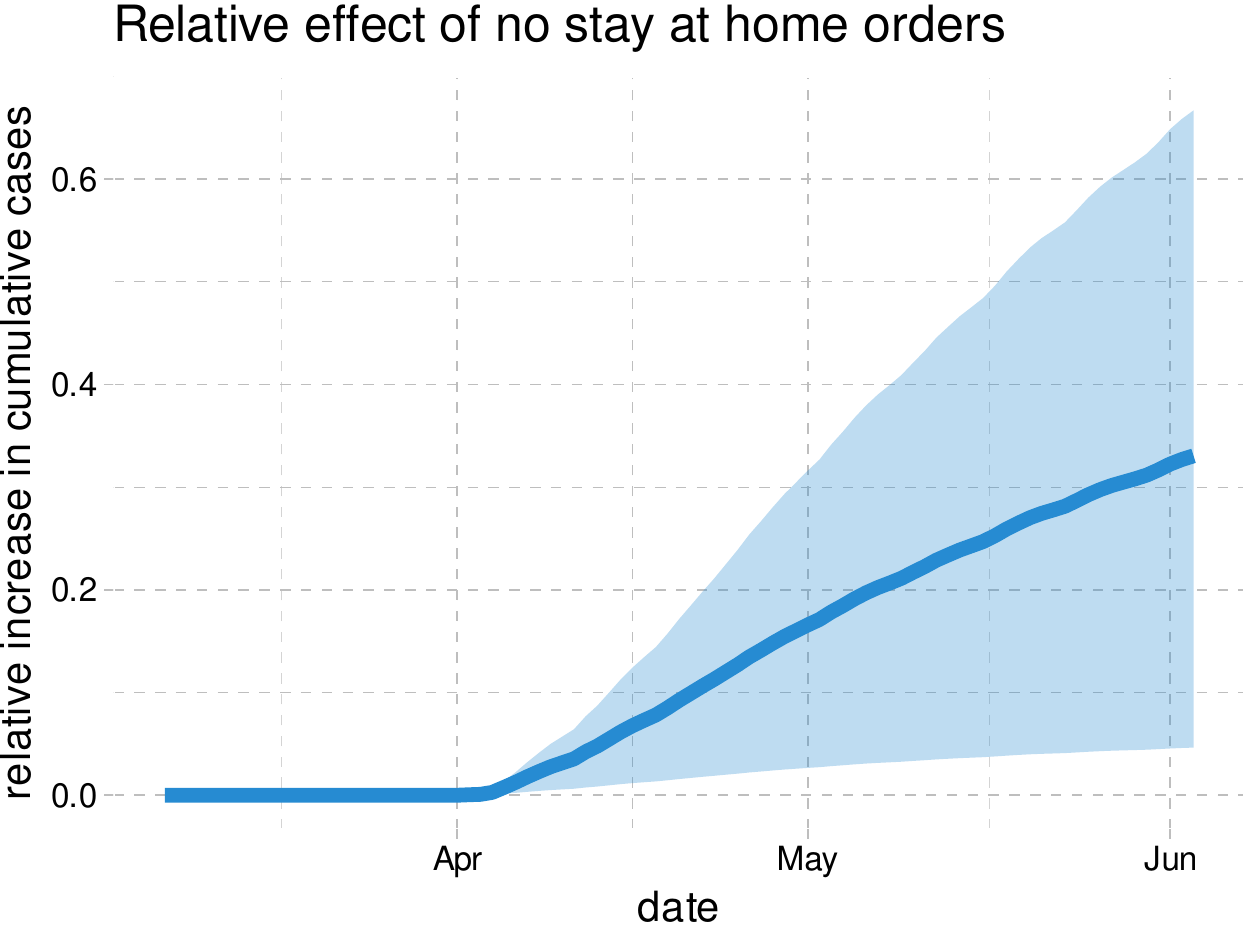}
      &
        \includegraphics[width=0.45\textwidth]{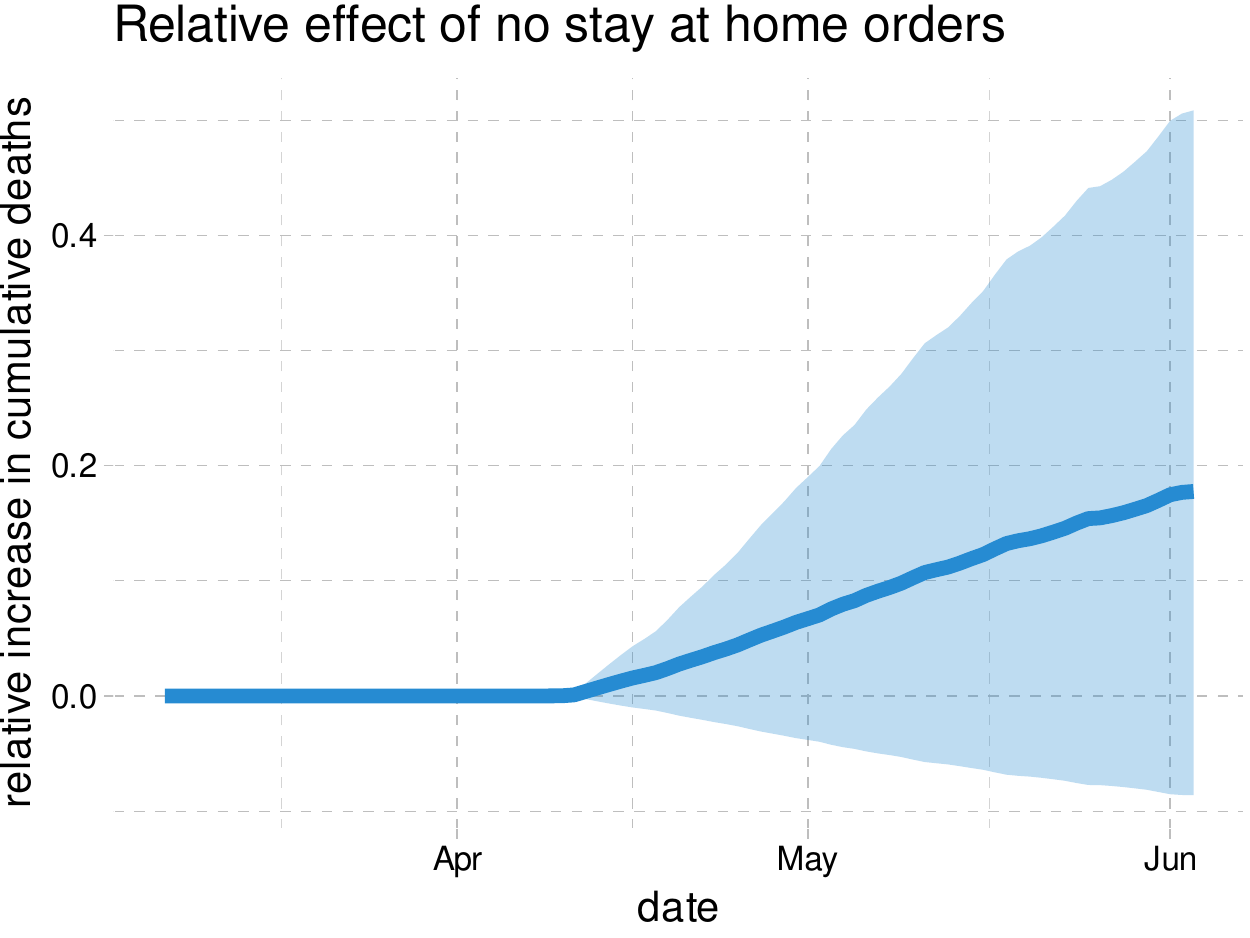}
    \end{tabular}
    \begin{flushleft}
      \footnotesize This figure shows the estimated relative change in
      cases and deaths if no states had ever issued stay at home
      orders. The thick blue line is the estimated change in cumulative
      cases or deaths relative to the observed number of cumulative cases
      or deaths. The shaded region is a point wise 90\% confidence
      band.
    \end{flushleft}
  \end{minipage}
\end{figure}

Figure \ref{fig:US-shelter-dgrowth} shows that stay-at-home orders had
effects of similar magnitude as business closures.  No stay-at-home
orders could have led to $37$\% more cases by the start of June with a
90 percent confidence interval given by $6$\% to $63$\%. The estimated
effect of no stay-at-home orders on deaths is a slightly smaller with
a 90 percent confidence interval of $-7$\% to $50$\%.  

We also conducted sensitivity analysis with respect to changes to our regression specification, sample selection, methodology, and assumptions about delays between policy changes and changes in recorded cases. In particular, we examined whether certain effect
sizes can be ruled out by various more flexible models or by using alternative timing assumptions
that define forward growth rates. The impact of mask mandates is more robustly and more precisely estimated than that of business closure policies or stay-at-home orders, and an undesirable effect of increasing the weekly death growth by 5 percentage points is ruled out by all of the  models we consider.\footnote{This null
hypothesis can be generated by looking at the meta-evidence from RCTs on the efficacy of masks in preventing other respiratory
cold-like deceases. Falsely rejecting this null is   costly in terms of potential loss of life, and so it is a reasonable null choice 
for the mask policy from decision-theoretic point of view.} This is largely due to the greater variation in the timing of mask mandates across states. The findings of shelter-in-place and business closures policies are considerably less robust. For example, for stay-at-home mandates, models with alternative timing and richer specification for information set suggested smaller effects. Albeit
after application of machine learning tools to reduce dimensionality, the range of effects $[0,0.15]$ could not be ruled out.  A similar
wide range of effects could  not be ruled out for business closures.
  
We also examine the impact of school closures. Unfortunately, there is
very little variation across states in the timing of
school closures.  Across robustness specifications, we obtain point
estimates of the effect of school closures as low as 0 and as high as
-0.6. In particular, we find that the results are sensitive to whether
the number of past national cases/deaths is included in a
specification or not. This highlights the uncertainty regarding the
impact of some policies versus private behavioral responses to
information. 

%


%
%

A growing number of other papers have examined the link between non-pharmaceutical interventions and Covid-19  cases.\footnote{We refer the reader to \cite{avery2020} for a comprehensive review of  a larger body of work researching Covid-19; here we focus on few quintessential comparisons on our work with other works that we are aware of.} \cite{hsiang2020} estimate the effect of policies on the growth rate of cases using
data from the United States, China, Iran, Italy, France, and South
Korea. In the United States, they find that the combined effect of all policies they consider on the
growth rate is $-0.347$ $(0.061)$. \cite{courtemanche2020} use US county level data to analyze the effect
of interventions on case growth rates. They find that the combination
of policies they study reduced growth rates by 9.1 percentage points
16-20 days after implementation, out of which 5.9 percentage points are attributable to shelter in place orders. Both \cite{hsiang2020} and \cite{courtemanche2020} adopt a reduced-form approach to estimate the total policy effect on case growth without using any social distancing behavior measures.\footnote{Using
a synthetic control approach, \cite{cho2020} finds that  the cases would have been lower by 75 percent  had Sweden adopted stricter lockdown policies.}

Existing evidence for the impact of social distancing policies on behavior in the US is mixed.
\cite{abouk2020} employ a difference-in-differences methodology to  find that statewide stay-at-home orders  have  strong causal impacts on reducing social interactions. In contrast, using data from Google Mobility Reports,
\cite{maloney2020}  find that the increase in social distancing is
largely voluntary and driven by information.\footnote{Specifically, they find
that of the 60 percentage point drop in workplace intensity, 40
percentage points can be explained by changes in information as
proxied by case numbers, while roughly 8 percentage points can be
explained by policy changes.} Another study by  \cite{gupta2020} also found little evidence that stay-at-home mandates induced distancing by
using mobility measures from PlaceIQ and SafeGraph. Using data from SafeGraph, \cite{anderson2020} shows that there has been substantial voluntary social distancing but also provide evidence that mandatory measures such as stay-at-home orders have  been effective at reducing the frequency of visits  outside of one's home.

 \cite{pei2020} use county-level observations of reported infections and deaths in conjunction with mobility  data from SafeGraph to conduct simulation of implementing all policies 1-2 weeks earlier and found that it would have resulted in reducing the number of cases and deaths by more than half. However, their study does not  explicitly analyze how policies are related to the effective reproduction numbers. 

Epidemiologists use model simulations to predict how cases and deaths evolve for the purpose of policy recommendation. As reviewed by \cite{avery2020}, there exists substantial uncertainty about the values of key epidimiological parameters \citep[see also][]{atkeson2020b,stock2020}. Simulations are often done under strong assumptions about the impact of social distancing policies without connecting to the relevant data  \citep[e.g.,][]{ferguson2020}. Furthermore, simulated models do not take into account that people may limit their contact with other people in response to higher transmission risks.\footnote{See \cite{atkeson2020a} and \cite{stock2020} for the implications of the SIR model for Covid-19 in the US. \cite{NBERw27128} estimate a SIRD model in which time-varying reproduction numbers depend on the daily deaths to capture feedback from daily deaths to future behavior and infections.}  When such a voluntary behavioral response is ignored, simulations would necessarily exhibit exponential spread and over-predict  cases and deaths. In contrast, as cases and deaths rise, a voluntary behavioral response may possibly reduce the effective reproduction number  below 1,  potentially preventing exponential spread. Our counterfactual experiments  illustrate the importance of this voluntary behavioral change.

Whether wearing  masks in public place should be mandatory or not has been one of the most contested policy issues with health authorities of different countries providing contradictory recommendations. Reviewing evidence, \cite{Greenhalghm2020}  recognize that there is no randomized controlled trial evidence for the effectiveness of face masks,  but they state ``indirect evidence exists to support the argument for the public wearing masks in the Covid-19 pandemic."\footnote{The virus remains viable in the air for several hours, for which surgical masks may be effective. Also, a substantial fraction of individual who are infected become infectious before showing symptom onset.}
\cite{howard2020} also review available medical evidence and conclude that ``mask wearing reduces the transmissibility per contact by reducing transmission of infected droplets in both laboratory and clinical contexts.''  The laboratory findings in \cite{hou2020} suggest that the nasal cavity may be the initial site of infection followed by aspiration to the lung, supporting the argument  ``for the widespread use of masks to prevent aersol, large droplet, and/or mechanical exposure to the nasal passages.''   \cite{He2020} examined temporal patterns of viral shedding in COVID-19 patients and found the highest viral load at the time of symptom onset; this suggests that a significant portion of transmission may have occurred before symptom onset and that universal face masks may be an effective control measure to reduce transmission.\footnote{\cite{Lee2020} find evidence that viral loads in asymptomatic patients are similar to those in symptomatic patients. Aerosol transmission of viruses may occur through aerosols particles released during breathing and speaking by asymptomatic infected individuals; masks reduce such airborne transmission \citep{Prather2020}. \cite{Anfinrud2020} provide visual evidence of speech-generated droplet as well as the effectiveness of cloth masks to reduce the emission of droplets.   \cite{Chu2020} conduct a meta-analysis of observational studies on transmission of the viruses that cause COVID-19 and related diseases and find the effectiveness of mask use for reducing transmission. } \cite{ollila2020} provide a meta-analysis of randomized controlled trials of non-surgical face masks in preventing viral respiratory infections in non-hospital and non-household settings, finding that face masks decreased infections across all five studies they reviewed.\footnote{Whether wearing masks creates a false sense of security and leads to decrease in social distancing is also a hotly debated topic.  A randomized field experiment in Berlin, Germany, conducted by \cite{seres2020}  finds that wearing masks actually increases social distancing, providing no evidence that mandatory masks leads to decrease in social distancing.}

Given the lack of experimental evidence on the effect of masks in the context of COVID-19, conducting observational studies  is useful and important. To the best of our knowledge, our paper is the first  empirical study that shows the effectiveness of  mask mandates on reducing the spread of Covid-19 by analyzing the US state-level data. This finding corroborates and is complementary to the medical observational evidence in \cite{howard2020}. Analyzing mitigation measures in New York, Wuhan, and Italy, \cite{zhangr2020} conclude that mandatory face coverings substantially reduced infections. \cite{abaluck2020}  find that the growth rates of cases and of deaths in countries with pre-existing norms that sick people should wear masks are lower by 8 to 10\% than those rates in countries with no pre-existing mask norms.\footnote{\cite{miyazawa2020} find that  country's COVID-19 death rates are negatively associated with mask wearing rates.} The Institute for Health Metrics and Evaluation at the University of Washington  assesses that, if 95\% of the people in the US were to start wearing masks from early August of 2020, 66,000 lives would be saved by December 2020 \citep{IHME2020}, which is largely consistent with our results.
Our finding is also independently corroborated by a completely different causal methodology based on synthetic control
using German data in \cite{Mitze2020}.\footnote{Our study was first released in ArXiv on May 28, 2020 whereas
 \cite{Mitze2020} was released at SSRN on June 8, 2020. }

Our empirical results contribute to informing the economic-epidemiological models that combine economic models with variants of SIR models to evaluate the efficiency of various economic policies aimed at  the gradual ``reopening" of various sectors of economy.\footnote{\cite{adda2016} analyzes the effect of policies reducing interpersonal contacts such as school closures or the closure of public transportation networks on the spread of  influenza, gastroenteritis, and chickenpox using high frequency data from France.} For example, the estimated effects of masks, stay-home mandates, and various other policies on behavior, and of behavior on infection can serve as useful inputs and validation checks in the calibrated macro, sectoral, and micro models (see, e.g., \cite{alvarez2020simple,baqaee2020reopening,NBERw27128,acemoglu2020multi,lsmith,nashSIR} and references therein). Furthermore, the causal framework developed in this paper could be applicable, with appropriate extensions, to the impact of policies on economic outcomes replacing health outcomes (see, e.g., \cite{chetty2020real,coibion2020labor}).

Finally, our causal model is framed using the language of structural equations models and causal diagrams of econometrics (\cite{pwright, haavelmo:1944, tinbergen:1940, wold:1954, pearl:biometrika}) and genetics \citep{wright1923},\footnote{The father and son, P. Wright (economist) and S. Wright (geneticist) collaborated to develop structural equation models and causal path diagrams; P. Wright's key work represented supply-demand system as a directed acyclical graph and established its identification using exclusion restrictions on instrumental variables. We view our work as following this classical tradition.} with natural unfolding potential/structural outcomes representation \citep{rubin1974,tinbergen1930,neyman:PO,imbens_rubin_2015}. The work on causal graphs has been modernized and developed by \cite{pearl:biometrika,pearl:robins,pearl:causality,pearl:why} and many others (e.g., \cite{pearl:why,white:chalak,richardson:mediation,peters2020book,bareinboim:H,hernanrobins2020book}), with applications in computer science, genetics, epidemiology, and econometrics (see, e.g., \cite{heckman:pinto,hunermund,white:chalak} for applications in econometrics). The particular causal diagram we use has several ``mediation" components, where variables affect outcomes directly and indirectly through other variables called mediators; these structures go back at least to \citet[][see Figure 6]{wright1923}; see, e.g., \cite{baron1986}, \cite{Hines2020}, \cite{richardson:mediation} for modern treatments. 

\section{The Causal Model for the Effect of Policies, Behavior, and Information on Growth of Infection}\label{sec:causal-model}

\subsection{The Causal Model and Its Structural Equation Form}
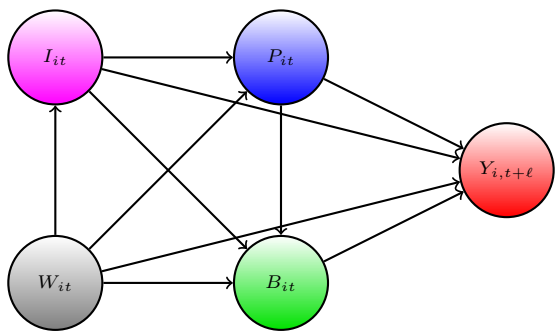
\begin{figure}[ht]
\begin{center}
\begin {tikzpicture}[-latex, auto, node distance =1.5cm and 3cm, on grid, thick,
  empty/.style ={circle, top color=white, bottom color = white, draw, white, text=white , minimum width =1.25 cm},
  policy/.style ={circle, top color=white, bottom color = blue, draw, black, text=black , minimum width =1.25 cm},
    behavior/.style ={circle, top color=white, bottom color = ForestGreen, draw, black, text=black , minimum width =1.25 cm},
  observed/.style ={circle, top color=white, bottom color = magenta, draw, black, text=black , minimum width =1.25 cm} ,
   confounder/.style ={circle, top color=white, bottom color =  gray, draw, black, text=black , minimum width =1.25 cm} ,
outcome/.style ={circle, top color=white, bottom color = red, draw, black, text=black , minimum width =1.25 cm} ]

\node[policy]   (P) {\tiny $P_{it}$};
\node[empty] (E) [below=of P] {\tiny $I_{it}$};
\node[outcome]  (Y)[ right =of E] {\tiny $Y_{i,t+\ell}$};
\node[behavior] (B) [below =of E] {\tiny $B_{it}$};
\node[observed] (I) [left =of P] {\tiny $I_{it}$};
\node[confounder] (W) [left=of B] {\tiny  $W_{it}$};
\path[->] (P) edge (Y);
\path[->] (B) edge (Y);
\path[->] (P) edge (B);
\path[->] (I) edge (B);
\path[->] (I) edge (P);
\path[->] (W) edge (Y);
\path[->] (W) edge (B);
\path[->] (W) edge (P);
\path[->] (W) edge (I);
\path[->] (I) edge (Y);

\end{tikzpicture} \end{center}
\caption{S. \& P. Wright type causal path diagram for our model. }\label{Wright}
\end{figure}

We introduce our approach through the Wright-style causal diagram shown in Figure \ref{Wright}.  The diagram describes how policies, behavior, and information interact together:
\begin{itemize}
\item The \textit{forward} health outcome,
$Y_{i,t+\ell}$, is determined last, after all other variables have been determined;
\item The  adopted policies, $P_{it}$,  affect health outcome $Y_{i,t+\ell}$ either directly, or indirectly by altering  human behavior  $B_{it}$;
\item  Information variables, $I_{it}$, such as lagged values of outcomes can affect human behavior and  policies, as well as  outcomes;
\item The confounding factors $W_{it}$, which vary across states and time, affect all other variables.
\end{itemize}
The index $i$ denotes observational unit, the state, and $t$ and $t+\ell$ denotes the time, where $\ell$ is a positive integer that represents the time lag  between infection and case confirmation or death.

Our main outcomes of interest are the growth rates in Covid-19 cases and deaths,  behavioral variables include proportion of time spent in transit, shopping, and workplaces, policy variables include mask mandates, stay-at-home orders, and school and business closures, and the information variables include lagged values of outcome. We provide a detailed description of these variables and their timing in the next section.

The causal structure  allows for  the effect of the policy to be either direct or indirect -- through behavior or through dynamics;  all of these effects are not mutually exclusive. The structure also allows for changes in behavior to be brought by change in policies and information. These are all realistic properties that we expect from the contextual knowledge of the problem. Policies such as closures of schools, non-essential business, and restaurants alter and constrain behavior in strong ways.  In contrast, policies such as mandating employees to wear masks can potentially affect the Covid-19 transmission directly.  The information variables, such as recent growth in the number of cases, can cause people to spend more time at home, regardless of adopted state policies; these changes in behavior in turn affect the transmission of Covid-19.  Importantly, policies can have the informational content as well, guiding behavior rather than constraining it.

The causal ordering induced by this directed acyclical graph is determined by the following
timing sequence: 
\begin{itemize}
\item[(1)]  information and confounders get determined at $t$,
\item[(2)] policies are set in place, given information and confounders at $t$;
\item[(3)] behavior is realized, given policies, information, and confounders at $t$;
\item[(4)] outcomes get realized at $t+\ell$ given policies, behavior, information, and confounders.
\end{itemize}

The model also allows for direct dynamic effects of information variables on the outcome through autoregressive structures that capture persistence in growth patterns. As  highlighted below, realized outcomes may become new information for future periods, inducing dynamics over multiple periods.

Our quantitative model for causal structure in Figure \ref{Wright} is given by the following econometric structural
 (or potential) outcomes model:\begin{equation} \label{eqn:SEM1} \tag{SO}
  \begin{aligned}
   &  Y_{i,t+\ell} (b,p,\iota) &:=& {\bcolor \alpha ' b}  +  {\pcolor \pi 'p} +
    {\icolor \mu'\iota } + {\wcolor \delta_Y 'W_{it}} + \varepsilon^y_{it}, \\
   &  B_{it} (p,\iota) &:= & {\pcolor \beta'p } + {\icolor \gamma'\iota} +      {\wcolor\delta_B 'W_{it} } + \varepsilon^b_{it},
      \end{aligned}
 \end{equation}
which is a collection of functional relations with stochastic shocks, decomposed into observable part $\delta' W$ and unobservable part $\varepsilon$.
The terms $\varepsilon^y_{it}$ and  $\varepsilon^b_{it} $  are the centered stochastic shocks that obey the orthogonality restrictions posed below.

The policies can be modeled via a linear form as well,
\begin{equation}\label{eq:Policy} \tag{P}
 P_{it}   (\iota) :=  {\icolor\eta'\iota} + {\wcolor \delta_P' W_{it}} +   \varepsilon^p_{it},   \end{equation}
although  linearity is not critical.\footnote{Under some additional independence conditions, this
can be replaced by an arbitrary non-additive function $P_{it}(\iota) = p (\iota, W_{it},  \varepsilon^p_{it})$, such that the unconfoundedness condition stated in the next footnote holds.}

The exogeneity restrictions on the stochastic shocks are as follows:
\begin{equation}\label{eqn:SEM2} \tag{E}
\begin{aligned}
   & \varepsilon^y_{it} &  \perp &  \quad (\varepsilon^b_{it}, \varepsilon^p_{it}, {\wcolor W_{it}}, {\icolor I_{it}}), \\
&  \varepsilon^b_{it}  & \perp & \quad  (\varepsilon^p_{it}, {\wcolor W_{it}}, {\icolor I_{it}}), \\
&   \varepsilon^p_{it} &  \perp &  \quad ({\wcolor W_{it}}, {\icolor I_{it}}),
\end{aligned}
\end{equation}
where we say that $V \perp U$ if $\Ep VU = 0$.\footnote{ An alternative useful
starting point is to impose the Rosenbaum-Rubin type unconfoundedness condition:
$$
Y_{i,t+\ell} (\cdot,\cdot,\cdot)  \indep (P_{it}, B_{it}, I_{it})  \mid W_{it}, \
B_{it} (\cdot,\cdot)  \indep (P_{it}, I_{it})  \mid W_{it}, \
 P_{it} (\cdot)  \indep  I_{it}  \mid W_{it},
$$
which imply, with treating stochastic errors as independent additive components, the orthogonal conditions stated above.
The same unconfoundedness restrictions can be formulated using formal causal DAGs, and also imply orthogonality restrictions stated above, once stochastic errors are modeled as independent additive components.} This is a standard way of representing restrictions on errors in structural equation modeling in econometrics.\footnote{The structural equations of this form are connected to triangular structural equation models, appearing in microeconometrics and macroeconometrics (SVARs), going back to the work of  \cite{strotz1960recursive}.}

The observed variables are generated by setting $\iota = I_{it}$ and propagating
the system from the last equation to the first:
\begin{equation} \label{eqn:SEMO}\tag{O}
\begin{aligned}
& {\ycolor Y_{i,t+\ell}}  & := & Y_{i,t+\ell} ( {\bcolor B_{it} } ,{\pcolor P_{it}}, {\icolor I_{it}}), \\
& {\bcolor B_{it} } & := &   B_{it}({\pcolor P_{it} } ,{\icolor I_{it}}), \\
& {\pcolor P_{it} }& := &  P_{it}({\icolor I_{it}}). \end{aligned}
\end{equation}


The specification of the model above grasps one-period responses.  The dynamics over multiple periods will be induced by the evolution
of information variables, which include time, lagged and integrated values of outcome:\footnote{Our empirical analysis also considers a specification in which information variables include lagged national cases/deaths as well as lagged behavior variables.}
\begin{equation}\label{eq:I}\tag{I}
 {\icolor I_{it} } := I_t( {\ycolor Y_{it}},  { \icolor  I_{i, t-\ell} }) := \Big (g(t) , {\ycolor Y_{it}},  \sum_{m=0}^{ \lfloor t/\ell \rfloor}
{\ycolor Y_{i,t - \ell m}} \Big )'  \end{equation}
 for  each  $t \in \{0,1,...,T\}$, where $g$ is deterministic function of time, e.g., month indicators,
 assuming that the log of new cases at time $t \leq 0$ is zero, for notational convenience.
 \footnote{The general formula  for  $I_{i, t-1}$ is $
S_{i, t, \ell } + \sum_{m=1}^{ \lfloor t/\ell \rfloor} Y_{i,t - \ell m}$, where $ S_{i, t, \ell }$ is the initial condition, the log of new cases at time $- t \ \mathrm{ mod } \ \ell$.} In this structure, people respond to both global information, captured by a function of time such as month dummies, and local information sources, captured by the local growth rate and the total number of cases. The local information also captures the persistence of the growth rate process.  We model the reaction of people's behavior via the term ${\icolor \gamma'I_{t}}$ in the behavior equation.  The  lagged values of behavior variable may be also included in the information set, but we postpone this discussion after the main empirical results are presented.

\begin{figure}[ht]
\begin{center}
\begin {tikzpicture}[-latex, auto, node distance =3cm and 3cm, on grid, thick,
  empty/.style ={circle, top color=white, bottom color = white, draw, white, text=white , minimum width =1.25 cm},
  policy/.style ={circle, top color=white, bottom color = blue, draw, black, text=black , minimum width =1.25 cm},
    behavior/.style ={circle, top color=white, bottom color = ForestGreen, draw, black, text=black , minimum width =1.25 cm},
  observed/.style ={circle, top color=white, bottom color = magenta, draw, black, text=black , minimum width =1.25 cm} ,
   confounder/.style ={circle, top color=white, bottom color =  gray, draw, black, text=black , minimum width =1.25 cm} ,
outcome/.style ={circle, top color=white, bottom color = red, draw, black, text=black , minimum width =1.25 cm},
myarrow/.style={-Stealth}]

\node[observed] (I)  {\tiny $I_{i,t-\ell}$};
\node[outcome]  (Y)[below=of I ]{\tiny $Y_{it}$};
\node[observed] (In)  [right=of I ]{\tiny $I_{it}$};
\node[outcome]  (Yn)[below=of In ]{\tiny $Y_{i,t+\ell}$};
\node[observed] (Inn)  [right=of In ]{\tiny $I_{i,t+\ell}$};
\node[outcome]  (Ynn)[below=of Inn ]{\tiny $Y_{i,t+2\ell}$};
 \draw [->] (I) -- node[sloped,font=\small,below] {\tiny SEM(t-$\ell$)} (Y);
 \draw [->] (I) --  (In);
 \draw [->] (In) -- node[sloped,font=\small,below] {\tiny SEM(t)} (Yn);
\draw [->] (Y) --  (In);
\draw [->] (In) --  (Inn);
 \draw [->] (Inn) -- node[sloped,font=\small,below] {\tiny SEM(t+$\ell$) } (Ynn);
\draw [->] (Yn) --  (Inn);

\end{tikzpicture} \end{center}
\caption{ Diagram for Information Dynamics in SEM}\label{fig:DynS}
\end{figure}
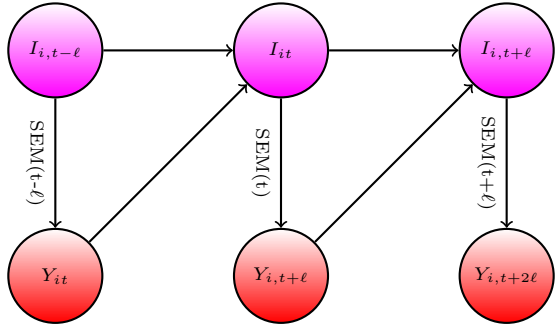

With any structure of this form, realized outcomes may become new information for future periods, inducing a dynamical system over multiple periods. We show the resulting dynamical system in a diagram of Figure \ref{fig:DynS}.  Specification
of this system is useful for studying delayed effects of policies and behaviors and in considering the counterfactual policy analysis.

Next we combine the above parts together with an appropriate
initializations to give a formal definition of the model we use.

\begin{quote}
\textsc{Structural Equations Model (SEM)}.  Let $i \in \{1,..., N\}$ denote the observational unit,  $t$ be the time periods, and $\ell$ be the time delay.  (1) For each $i$ and $t \leq -\ell$, the confounder, information, behavior, and policy variables $W_{it}, I_{it},  B_{it}, P_{it}$ are determined outside of the model, and the outcome variable $Y_{i, t+\ell}$ is determined by factors outside of the model for $t \leq 0$.  (2) For each $i$ and $t \geq -\ell$, confounders $W_{it}$ are  determined by factors outside of the model, and information variables $I_{it}$ are determined  by (I); policy variables $P_{it}$ are determined by setting $\iota = I_{it}$  in (P) with  a realized stochastic shock $\varepsilon^p_{it}$ that obeys the exogeneity condition (E);  behavior variables $B_{it}$ are determined by setting $\iota = I_{it}$ and $p= P_{it}$ in  (SO) with a shock $\varepsilon^b_{it}$ that obeys (E); finally, the outcome $Y_{i, t + \ell}$ is realized by setting $\iota = I_{it}$, $p= P_{it}$, and $b = B_{it}$ in  (SO) with a shock $\varepsilon^y_{it}$ that obeys (E). \end{quote}


\subsection{Main Testable Implication, Identification, Parameter Estimation}

The system above together with  orthogonality restrictions (\ref{eqn:SEM2}) implies the following collection of projection equations for realized variables:
\begin{align}
   &  {\ycolor  Y_{i,t+\ell}}
    = {\bcolor\alpha ' B_{it}} + {\pcolor\pi 'P_{it}} + {\icolor\mu'I_{it}} + {\wcolor\delta_Y 'W_{it}}  + \varepsilon^y_{it},
    &  & \varepsilon^y_{it} \perp {\bcolor B_{it}}, {\pcolor P_{it}}, {\icolor I_{it}}, {\wcolor W_{it}} \label{eq:R1} \tag{BPI$\to$Y} \\
    &  {\bcolor B_{it}}
     =  {\pcolor \beta' P_{it}} + {\icolor \gamma'I_{it}} +  {\wcolor \delta_B' W_{it}} + \varepsilon^b_{it},
   & & \varepsilon^b_{it} \perp {\pcolor P_{it}}, {\icolor I_{it}}, {\wcolor W_{it}}  \label{eq:R2} \tag{PI$\to$B}  \\
    & {\pcolor P_{it}}
    =  {\icolor\eta'I_{it}} + {\wcolor \delta_P' W_{it}} +   \varepsilon^p_{it},   & & \varepsilon^p_{it} \perp   {\icolor I_{it}}, {\wcolor W_{it}}  \label{eq:R3}  \tag{I$\to$P} \\
    &  {\ycolor  Y_{i,t+\ell}}
   =     ( {\bcolor\alpha '}  {\pcolor \beta' }+{\pcolor\pi'} )   {\pcolor P_{it}} +    ( {\bcolor\alpha '}  {\icolor \gamma'} + {\icolor \mu'}){\icolor I_{it} }+ {\wcolor \bar{\delta} '}{\wcolor W_{it}}  + {\bar \varepsilon}_{it},  &&  {\bar \varepsilon}_{it} \perp
  {\pcolor P_{it}},  {\icolor I_{it}}, {\wcolor W_{it}}.  \label{eq:R4} \tag{PI$\to$Y}
      \end{align}

Therefore the projection equation:
\begin{equation}
   {\ycolor  Y_{i,t+\ell}}
   = \mathsf{a}'
    {\pcolor P_{it}} + \mathsf{b}'    {\icolor I_{it} }+ {\wcolor \tilde {\delta} '}{\wcolor W_{it}}  + {\bar \varepsilon}_{it},  \quad   {\bar \varepsilon}_{it} \perp
  {\pcolor P_{it}},  {\icolor I_{it}}, {\wcolor W_{it}}.  \label{eq:PR} \tag{Y$\sim$PI}
\end{equation}
should obey:
\begin{equation}\label{eq:TI} \tag{TR}
 \mathsf{a}'  = ( {\bcolor\alpha '}  {\pcolor \beta' }+{\pcolor\pi'} ) \text{ and }
\mathsf{b}'  = ( {\bcolor\alpha '}  {\icolor \gamma'} + {\icolor \mu'}).
\end{equation}

Without any exclusion restrictions, this equality is just a decomposition of total effects
into direct and indirect components and is not a testable restriction. However, in our case we rely on the SIR
model with testing to motivate the presence of change in testing rate as a confounder in the outcome
equations but not in the behavior equation (therefore, a component of $\delta_B$ and $\delta_P$ is set to 0), implying that (TR) does not necessarily hold and is testable. Furthermore, we estimate (\ref{eq:R2}) on the data set that has many more observations than the data set used to estimate the outcome equations, implying that (TR) is again testable.  Later we  shall also try to
utilize the contextual knowledge that mask mandates  only affect the outcome directly and not by changing mobility (i.e., $\beta=0$ for mask policy), implying again that (TR) is testable. If not rejected by the data, (TR) can be used to sharpen the estimate of the causal effect of mask policies on the outcomes.

Validation of the model by (TR) allows us to check exclusion restrictions brought by contextual knowledge
and check stability of the model by using different data subsets.  However, passing the (TR) does
not guarantee that the model is necessarily valid for recovering causal effects. The only fundamental way to truly validate a causal model for observational data is through a controlled experiment, which is impossible to carry out in our setting. 

The parameters of the SEM are identified by the projection equation set above, provided the latter are identified by sufficient joint variation of these variables across states and time. We can develop this point formally as follows. Consider the previous system of equations, after partialling out the confounders:
\begin{equation}\label{eqn:SEM-PO}
 \begin{aligned}
   & {\ycolor { \tilde Y_{i,t+\ell}}}  & =& {\bcolor {\alpha ' \tilde B_{it}}} +{\pcolor {\pi '\tilde P_{it}}} +{\icolor \mu'\tilde I_{it} } + \varepsilon^y_{it},&\varepsilon^y_{it} &\perp {\tilde B_{it}}, {\tilde  P_{it}}, { \tilde I_{it}},  \\
   & {\bcolor { \tilde B_{it}}} & = &{\pcolor { \beta' \tilde P_{it}} } + {\icolor { \gamma' \tilde I_{it}} } + \varepsilon^b_{it},
   &\quad \varepsilon^b_{it} &\perp { \tilde P_{it}}, { \tilde I_{it}},  \\
   & {\pcolor { \tilde P_{it}} } &= &  {\icolor {\eta' \tilde I_{it}} } +   \varepsilon^p_{it}, &\quad \varepsilon^p_{it} &\perp {\tilde I_{it}}  \\
     & {\ycolor { \tilde Y_{i,t+\ell}}}  & =&  \mathsf{a} ' {\pcolor {\tilde P_{it}}} + \mathsf{b}' {\icolor \tilde I_{it} } + \bar \varepsilon^y_{it},&\bar \varepsilon^y_{it} &\perp  {\tilde  P_{it}}, { \tilde I_{it}},  \\
   \end{aligned}
\end{equation}
where $ \tilde V_{it} = V_{it}   -     {\wcolor W_{it}'} \Ep[{\wcolor W_{it}W_{it}'}]^{-} \Ep[{\wcolor W_{it}} V_{it}]$ denotes
the residual after removing the orthogonal projection of $V_{it}$ on ${\wcolor W_{it}}$. The residualization is a linear operator, implying that (\ref{eqn:SEM-PO}) follows immediately from the above. The parameters of (\ref{eqn:SEM-PO})  are identified as projection coefficients in these equations, provided that residualized vectors appearing in each of the equations have non-singular variance, that is
 \begin{equation}
 \Var ( {\pcolor \tilde P_{it}'} ,{\bcolor \tilde B_{it}'},{\icolor \tilde I_{it}'})>0,
 \ \Var ({\pcolor \tilde P_{it}'}, {\icolor \tilde I_{it}'})> 0 , \ \text{ and  }  \Var ( {\icolor \tilde I_{it}'}) >0.
 \end{equation}

Our main estimation method is the standard correlated random effects estimator, where the random effects
are parameterized as functions of observable characteristic, ${\wcolor W_{it}}$, which include both state-level and time random effects.  The state-level random effects
are modeled as a function of state level characteristics, and the time random effects are modeled
by including month dummies and their interactions with state level characteristics (in the sensitivity analysis,
we also add weekly dummies).  The stochastic shocks $\{ \varepsilon_{it}\}_{t=1}^T$ are treated as independent across states $i$ and can be arbitrarily dependent across time $t$ within a state.

Another modeling approach is the fixed effects panel data model, where ${\wcolor W_{it}}$ includes
\text{latent} (unobserved) state level confonders ${\wcolor W_i}$ and  and time level effects ${\wcolor W_t}$,
which must be estimated from the data.  This approach is much more demanding of the data and relies on long time and cross-sectional histories to estimate ${\wcolor W_i}$ and ${\wcolor W_t}$, resulting in amplification of uncertainty. In addition, when histories are relatively short,  large biases emerge and they need to be removed using debiasing methods, see e.g., \cite{chen2019mastering} for overview. We present the results
on debiased fixed effect estimation with weekly dummies as parts of our sensitivity analysis. Our sensitivity analysis also considers a debiased machine learning approach  using Random Forest in which observed confounders enter the model nonlinearly.

With exclusion restrictions there are multiple approaches to estimation, for example,
via generalized method of moments. We shall take a more pragmatic approach where we estimate
the parameters of equations separately and then compute
\[
 \frac{1}{2} \hat {\mathsf{a}}'  + \frac{1}{2}  ( {\bcolor \hat \alpha '}  {\pcolor \hat \beta' }+{\pcolor\hat \pi'} ) \text{ and }
 \frac{1}{2} \hat{\mathsf{b}}'  + \frac{1}{2}  ( {\bcolor \hat \alpha '}  {\icolor \hat \gamma'} + {\icolor \hat \mu'}),
\]
as the estimator of the total policy effect. Under standard regularity conditions,
these estimators concentrate around  their population analogues
\[
 \frac{1}{2}  {\mathsf{a}}'  + \frac{1}{2}  ( {\bcolor \alpha '}  {\pcolor \beta' }+{\pcolor \pi'} ) \text{ and }
 \frac{1}{2} {\mathsf{b}}'  + \frac{1}{2}  ( {\bcolor  \alpha '}  {\icolor \gamma'} + {\icolor \mu'}),
\]
with approximate deviations controlled by the normal laws, with standard deviations that can be approximated by the bootstrap
resampling of observational units $i$. Under correct specification the target quantities
reduce to $$  {\bcolor \alpha '}  {\pcolor  \beta' }+{\pcolor \pi'} \text{ and } {\bcolor  \alpha '}  {\icolor  \gamma'} + {\icolor  \mu'},$$ respectively.\footnote{This construction is not as efficient as generalized method of moments
but has a nicer interpretation under possible misspecification of the model: we are combining predictions
from two models, one motivated via the causal path, reflecting contextual knowledge, and another from a ``reduced form" model not exploiting the path. The combined estimator can improve on precision of either estimator.}

\subsection{Counterfactual Policy Analysis}

We  also consider simple counterfactual exercises, where we examine the effects of setting
a sequence of counterfactual policies for each state:
\begin{equation} \label{p-cf}\tag{CF-P}
\{{\pcolor P^\star_{it}} \}_{t=1}^T, \quad i=1, \ldots N.
\end{equation}
We assume that the SEM remains invariant, except for the policy equation.\footnote{It is possible to consider counterfactual exercises in which policy  responds to information through the policy equation if we are interested in endogenous policy responses to information. Counterfactual experiments with endogenous government policy would be important, for example, to understand the issues related to the lagged response of government policies to higher infection rates due to incomplete information. } 
The assumption  of invariance captures the idea that counterfactual policy interventions would not change the structural functions within the period of the study.  The assumption is strong but is necessary to conduct
counterfactual experiments, e.g. \cite{sims1972} and \cite{strotz1960recursive}. To make the assumption more plausible we limited our study to the early pandemic period.\footnote{Furthermore, we conducted several stability checks, for example, checking if the coefficients on mask policies remain stable (reported in the previous version) and also looking at more recent data during reopening, beyond early pandemic, to examine the stability of the model.} 

Given the policies, we generate the counterfactual
outcomes, behavior, and information by propagating the dynamic equations:
\begin{equation} \label{eqn:CF-SEM} \tag{CF-SEM}
  \begin{aligned}
& {\ycolor Y^\star_{i,t+\ell}}  & := & Y_{i,t+\ell} ( {\bcolor B^\star_{it} } , {\pcolor P^\star_{it}}, {\icolor I^\star_{it}}), \\
& {\bcolor B^\star_{it} } & := &   B_{it}({\pcolor P^\star_{it} } ,{\icolor I^\star_{it}}), \\
& {\icolor I^\star_{it} }& := &  I_t ( {\ycolor Y^\star_{it}}, \icolor I^\star_{i,t-\ell}),  \end{aligned}
\end{equation}
with the same initialization as the factual system up to $t \leq 0$. In stating this counterfactual system of equations, we assume that structural outcome equations (SO) and information equations (I) remain invariant and so do the stochastic shocks, decomposed
into observable and unobservable parts. Formally, we record this assumption and above discussion as follows.

\begin{quote}
\textsc{Counterfactual Structural Equations Model (CF-SEM)}.  Let $i \in \{1,..., N\}$ be the observational unit, $t$ be time periods, and $\ell$ be the time delay. (1) For each $i$ and $t \leq 0$, the confounder, information, behavior, policy, and outcome variables are determined as previously stated in SEM: $W^\star_{it}= W_{it}$, $I^\star_{it}= I_{it}$,  $B^\star_{it} = B_{it}$, $P^\star_{it}=P_{it}$, $Y^\star_{it} =Y_{it}$.  (2) For each $i$ and $t \geq 0$, confounders $W^\star_{it} =W_{it}$ are determined as in SEM, and information variables $I^\star_{it}$ are determined  by (I);  policy variables $P^\star_{it}$ are set in (\ref{p-cf}); behavior variables $B^\star_{it}$ are determined by setting $\iota = I^\star_{it}$ and $p= P^\star_{it}$ in  (SO) with  the same stochastic shock $\varepsilon^b_{it}$ in (SO); the  counterfactual outcome $Y^\star_{i, t + \ell}$ is realized by setting $\iota = I^\star_{it}$, $p= P^\star_{it}$, and $b = B^\star_{it}$ in  (SO) with the same stochastic shock $\varepsilon^y_{it}$ in  (SO).
\end{quote}

Figures \ref{Wright2} and \ref{fig:DynS2} present the causal path diagram for CF-SEM as well as the dynamics
of counterfactual information in CF-SEM.

\begin{figure}[ht]
\begin{center}
\begin {tikzpicture}[-latex, auto, node distance =1.5cm and 3cm, on grid, thick,
  empty/.style ={circle, top color=white, bottom color = white, draw, white, text=white , minimum width =1.25 cm},
  policy/.style ={circle, top color=white, bottom color = blue, draw, black, text=black , minimum width =1.25 cm},
    behavior/.style ={circle, top color=white, bottom color = ForestGreen, draw, black, text=black , minimum width =1.25 cm},
  observed/.style ={circle, top color=white, bottom color = magenta, draw, black, text=black , minimum width =1.25 cm} ,
   confounder/.style ={circle, top color=white, bottom color =  gray, draw, black, text=black , minimum width =1.25 cm} ,
outcome/.style ={circle, top color=white, bottom color = red, draw, black, text=black , minimum width =1.25 cm} ]

\node[policy]   (P) {\tiny $P^\star_{it}$};
\node[empty] (E) [below=of P] {\tiny $I^\star_{it}$};
\node[outcome]  (Y)[ right =of E] {\tiny $Y^\star_{i,t+\ell}$};
\node[behavior] (B) [below =of E] {\tiny $B^\star_{it}$};
\node[observed] (I) [left =of P] {\tiny $I^\star_{it}$};
\node[confounder] (W) [left=of B] {\tiny  $W_{it}$};
\path[->] (P) edge (Y);
\path[->] (B) edge (Y);
\path[->] (P) edge (B);
\path[->] (I) edge (B);
\path[->] (W) edge (Y);
\path[->] (W) edge (B);
\path[->] (W) edge (I);
\path[->] (I) edge (Y);

\end{tikzpicture} \end{center}
\caption{Causal path diagram for CF-SEM. }\label{Wright2}
\end{figure}

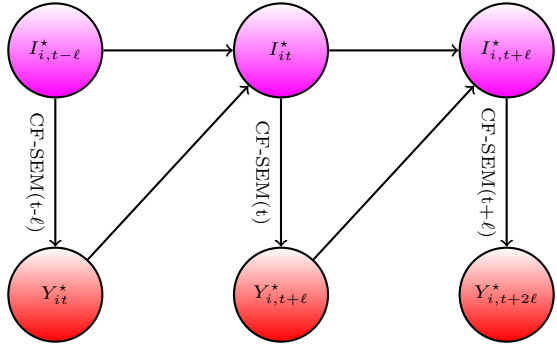
\begin{figure}[ht]
\begin{center}
\begin {tikzpicture}[-latex, auto, node distance =3.25 cm and 3cm, on grid, thick,
  empty/.style ={circle, top color=white, bottom color = white, draw, white, text=white , minimum width =1.25 cm},
  policy/.style ={circle, top color=white, bottom color = blue, draw, black, text=black , minimum width =1.25 cm},
    behavior/.style ={circle, top color=white, bottom color = ForestGreen, draw, black, text=black , minimum width =1.25 cm},
  observed/.style ={circle, top color=white, bottom color = magenta, draw, black, text=black , minimum width =1.25 cm} ,
   confounder/.style ={circle, top color=white, bottom color =  gray, draw, black, text=black , minimum width =1.25 cm} ,
outcome/.style ={circle, top color=white, bottom color = red, draw, black, text=black , minimum width =1.25 cm},
myarrow/.style={-Stealth}]

\node[observed] (I)  {\tiny $I^\star_{i,t-\ell}$};
\node[outcome]  (Y)[below=of I ]{\tiny $Y^\star_{it}$};
\node[observed] (In)  [right=of I ]{\tiny $I^\star_{it}$};
\node[outcome]  (Yn)[below=of In ]{\tiny $Y^\star_{i,t+\ell}$};
\node[observed] (Inn)  [right=of In ]{\tiny $I^\star_{i,t+\ell}$};
\node[outcome]  (Ynn)[below=of Inn ]{\tiny $Y^\star_{i,t+2\ell}$};
 \draw [->] (I) -- node[sloped,font=\small,below] {\tiny CF-SEM(t-$\ell$)} (Y);
 \draw [->] (I) --  (In);
 \draw [->] (In) -- node[sloped,font=\small,below] {\tiny CF-SEM(t)} (Yn);
\draw [->] (Y) --  (In);
\draw [->] (In) --  (Inn);
 \draw [->] (Inn) -- node[sloped,font=\small,below] {\tiny CF-SEM(t+$\ell$) } (Ynn);
\draw [->] (Yn) --  (Inn);

\end{tikzpicture} \end{center}
\caption{ A Diagram for Counterfactual Information Dynamics in CF-SEM}\label{fig:DynS2}
\end{figure}

The counterfactual outcome $Y^\star_{i,t+\ell}$  and  factual outcome $Y_{i,t+\ell}$  are given by:
$$
 {\ycolor Y^\star_{i,t+\ell}} = ( {\bcolor\alpha '} {\pcolor \beta'}  + {\pcolor\pi'})
    {\pcolor P^\star_{it}} + ({\bcolor\alpha '}  {\icolor \gamma' + \mu'})
    {\icolor I^\star_{it} }+ {\wcolor \bar \delta 'W_{it}} + \bar \varepsilon_{it}^y, $$$$\ {\ycolor Y_{i,t+\ell}} \ =
    ( {\bcolor\alpha '}  {\pcolor \beta'} + {\pcolor\pi'})
    {\pcolor P_{it}} + ({\bcolor\alpha '}  {\icolor \gamma' + \mu'})
    {\icolor I_{it} }+ {\wcolor \bar \delta 'W_{it}} + \bar \varepsilon_{it}^y.
$$
In generating these predictions, we explore the assumption of invariance stated above.
We can write  the counterfactual contrast into the sum of three components:
  \begin{align}
 \underbracket{{\ycolor Y^\star_{i,t+\ell} -  Y_{i,t+\ell} }}_{\text{CF Change}} & =
\underbracket{{\bcolor\alpha}' {\pcolor\beta}' {\pcolor\left(P^\star_{it} - P_{it} \right)}}_{\text{CF Policy Effect via Behavior}}  +    {\underbracket{\pcolor\pi' \left( P^\star_{it} -  P_{it} \right)}_{\text{CF Direct  Effect}} } \nonumber\\
&\qquad +  {\underbracket{{\bcolor\alpha}' {\icolor{\gamma}'  \left( I^\star_{it} - I_{it} \right)}+ {\icolor{\mu}'  \left( I^\star_{it} -  I_{it} \right)}}_{\text{ CF Dynamic Effect}}} \nonumber \\
 & =:   \mathrm{PEB}^\star_{it} + \mathrm{PED}^\star_{it} + \mathrm{DynE}^\star_{it}, \label{eqn:DC2}
  \end{align}
 which describe the immediate indirect effect of the policy via behavior, the direct effect of the policy, and the dynamic effect
 of the policy.  By recursive substitutions the dynamic effect  can be further decomposed into a weighted sum of delayed policy effects via behavior and a weighted sum of delayed policy effects via direct impact.

 
All counterfactual quantities and contrasts can be computed from the expressions given
above. For examples, given $\Delta C_{i0}>0$, new confirmed cases are linked to growth rates via relation
(taking $t$ divisible by $\ell$ for simplicity):
$$
\frac{\Delta C^{\star}_{it}}{\Delta C_{i0}} = \exp\left (  \sum_{m=1}^{t/\ell}  Y^\star_{i,m\ell} \right) \text{ and }\frac{ \Delta C_{it}}{\Delta C_{i0}} = \exp
\left (  \sum_{m=1}^{t/\ell}   Y_{i, m \ell }\right).
$$
The cumulative cases can be constructed by summing over the new cases. Various contrasts are
then calculated from these quantities. For example, the relative contrast of counterfactual new confirmed cases
to the  factual confirmed cases is given by:
$$
\Delta C^{\star}_{it}/\Delta C_{it} = \exp\left(  \sum_{m=1}^{t/\ell} ( Y^\star_{i,m\ell} -  Y_{i,m \ell}) \right).
$$
We refer to the appendix for further details. Similar calculations apply for fatalities. 
Note that our analysis is conditional on the factual history and structural stochastic 
shocks.\footnote{For unconditional counterfactual,
we need to make assumptions about the evolution of stochastic shocks appearing in $Y_{it}$. See, e.g, previous
versions of our paper in ArXiv, where unconditional counterfactuals were calculated by assuming stochastic shocks
are i.i.d. and resampling them from the empirical distribution. The differences between conditional and unconditional
contrasts were small in our empirical analysis.}

The estimated counterfactuals are smooth functionals of the underlying parameter estimates. Therefore, we
construct the confidence intervals for counterfactual quantities and contrasts by bootstrapping the parameter
estimates. We refer to the appendix for further details.

\subsection{Outcome and Key Confounders via SIRD model}\label{sec:sirmodel} We next provide details
of our key measurement equations, defining the outcomes and key confounders. We motivate the structural
outcome equations via the fundamental epidemiological model for the spread of infectious decease called
the Susceptible-Infected-Recovered-Dead (SIRD) model with testing.

Letting $C_{it}$ denote the cumulative number of  confirmed cases in state $i$ at time $t$, our outcome
\begin{equation} \label{eq:y}
{\ycolor Y_{it}} =\Delta \log(\Delta C_{it}):= \log( \Delta C_{it} ) -
\log( \Delta C_{i,t-7})
\end{equation}
approximates the weekly growth rate in new cases from $t-7$ to $t$. Here $\Delta$ denotes the differencing operator over 7 days from $t$ to $t-7$, so that $\Delta C_{it}:=C_{it}-C_{i,t-7}$ is the number of new confirmed cases in the past 7 days.

We chose this metric as this is the key metric for policy makers deciding when to relax Covid mitigation policies.  The U.S. government's guidelines for state reopening
recommend that states display a
``downward trajectory of documented cases within a 14-day period''
\citep{whitehouse2020}. A negative value of
$Y_{it}$ is an indication of meeting this
criteria for reopening. By focusing on weekly cases  rather than daily cases, we smooth idiosyncratic daily fluctuations as well as periodic fluctuations associated with days of the week.

Our measurement equation for estimating equations (\ref{eq:R1}) and (\ref{eq:R4}) will take the form:
\begin{align}
{\ycolor \Delta \log(\Delta C_{it})}  =    X_{i,t-14} '   \theta  -\gamma +  \delta_T \Delta   \log(T_{it})  + \epsilon_{it},
 \label{eq:M} \tag{M-C}
\end{align}
where $i$ is state, $t$ is day, $C_{it}$ is cumulative confirmed
cases, $T_{it}$ is the number of tests over 7 days, $\Delta$ is
a 7-days differencing operator, and $\epsilon_{it}$ is an unobserved error term.
 $X_{i,t-14}$  collects other behavioral, policy, and confounding variables, depending
on whether we estimate (\ref{eq:R1}) or (\ref{eq:R4}), where the lag of $14$ days captures the time lag between infection and confirmed case (see the Appendix \ref{sec:timing}). 
   Here
$$\Delta   \log(T_{it} ):=  \log(T_{it}) - \log(T_{i,t-7})  $$ 
is the key confounding variable,
derived from considering the SIRD model below. We are treating the change in testing
rate as exogenous.\footnote{To check sensitivity to this assumption
we performed robustness checks, where we used
the further lag of $\Delta   \log(T_{it} )$ as a proxy for exogenous change in the testing rate, and we also
used that as an instrument for $\Delta   \log(T_{it} )$; this did not affect the results on policy effects, although the instrument
was not sufficiently strong.} We describe other confounders in the empirical section.


 Our main measurement equation (\ref{eq:M}) is motivated by a variant of SIRD
model, where we add confirmed cases and infection detection via testing.
Let $S$, $\Infected$, $\Recovered$, and $D$ denote the number of susceptible,
infected, recovered, and deceased individuals in a given state. Each of these variables are a function of time. We model
them as evolving as
\begin{align}
  \dot{S}(t) & = -\frac{S(t)}{N} \beta(t) \Infected(t) \label{eq:s} \\
  \dot{\Infected}(t) & = \frac{S(t)}{N} \beta(t) \Infected(t) - \gamma  \Infected(t) \label{eq:i}\\
  \dot{\Recovered}(t) & = (1-\kappa) \gamma  \Infected(t) \label{eq:r}\\ 
  \dot{D}(t) & = \kappa \gamma \Infected(t) 
  \label{eq:d}
\end{align}
where $N$ is the population, $\beta(t)$ is the rate of infection
spread, $\gamma$ is the rate of recovery or death, and $\kappa$ is the
probability of death conditional on infection.

Confirmed cases, $C(t)$, evolve as
\begin{equation}
  \dot{C}(t) = \tau(t) \Infected(t), \label{eq:c}
\end{equation}
where $\tau(t)$ is the rate that infections are detected.

Our goal is to examine how the rate of infection $\beta(t)$ varies with observed policies
and measures of social distancing behavior. A key challenge is that we only
observed $C(t)$ and $D(t)$, but not $\Infected(t)$. The unobserved $\Infected(t)$ can
be eliminated by differentiating (\ref{eq:c}) and using (\ref{eq:i})  as
\begin{align}
  \frac{\ddot{C}(t)}{\dot{C}(t)}
              & =
                \frac{S(t)}{N} \beta(t) -\gamma  + \frac{\dot{\tau}(t)}{\tau(t)}. \label{eq:c2}
\end{align}
We consider a discrete-time analogue of equation (\ref{eq:c2}) to motivate our empirical
specification by relating the detection rate $\tau(t)$  to the number of tests $T_{it}$ while specifying $\frac{S(t)}{N}\beta(t)$ as a linear function of variables $X_{i,t-14}$.
This results in
\begin{align}
  \underbracket{\Delta \log(\Delta C_{it})}_{\frac{\ddot{C}(t)}{\dot{C}(t)}}
  =
      \underbracket{X_{i,t-14}' \theta + \epsilon_{it}}_{\frac{S(t)}{N}\beta(t) -\gamma}
       +
       & \underbracket{\delta_T \Delta
      \log(T)_{it}}_{\frac{\dot{\tau}(t)}{\tau(t)} } \nonumber
\end{align}
which is equation (\ref{eq:M}), where $X_{i,t-14}$ captures a vector of variables related to $\beta(t)$.

\begin{quote}
\textsc{Structural Interpretation}. Early in the pandemic, when
the number of susceptibles is approximately the same as the entire population, i.e. $S_{it}/N_{it} \approx 1$, the component $X_{i,t-14}' \theta$
is the projection of infection rate $ \beta_i(t)$ on   $X_{i,t-14}$ (policy, behavioral, information, and confounders other than testing rate), provided
the stochastic component $\epsilon_{it}$ is orthogonal
to $X_{i,t-14}$ and the testing variables:
$$
\beta_i(t)S_{it}/N_{it}  - \gamma = X_{i,t-14}' \theta + \epsilon_{it}, \quad \epsilon_{it} \perp X_{i,t-14}.
$$
\end{quote}

The specification for growth rate in deaths as the outcome  is motivated by SIRD as follows. By differentiating (\ref{eq:d}) and (\ref{eq:c}) with respect to $t$ and using (\ref{eq:c2}), we obtain
\begin{align}
\frac{\ddot{D}(t) }{\dot D(t)}& = \frac{\ddot{C}(t) }{\dot C(t)}  - \frac{\dot{\tau}(t) }{ \tau(t)}    =  \frac{S(t)}{N}\beta(t)  -   \gamma.\label{eq:d2}
\end{align}
Our measurement equation for the growth rate of deaths is based on equation (\ref{eq:d2}) but   accounts for a $21$ day lag between infection and death as
\begin{align}
{\ycolor \Delta \log(\Delta D_{it})}  = X_{i,t-21}' \theta + \epsilon_{it},\label{eq:M-D} \tag{M-D}
\end{align}
where
\begin{equation} \label{eq:y-d}
 \Delta \log(\Delta D_{it}):= \log( \Delta D_{it} ) -
\log( \Delta D_{i,t-7})
\end{equation}
approximates the weekly growth rate in deaths from $t-7$ to $t$ in state $i$. 

\section{Empirical Analysis}
\subsection{Data}

Our baseline measures for daily Covid-19   cases and deaths are from
\href{https://github.com/nytimes/covid-19-data}{The New York Times (NYT)}. When there are missing values in NYT, we use reported cases and deaths from \href{https://github.com/CSSEGISandData/COVID-19}{JHU CSSE}, and then the
\href{https://github.com/COVID19Tracking/covid-tracking-data}{Covid
Tracking Project}.  The number of tests for each state is from  \href{https://github.com/COVID19Tracking/covid-tracking-data}{Covid
 Tracking Project}. As shown in the lower right panel of Figure \ref{fig:casedeathtest} in the appendix, there was a rapid increase in testing in the second half of March and then the
number of tests increased very slowly in each state in April.

We use the database on US state policies created by
\cite{raifman2020}.
In our analysis, we focus on 6
policies:  stay-at-home, closed nonessential
businesses, closed K-12 schools, closed restaurants except takeout, closed movie theaters, and face mask mandates for employees in public facing businesses.
We believe that the first four of these
policies are the most widespread and important.
Closed movie theaters is included because it captures common bans on gatherings of more than
a handful of people. We also include mandatory face mask use by employees because its effectiveness on slowing down Covid-19 spread is a controversial policy issue \citep{howard2020,Greenhalghm2020,zhangr2020}.
Table \ref{tab:policies_inreg} provides summary statistics, where $N$ is the number of states that have ever implemented the policy. We also obtain information on state-level covariates mostly from \citet{raifman2020}, which include population size, total area, unemployment rate, poverty rate,  a percentage of people who are subject to illness, and state governor's party affiliations.  These confounders are motivated by \cite{wheaton2020} who find that case growth is associated  with residential density and per capita income.

\begin{table}[ht]
\centering
\begin{tabular}{rllll}
  \hline
 & N & Min & Median & Max \\ 
  \hline
Date closed K 12 schools & 51 & 2020-03-13 & 2020-03-17 & 2020-04-03 \\ 
  Stay at home  shelter in place & 42 & 2020-03-19 & 2020-03-28 & 2020-04-07 \\ 
  Closed movie theaters & 49 & 2020-03-16 & 2020-03-21 & 2020-04-06 \\ 
  Closed restaurants except take out & 48 & 2020-03-15 & 2020-03-17 & 2020-04-03 \\ 
  Closed non essential businesses & 43 & 2020-03-19 & 2020-03-25 & 2020-04-06 \\ 
  Mandate face mask use by employees & 44 & 2020-04-03 & 2020-05-01 & 2020-08-03 \\ 
   \hline
\end{tabular}
\caption{State Policies \label{tab:policies_inreg}} 
\end{table}

We obtain social distancing behavior measures from``\href{https://www.google.com/covid19/mobility/}{Google COVID-19 Community Mobility Reports}''
\citep{google2020}.  The dataset provides six measures of ``mobility trends''  that report a percentage change in visits and length of stay at different places relative to a baseline computed by their median values of the same day of the week from January 3 to February 6, 2020.  Our analysis focuses on the following four measures: ``Grocery \& pharmacy," ``Transit stations,'' ``Retail \& recreation,'' and ``Workplaces.''\footnote{The other two measures are ``Residential'' and  ``Parks.''  We drop ``Residential'' because it is highly correlated with  ``Workplaces'' and  ``Retail \& recreation''  at correlation coefficients of -0.98 and -0.97, respectively. We also drop ``Parks'' because it does not have clear implication on the spread of Covid-19.}

\begin{figure}\caption{The Evolution of Google Mobility Measures: Transit stations and Workplaces\label{fig:transit-workplaces}}\vspace{0.1cm}
    \begin{tabular}{cc}
      \includegraphics[width=0.5\linewidth]{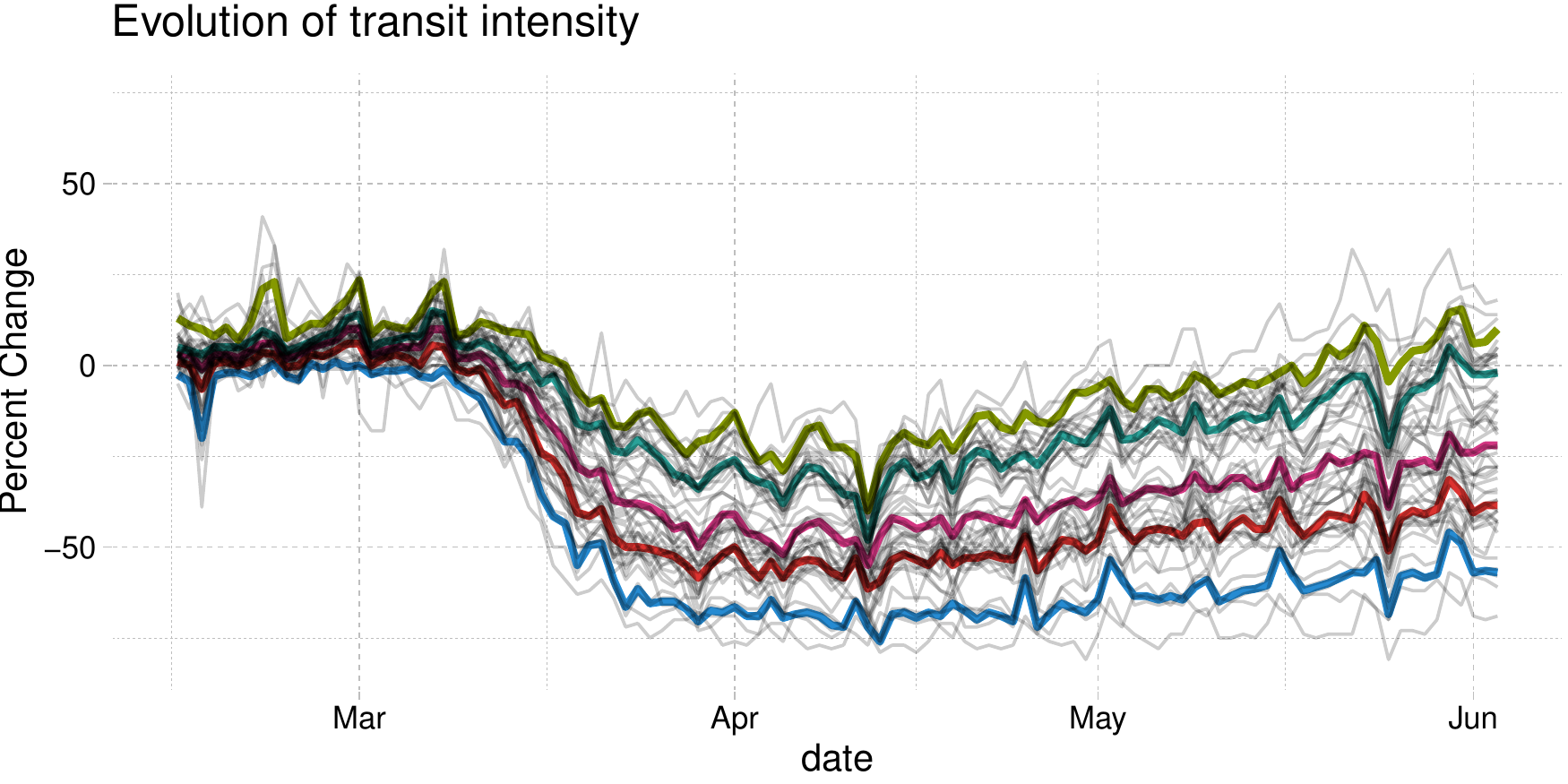}
      \includegraphics[width=0.5\linewidth]{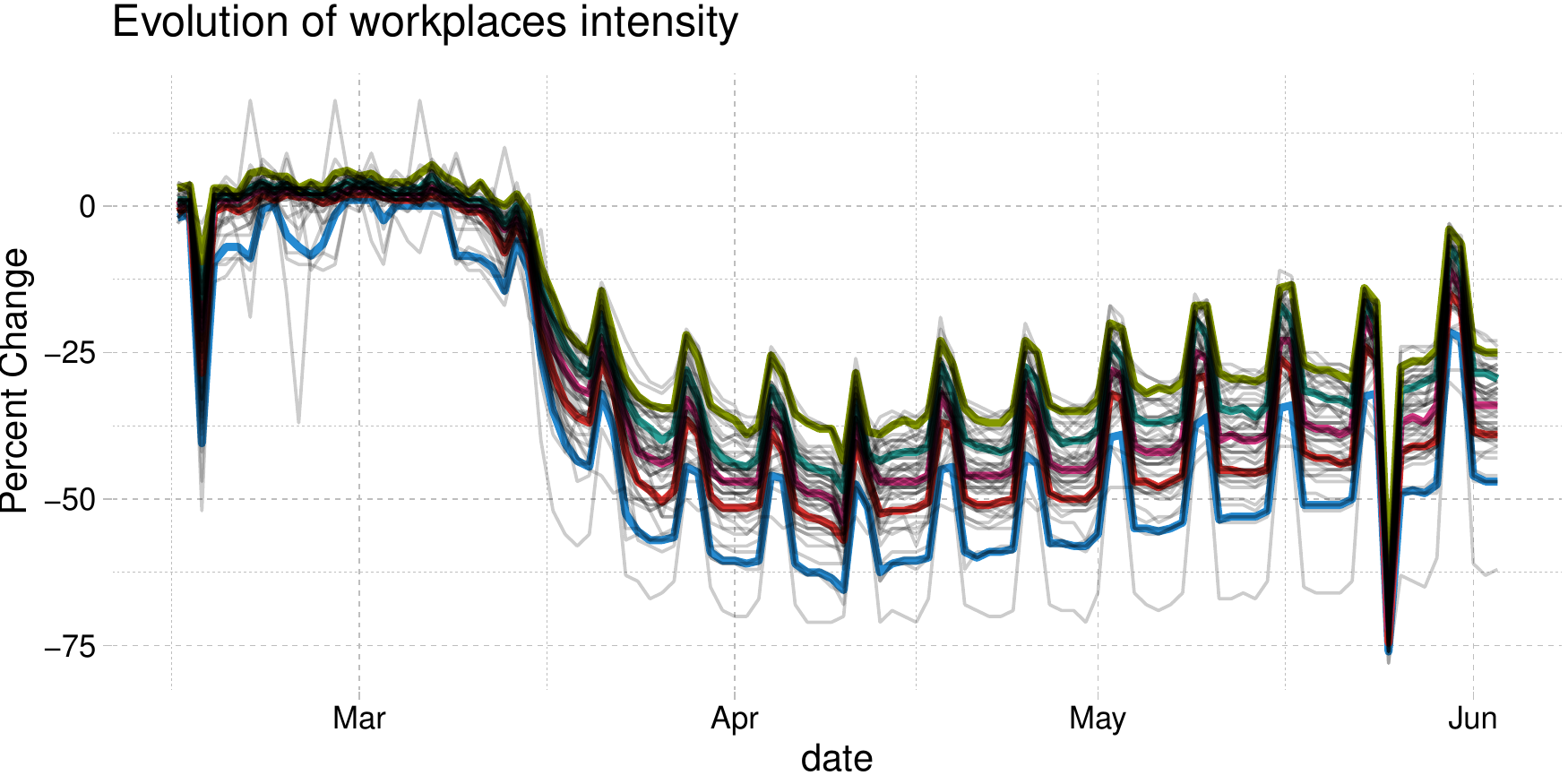}
       \end{tabular}
        \begin{flushleft}
        \scriptsize{This figure shows the evolution of ``Transit stations'' and  ``Workplaces'' of Google Mobility
      Reports. Thin gray lines are the value in each state and date. Thicker colored lines are
      quantiles of the variables conditional on date.}
      \end{flushleft}
\end{figure}

Figure \ref{fig:transit-workplaces} shows the evolution of  ``Transit stations'' and ``Workplaces,'' where thin lines are the value in each state and date while thicker colored lines are quantiles  conditional on date. The figures illustrate a sharp decline in people's movements starting from mid-March as well as differences in their evolutions across states. They also reveal periodic fluctuations associated with days of the week, which motivates our use of weekly measures.

In our empirical analysis, we use weekly measures for cases, deaths,
and tests by summing up their daily measures from day \(t\) to
\(t-6\).  We focus on weekly cases and deaths because daily new cases and deaths are
affected by the timing of reporting and testing and are quite
volatile as shown in the upper right panel of Figure \ref{fig:casedeathtest} in the
appendix.  Aggregating to weekly new cases/deaths/tests smooths
out idiosyncratic daily noises as well as periodic fluctuations
associated with days of the week.
We also construct weekly policy and
behavior variables by taking 7 day moving averages from day \(t-14\) to
\(t-21\) for case growth, where the  delay reflects the time lag between infection and case confirmation. The four weekly behavior variables are referred  to ``Transit
Intensity,'' ``Workplace Intensity,'' ``Retail Intensity,'' and ``Grocery
Intensity.'' Consequently, our empirical analysis uses 7 day moving averages of all variables recorded at daily frequencies. Our sample period is from March 7, 2020 to June 3, 2020.

\begin{table}\caption{Correlations among Policies and Behavior \label{tab:correlation}}\vspace{-0.2cm}
  \begin{minipage}{\linewidth}
    \resizebox{\linewidth}{!}{
      
\begin{tabular}{lccccccccccc}
\toprule
\rotatebox{90}{ } & \rotatebox{90}{workplaces} & \rotatebox{90}{retail} & \rotatebox{90}{grocery} & \rotatebox{90}{transit} & \rotatebox{90}{masks for employees} & \rotatebox{90}{closed K-12 schools} & \rotatebox{90}{stay at home} & \rotatebox{90}{closed movie theaters} & \rotatebox{90}{closed restaurants} & \rotatebox{90}{closed non-essent bus} & \rotatebox{90}{business closure policies}\\
\midrule
workplaces & 1.00 &  &  &  &  &  &  &  &  &  & \\
retail & 0.93 & 1.00 &  &  &  &  &  &  &  &  & \\
grocery & 0.75 & 0.83 & 1.00 &  &  &  &  &  &  &  & \\
transit & 0.89 & 0.92 & 0.83 & 1.00 &  &  &  &  &  &  & \\
masks for employees & -0.32 & -0.17 & -0.15 & -0.29 & 1.00 &  &  &  &  &  & \\
\addlinespace
closed K-12 schools & -0.91 & -0.79 & -0.55 & -0.72 & 0.43 & 1.00 &  &  &  &  & \\
stay at home & -0.69 & -0.69 & -0.70 & -0.71 & 0.28 & 0.62 & 1.00 &  &  &  & \\
closed movie theaters & -0.81 & -0.76 & -0.64 & -0.71 & 0.34 & 0.82 & 0.72 & 1.00 &  &  & \\
closed restaurants & -0.77 & -0.82 & -0.68 & -0.76 & 0.21 & 0.74 & 0.72 & 0.82 & 1.00 &  & \\
closed non-essent bus & -0.65 & -0.68 & -0.68 & -0.64 & 0.08 & 0.56 & 0.76 & 0.68 & 0.71 & 1.00 & \\
\addlinespace
business closure policies & -0.84 & -0.84 & -0.75 & -0.79 & 0.24 & 0.78 & 0.81 & 0.92 & 0.93 & 0.87 & 1.00\\
\bottomrule
\end{tabular}
    }\smallskip
       \begin{flushleft}
         \scriptsize
         Each off-diagonal entry reports a correlation coefficient of
         a pair of policy and behavior variables. ``business closure policies'' is defined by the average of  closured movie theaters, closured restaurants, and closured non-essential businesses.
       \end{flushleft}
  \end{minipage}
\end{table}

Table \ref{tab:correlation} reports that weekly policy and behavior
variables are highly correlated with each other, except for the``masks
for employees'' policy.  High correlations may cause multicolinearity
problems and could limit our ability to separately identify the
effect of each policy or behavior variable on case growth.
For this reason, we define  the ``business closure policies'' variable by the average of closed movie theaters, closed  restaurants, and closed non-essential businesses variables and consider a specification that includes business closure policies   in place of these three policy variables separately.

\begin{figure}[ht]\caption{Portion of states with each
    policy \label{fig:policyportion}}
  \begin{minipage}{\linewidth}
    \begin{tabular}{ccc}
      \includegraphics[width=0.31\textwidth]{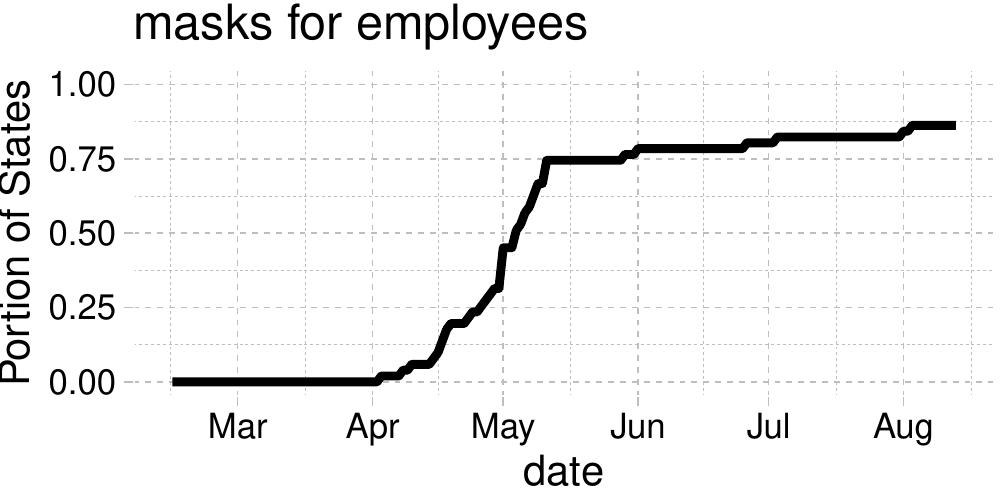}
      &
        \includegraphics[width=0.31\textwidth]{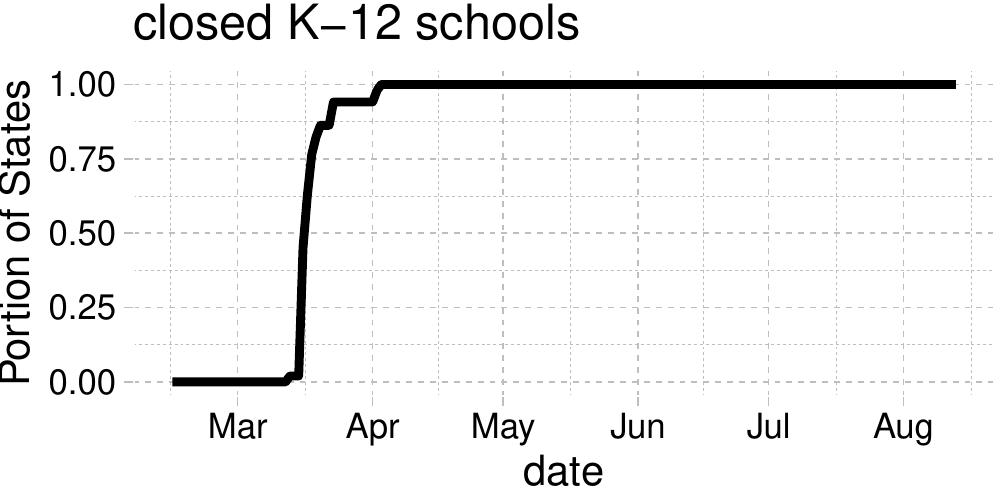}
      &
        \includegraphics[width=0.31\textwidth]{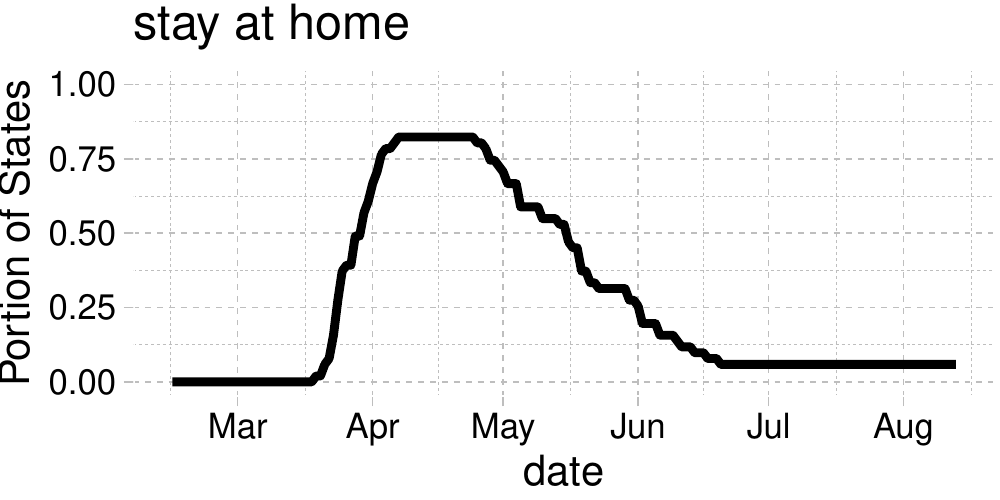}
      \\
      \includegraphics[width=0.31\textwidth]{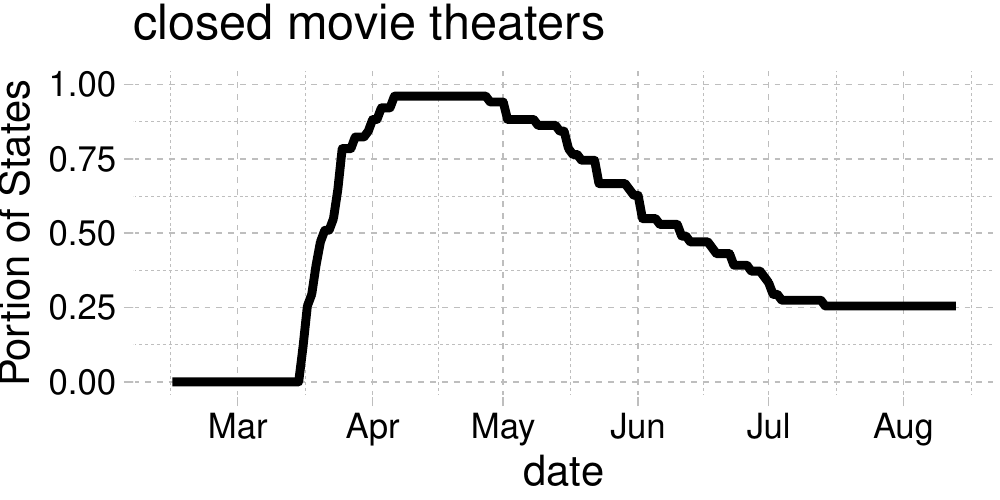}
      &
        \includegraphics[width=0.31\textwidth]{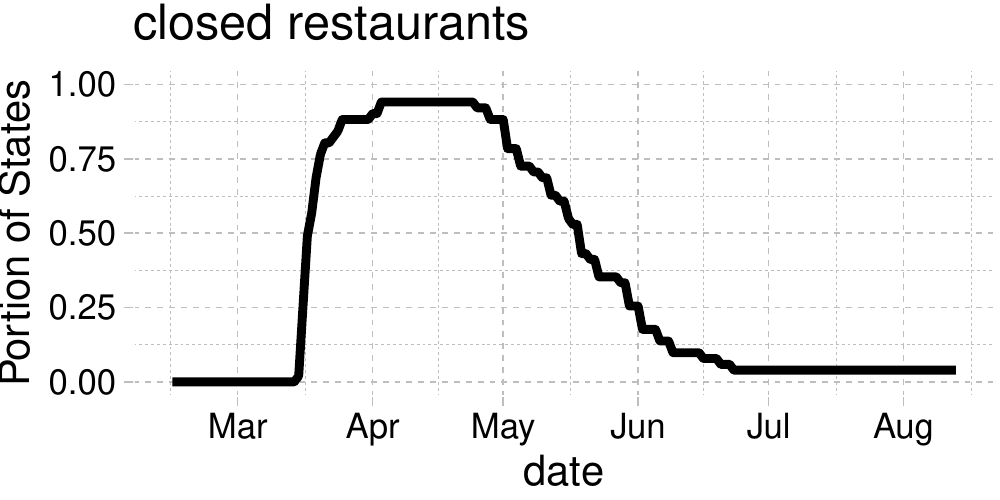}
      &
        \includegraphics[width=0.31\textwidth]{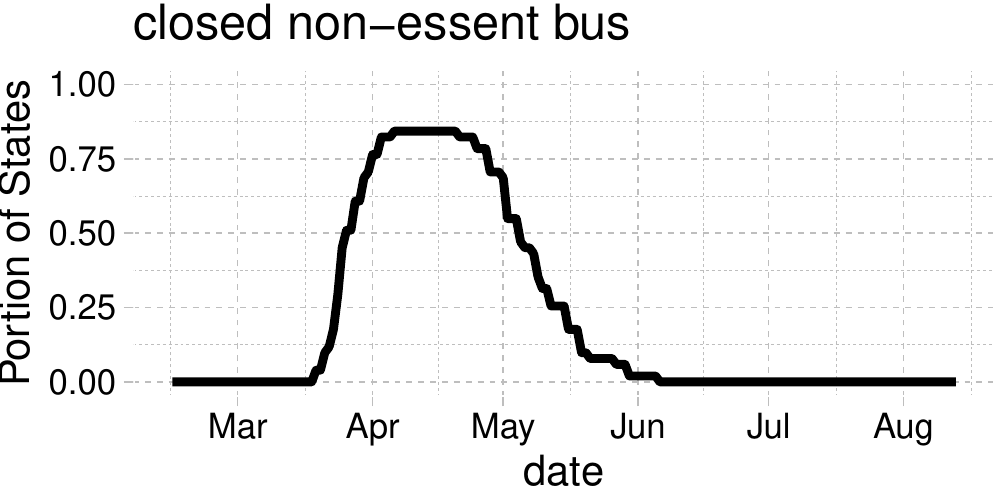}
    \end{tabular}
  \end{minipage}
\end{figure}

Figure \ref{fig:policyportion} shows the portion of states that have
each policy in place at each date. For most policies, there is
considerable variation across states in the time in which the policies
are active. The one exception is K-12 school closures. About 80\% of
states closed schools within a day or two of March 15th, and all
states closed schools by early April. This makes the effect of school
closings difficult to separate from aggregate time series variation.

\subsection{The Effect of Policies and Information on Behavior\label{policies-and-behavior}}

We first examine how policies and information affect social distancing behaviors by estimating a version of (\ref{eq:R2}):
\begin{align}
  {\bcolor B_{it}^j}
  & = {\pcolor (\beta^j)' P_{it}} + {\icolor (\gamma^j)' I_{it}} +
    {\wcolor (\delta_B^j)' W_{it}} + \varepsilon_{it}^{bj}, \notag
\end{align}
where ${\bcolor B_{it}^j}$ represents behavior variable $j$  in state  $i$ at time $t$.
${\pcolor P_{it}}$ collects the Covid related policies  in state $i$ at $t$.
Confounders, ${\wcolor W_{it}}$, include state-level covariates, month
indicators, and their interactions.
${\icolor I_{it}}$ is a set of information variables that affect
people's behaviors at $t$. As information, we include each state's
growth of cases (in panel A) or deaths (in panel B), and log cases or deaths.  Additionally, in
columns (5)-(8) of Table \ref{tab:PItoB}, we include national growth and log
of cases or deaths.

\afterpage{\clearpage
  \begin{landscape}
    \begin{table}[!htbp] \centering
      \caption{The Effect of Policies and Information on Behavior ($P
        I \to B$)}\vspace{-0.3cm}\label{tab:PItoB}

      \begin{subtable}{\linewidth}\caption{Cases as Information\label{tab:PItoBcases}}
      \begin{minipage}{\linewidth}
        \centering
        \resizebox{\textwidth}{!}{
          \footnotesize
          \begin{tabular}{@{\extracolsep{1pt}}lcccccccc} 
\\[-1.8ex]\hline 
\hline \\[-1.8ex] 
 & \multicolumn{8}{c}{\textit{Dependent variable:}} \\ 
\cline{2-9} 
 & workplaces & retail & grocery & transit & workplaces & retail & grocery & transit \\ 
\\[-1.8ex] & (1) & (2) & (3) & (4) & (5) & (6) & (7) & (8)\\ 
\hline \\[-1.8ex] 
 masks for employees & 0.023$^{*}$ & 0.033$^{*}$ & 0.012 & 0.015 & 0.005 & 0.0005 & 0.003 & $-$0.010 \\ 
  & (0.012) & (0.020) & (0.012) & (0.025) & (0.008) & (0.015) & (0.011) & (0.023) \\ 
  closed K-12 schools & $-$0.196$^{***}$ & $-$0.253$^{***}$ & $-$0.143$^{***}$ & $-$0.243$^{***}$ & $-$0.044$^{***}$ & $-$0.025 & $-$0.069$^{**}$ & $-$0.047 \\ 
  & (0.030) & (0.048) & (0.027) & (0.050) & (0.013) & (0.018) & (0.027) & (0.041) \\ 
  stay at home & $-$0.028$^{**}$ & $-$0.024 & $-$0.068$^{***}$ & $-$0.062$^{**}$ & $-$0.034$^{***}$ & $-$0.047$^{***}$ & $-$0.071$^{***}$ & $-$0.074$^{***}$ \\ 
  & (0.013) & (0.017) & (0.015) & (0.028) & (0.011) & (0.013) & (0.015) & (0.028) \\ 
  business closure policies & $-$0.081$^{***}$ & $-$0.136$^{***}$ & $-$0.088$^{***}$ & $-$0.080$^{**}$ & $-$0.049$^{***}$ & $-$0.094$^{***}$ & $-$0.073$^{***}$ & $-$0.042 \\ 
  & (0.017) & (0.028) & (0.017) & (0.038) & (0.012) & (0.020) & (0.018) & (0.036) \\ 
  $\Delta \log \Delta C_{it}$ & 0.015$^{***}$ & $-$0.002 & 0.016$^{***}$ & 0.014$^{***}$ & 0.017$^{***}$ & 0.014$^{***}$ & 0.018$^{***}$ & 0.020$^{***}$ \\ 
  & (0.003) & (0.005) & (0.004) & (0.005) & (0.002) & (0.004) & (0.004) & (0.005) \\ 
  $\log \Delta C_{it}$ & $-$0.024$^{***}$ & $-$0.022$^{***}$ & $-$0.0002 & $-$0.018$^{*}$ & $-$0.005 & $-$0.001 & 0.009 & 0.004 \\ 
  & (0.005) & (0.008) & (0.005) & (0.010) & (0.004) & (0.007) & (0.006) & (0.011) \\ 
  $\Delta \log \Delta C_{it}$.national &  &  &  &  & $-$0.033$^{***}$ & $-$0.086$^{***}$ & $-$0.018$^{**}$ & $-$0.053$^{***}$ \\ 
  &  &  &  &  & (0.005) & (0.008) & (0.008) & (0.012) \\ 
  $\log \Delta C_{it}$.national &  &  &  &  & $-$0.072$^{***}$ & $-$0.103$^{***}$ & $-$0.035$^{***}$ & $-$0.091$^{***}$ \\ 
  &  &  &  &  & (0.004) & (0.008) & (0.008) & (0.012) \\ 
 \hline \\[-1.8ex] 
state variables & Yes & Yes & Yes & Yes & Yes & Yes & Yes & Yes \\ 
Month $\times$ state variables & Yes & Yes & Yes & Yes & Yes & Yes & Yes & Yes \\ 
\hline \\[-1.8ex] 
$\sum_j \mathrm{Policy}_j$ & -0.282$^{***}$ & -0.380$^{***}$ & -0.287$^{***}$ & -0.371$^{***}$ & -0.122$^{***}$ & -0.166$^{***}$ & -0.211$^{***}$ & -0.172$^{***}$ \\ 
 & (0.041) & (0.066) & (0.040) & (0.078) & (0.022) & (0.035) & (0.037) & (0.060) \\ 
Observations & 3,825 & 3,825 & 3,825 & 3,825 & 3,825 & 3,825 & 3,825 & 3,825 \\ 
R$^{2}$ & 0.913 & 0.829 & 0.749 & 0.818 & 0.953 & 0.906 & 0.765 & 0.856 \\ 
Adjusted R$^{2}$ & 0.912 & 0.829 & 0.748 & 0.817 & 0.952 & 0.905 & 0.764 & 0.855 \\ 
\hline 
\hline \\[-1.8ex] 
\textit{Note:}  & \multicolumn{8}{r}{$^{*}$p$<$0.1; $^{**}$p$<$0.05; $^{***}$p$<$0.01} \\ 
\end{tabular} 
        }
      \begin{flushleft}
        \scriptsize
        Dependent variables are   ``Transit
        Intensity,'' ``Workplace Intensity,'' ``Retail Intensity,'' and ``Grocery
        Intensity" defined as 7 days moving averages of corresponding daily measures obtained from Google Mobility Reports.  All specifications include state-level characteristics (population, area, unemployment rate, poverty rate,  a percentage of people subject to illness, and governor's party) as well as their interactions with the log of days since Jan 15, 2020. The row ``$\sum_j \text{Policy}_j$'' reports the sum of four policy coefficients. Standard errors are clustered at the state level.
        \end{flushleft}
      \end{minipage}
    \end{subtable}
  \end{table}
  \pagebreak
  \begin{table}[!htbp] \centering
    \ContinuedFloat
    \begin{subtable}{\linewidth}\caption{Deaths as Information \label{tab:PItoBdeaths}}
      \begin{minipage}{\linewidth}
        \centering
        \resizebox{\textwidth}{!}{
          \footnotesize
            \begin{tabular}{@{\extracolsep{1pt}}lcccccccc} 
\\[-1.8ex]\hline 
\hline \\[-1.8ex] 
 & \multicolumn{8}{c}{\textit{Dependent variable:}} \\ 
\cline{2-9} 
 & workplaces & retail & grocery & transit & workplaces & retail & grocery & transit \\ 
\\[-1.8ex] & (1) & (2) & (3) & (4) & (5) & (6) & (7) & (8)\\ 
\hline \\[-1.8ex] 
 masks for employees & 0.018 & 0.027 & 0.006 & 0.006 & 0.007 & 0.012 & 0.002 & $-$0.005 \\ 
  & (0.011) & (0.017) & (0.014) & (0.026) & (0.009) & (0.016) & (0.012) & (0.025) \\ 
  closed K-12 schools & $-$0.226$^{***}$ & $-$0.261$^{***}$ & $-$0.114$^{***}$ & $-$0.240$^{***}$ & $-$0.045$^{***}$ & $-$0.033$^{*}$ & $-$0.024 & $-$0.050 \\ 
  & (0.024) & (0.033) & (0.018) & (0.035) & (0.014) & (0.019) & (0.020) & (0.039) \\ 
  stay at home & $-$0.031$^{**}$ & $-$0.035$^{**}$ & $-$0.082$^{***}$ & $-$0.072$^{**}$ & $-$0.024$^{*}$ & $-$0.032$^{**}$ & $-$0.073$^{***}$ & $-$0.066$^{**}$ \\ 
  & (0.012) & (0.017) & (0.016) & (0.031) & (0.012) & (0.015) & (0.017) & (0.032) \\ 
  business closure policies & $-$0.101$^{***}$ & $-$0.156$^{***}$ & $-$0.101$^{***}$ & $-$0.111$^{***}$ & $-$0.044$^{***}$ & $-$0.091$^{***}$ & $-$0.067$^{***}$ & $-$0.054 \\ 
  & (0.022) & (0.033) & (0.020) & (0.041) & (0.012) & (0.021) & (0.020) & (0.039) \\ 
  $\Delta \log \Delta D_{it}$ & $-$0.012$^{***}$ & $-$0.031$^{***}$ & $-$0.001 & $-$0.021$^{***}$ & $-$0.003 & $-$0.012$^{***}$ & $-$0.002 & $-$0.009$^{*}$ \\ 
  & (0.005) & (0.006) & (0.004) & (0.006) & (0.002) & (0.004) & (0.004) & (0.005) \\ 
  $\log \Delta D_{it}$ & $-$0.013$^{***}$ & $-$0.001 & $-$0.005 & $-$0.006 & $-$0.007 & 0.003 & 0.002 & $-$0.0003 \\ 
  & (0.004) & (0.006) & (0.005) & (0.009) & (0.004) & (0.007) & (0.005) & (0.010) \\ 
  $\Delta \log \Delta D_{it}$.national &  &  &  &  & $-$0.058$^{***}$ & $-$0.095$^{***}$ & $-$0.008 & $-$0.067$^{***}$ \\ 
  &  &  &  &  & (0.005) & (0.008) & (0.006) & (0.010) \\ 
  $\log \Delta D_{it}$.national &  &  &  &  & $-$0.056$^{***}$ & $-$0.065$^{***}$ & $-$0.033$^{***}$ & $-$0.057$^{***}$ \\ 
  &  &  &  &  & (0.004) & (0.007) & (0.006) & (0.010) \\ 
 \hline \\[-1.8ex] 
state variables & Yes & Yes & Yes & Yes & Yes & Yes & Yes & Yes \\ 
Month $\times$ state variables & Yes & Yes & Yes & Yes & Yes & Yes & Yes & Yes \\ 
\hline \\[-1.8ex] 
$\sum_j \mathrm{Policy}_j$ & -0.340$^{***}$ & -0.425$^{***}$ & -0.292$^{***}$ & -0.417$^{***}$ & -0.105$^{***}$ & -0.144$^{***}$ & -0.162$^{***}$ & -0.176$^{***}$ \\ 
 & (0.026) & (0.036) & (0.033) & (0.060) & (0.024) & (0.039) & (0.034) & (0.068) \\ 
Observations & 3,468 & 3,468 & 3,468 & 3,468 & 3,468 & 3,468 & 3,468 & 3,468 \\ 
R$^{2}$ & 0.908 & 0.836 & 0.760 & 0.826 & 0.953 & 0.899 & 0.783 & 0.856 \\ 
Adjusted R$^{2}$ & 0.907 & 0.835 & 0.759 & 0.825 & 0.953 & 0.898 & 0.782 & 0.855 \\ 
\hline 
\hline \\[-1.8ex] 
\textit{Note:}  & \multicolumn{8}{r}{$^{*}$p$<$0.1; $^{**}$p$<$0.05; $^{***}$p$<$0.01} \\ 
\end{tabular} 
          }
        \begin{flushleft}
          \scriptsize
          Dependent variables are   ``Transit
          Intensity,'' ``Workplace Intensity,'' ``Retail Intensity,'' and ``Grocery
          Intensity" defined as 7 days moving averages of corresponding daily measures obtained from Google Mobility Reports.  All specifications include state-level characteristics (population, area, unemployment rate, poverty rate,  a percentage of people subject to illness, and governor's party) as well as their interactions with the log of days since Jan 15, 2020. The row ``$\sum_j \text{Policy}_j$'' reports the sum of four policy coefficients. Standard errors are clustered at the state level.
        \end{flushleft}
      \end{minipage}
    \end{subtable}
  \end{table}
\end{landscape}
\clearpage}

Table \ref{tab:PItoB} reports the estimates  with standard errors clustered at the state level.  Across different
specifications, our results imply that policies have large effects on
behavior. Comparing columns (1)-(4) with columns (5)-(8),
the magnitude of policy effects are sensitive to whether national
cases or deaths are included as information. The coefficient on school
closures is particularly sensitive to the inclusion of national
information variables.  As shown in Figure
\ref{fig:policyportion}, there is little variation across states in
the timing of school closures. Consequently, it is difficult to
separate the effect of school closures from a behavioral response to the national
trend in cases and deaths.

The other policy coefficients are less sensitive to the inclusion of
national case/death variables. After school closures, business closure policies have the next largest effect followed by stay-at-home
orders.  The effect of masks for employees is
small.\footnote{Similar to our finding, \cite{kovacs2020} find no evidence that introduction of compulsory face mask policies affect community mobility in Germany.}


The row ``$\sum_j \mathrm{Policy}_j$'' reports the sum of the estimated
effect of all policies, which is substantial and can account for a
large fraction of the observed declines in behavior variables.  For
example,  in Figure \ref{fig:transit-workplaces}, transit intensity for a median state was approximately -50\% at its lowest point
 in early April. The estimated policy coefficients in columns imply
that imposing all policies would lead to roughly 75\% (in column
4) or roughly 35\% (in column 8) of the observed decline. The large impact of policies on transit intensity suggests that the policies may have reduced the Covid-19 infection by reducing people's use of public transportation.\footnote{Analyzing the New York City's subway ridership,  \cite{NBERw27021} finds  a strong link between public transit and spread of infection.}


In Table \ref{tab:PItoB}(B), estimated coefficients of deaths and
death growth are generally negative. This suggests that the higher
number of deaths reduces social interactions measured by Google
Mobility Reports perhaps because people are increasingly aware of
prevalence of Covid-19 \citep{maloney2020}.  The coefficients on log
cases and case growth in Table \ref{tab:PItoB}(A) are more mixed.\footnote{Rewrite a regression specification after omitting other variables as
$B_{it} =  \gamma_1 \Delta \log\Delta C_{it} + \gamma_2   \log\Delta C_{it} =  (\gamma_1 +\gamma_2) \log\Delta C_{it} - \gamma_1   \log\Delta C_{i,t-7}$.
In columns (1)-(4) of Table \ref{tab:PItoB}(A),  the estimated values of both $(\gamma_1 +\gamma_2)$ and $-\gamma_1$ are negative except for grocery. This suggests that  a higher \textit{level} of confirmed cases reduces people's mobility in workplaces, retails, and transit. For grocery, the positive estimated coefficient of $(\gamma_1 +\gamma_2)$ may reflect stock-piling behavior in early pandemic periods.} In
columns (5)-(8) of both panels, we see that national case/death
variables have large, negative coefficients. This suggests that behavior responded to national conditions although it is also likely that national case/death variables capture unobserved aggregate time effects beyond information which are not fully controlled by month dummies (e.g., latent policy variables and time-varying confounders that are common across states).


  
\begin{figure}[ht]
  \caption{Case and death growth conditional on mask mandates \label{fig:masks}}\vspace{0.2cm}
 \begin{minipage}{\linewidth}
    \centering
    \begin{tabular}{cc}
      \includegraphics[width=0.5\textwidth]{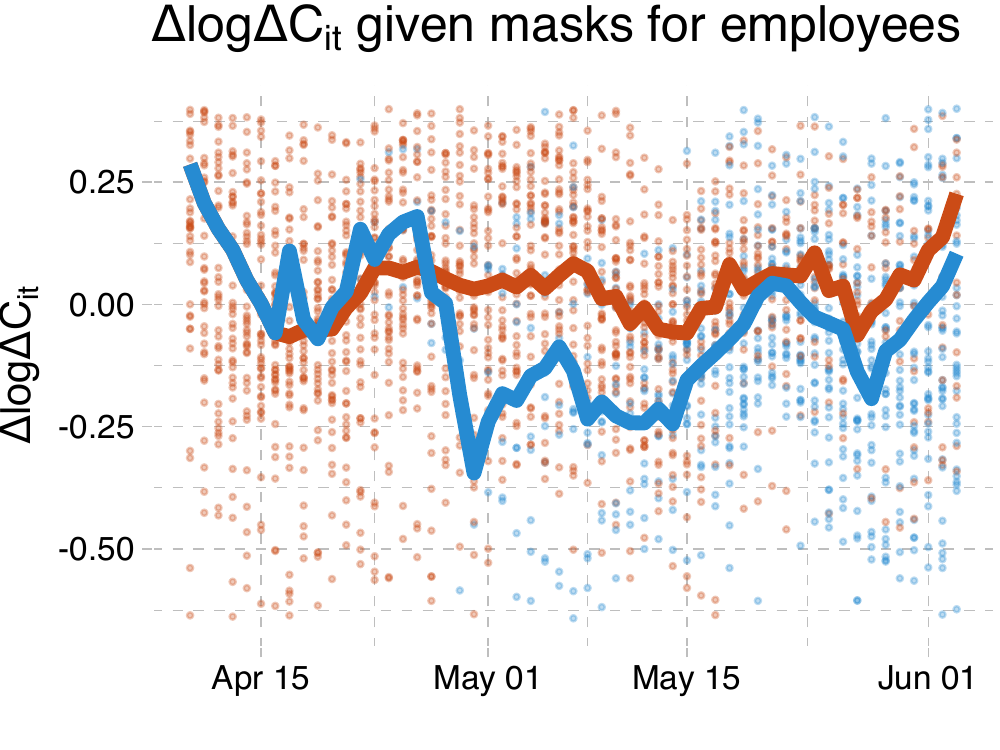}
      &
        \includegraphics[width=0.5\textwidth]{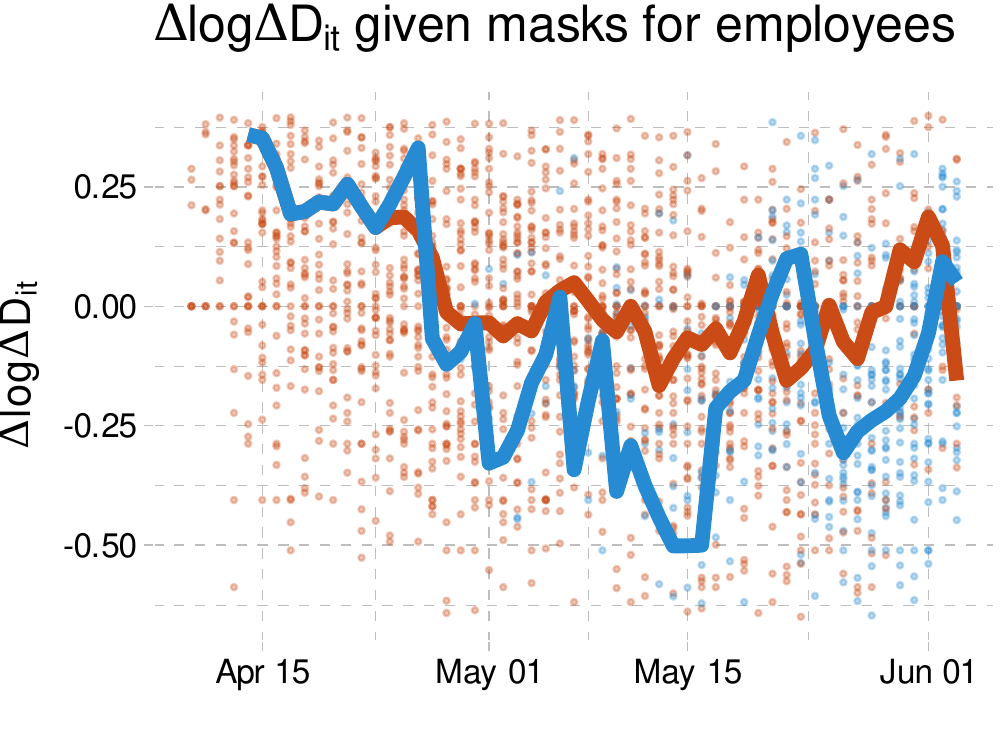}
    \end{tabular}
    \begin{flushleft}
      \scriptsize In these figures, red points are the case or death
      growth rate in states without a mask mandate. Blue points are
      states with a mask mandate 14 (21 for deaths) days prior. The
      red line is the average across states without a mask mandate 14
      (21 for deaths) days earlier. The blue line is the average
      across states with a mask mandate 14 (21 for deaths) earlier.
    \end{flushleft}
  \end{minipage}
\end{figure}

\subsection{The Direct Effect of Policies and Behavior on Case
and Death Growth\label{policy-behavior-and-case-growth}}

We now analyze how behavior and policies together influence case and death
growth rates. We begin with some simple graphical evidence of the
effect of policies on case and death growth. Figure \ref{fig:masks}
shows average case and death growth conditional on date and whether
masks are mandatory for employees.\footnote{We take 14 and 21 day lags of mask policies for case and death growths, respectively, to  identify the states with a mask mandate because policies affect cases and deaths with time lags. See our discussion in the Appendix \ref{sec:timing}.} The   left panel of the figure shows that states with a mask mandate consistently have 0-0.2 lower case
growth than states without. The  right panel also illustrates that
states with a mask mandate tend to have lower average death growth than states without a mask mandate. 


Similar plots are shown for other policies in Figures
\ref{fig:growthpolicies1} and \ref{fig:growthpolicies2} in the
appendix. The  figures for stay-at-home orders and closure
of nonessential businesses are qualitatively similar to that for
masks. States with these two policies appear to have about 0.1 percentage point lower
case growth  than states without. The effects of school closures, movie theater closures, and
restaurant closures are not clearly visible in these figures. These
figures are merely suggestive; the patterns observed in them may be
driven by confounders.

\afterpage{
 \clearpage
 \begin{landscape}
\begin{table}[!htbp] \centering
 \caption{The Direct Effect of Behavior and Policies on Case and
   Death Growth ($BPI \to Y$)}\vspace{-0.3cm}
 \label{tab:BPItoY}
 \begin{minipage}{\linewidth}
   \centering
   \resizebox{\textwidth}{!}{
   \tiny
   \begin{tabular}{c|c}
   \begin{minipage}{0.51\linewidth}
     \centering
     \begin{tabular}{@{\extracolsep{1pt}}lcccc} 
\\[-1.8ex]\hline 
\hline \\[-1.8ex] 
 & \multicolumn{4}{c}{\textit{Dependent variable:}} \\ 
\cline{2-5} 
 & \multicolumn{4}{c}{$\Delta \log \Delta C_{it}$} \\ 
\\[-1.8ex] & (1) & (2) & (3) & (4)\\ 
\hline \\[-1.8ex] 
 lag(masks for employees, 14) & $-$0.090$^{***}$ & $-$0.091$^{***}$ & $-$0.100$^{***}$ & $-$0.100$^{***}$ \\ 
  & (0.031) & (0.032) & (0.029) & (0.030) \\ 
  lag(closed K-12 schools, 14) & $-$0.074 & $-$0.083 & 0.043 & 0.031 \\ 
  & (0.080) & (0.090) & (0.096) & (0.103) \\ 
  lag(stay at home, 14) & $-$0.063 & $-$0.058 & $-$0.079 & $-$0.071 \\ 
  & (0.050) & (0.048) & (0.052) & (0.050) \\ 
  lag(business closure policies, 14) & 0.051 &  & 0.045 &  \\ 
  & (0.062) &  & (0.060) &  \\ 
  lag(closed movie theaters, 14) &  & 0.032 &  & 0.045 \\ 
  &  & (0.050) &  & (0.049) \\ 
  lag(closed restaurants, 14) &  & 0.023 &  & 0.022 \\ 
  &  & (0.044) &  & (0.043) \\ 
  lag(closed non-essent bus, 14) &  & $-$0.001 &  & $-$0.016 \\ 
  &  & (0.040) &  & (0.040) \\ 
  lag(workplaces, 14) & 1.055$^{*}$ & 1.042$^{*}$ & 0.391 & 0.355 \\ 
  & (0.543) & (0.556) & (0.610) & (0.618) \\ 
  lag(retail, 14) & 0.594$^{*}$ & 0.611$^{**}$ & 0.316 & 0.342 \\ 
  & (0.303) & (0.309) & (0.316) & (0.317) \\ 
  lag(grocery, 14) & $-$0.471$^{*}$ & $-$0.478$^{*}$ & $-$0.259 & $-$0.266 \\ 
  & (0.284) & (0.288) & (0.282) & (0.284) \\ 
  lag(transit, 14) & 0.347 & 0.339 & 0.355 & 0.339 \\ 
  & (0.258) & (0.268) & (0.247) & (0.253) \\ 
  lag($\Delta \log \Delta C_{it}$, 14) & 0.015 & 0.015 & 0.024 & 0.024 \\ 
  & (0.026) & (0.025) & (0.028) & (0.028) \\ 
  lag($\log \Delta C_{it}$, 14) & $-$0.105$^{***}$ & $-$0.105$^{***}$ & $-$0.088$^{***}$ & $-$0.087$^{***}$ \\ 
  & (0.019) & (0.019) & (0.021) & (0.021) \\ 
  lag($\Delta \log \Delta C_{it}$.national, 14) &  &  & $-$0.095$^{**}$ & $-$0.095$^{**}$ \\ 
  &  &  & (0.042) & (0.043) \\ 
  lag($\log \Delta C_{it}$.national, 14) &  &  & $-$0.177$^{***}$ & $-$0.180$^{***}$ \\ 
  &  &  & (0.049) & (0.050) \\ 
  $\Delta \log T_{it}$ & 0.152$^{***}$ & 0.153$^{***}$ & 0.155$^{***}$ & 0.156$^{***}$ \\ 
  & (0.043) & (0.043) & (0.042) & (0.041) \\ 
 \hline \\[-1.8ex] 
state variables & Yes & Yes & Yes & Yes \\ 
Month $\times$ state variables & Yes & Yes & Yes & Yes \\ 
\hline \\[-1.8ex] 
$\sum_j \mathrm{Policy}_j$ & -0.176 & -0.178 & -0.091 & -0.090 \\ 
 & (0.128) & (0.133) & (0.153) & (0.158) \\ 
$\sum_k w_k \mathrm{Behavior}_k$ & -0.804$^{***}$ & -0.801$^{***}$ & -0.425$^{***}$ & -0.413$^{***}$ \\ 
 & (0.140) & (0.140) & (0.157) & (0.160) \\ 
Observations & 3,825 & 3,825 & 3,825 & 3,825 \\ 
R$^{2}$ & 0.761 & 0.761 & 0.766 & 0.766 \\ 
Adjusted R$^{2}$ & 0.759 & 0.759 & 0.763 & 0.764 \\ 
\hline 
\hline \\[-1.8ex] 
\textit{Note:}  & \multicolumn{4}{r}{$^{*}$p$<$0.1; $^{**}$p$<$0.05; $^{***}$p$<$0.01} \\ 
\end{tabular} 
   \end{minipage}
     &
   \begin{minipage}{0.51\linewidth}
     \centering
     \begin{tabular}{@{\extracolsep{1pt}}lcccc} 
\\[-1.8ex]\hline 
\hline \\[-1.8ex] 
 & \multicolumn{4}{c}{\textit{Dependent variable:}} \\ 
\cline{2-5} 
 & \multicolumn{4}{c}{$\Delta \log \Delta D_{it}$} \\ 
\\[-1.8ex] & (1) & (2) & (3) & (4)\\ 
\hline \\[-1.8ex] 
 lag(masks for employees, 21) & $-$0.146$^{***}$ & $-$0.150$^{***}$ & $-$0.147$^{***}$ & $-$0.150$^{***}$ \\ 
  & (0.050) & (0.050) & (0.049) & (0.049) \\ 
  lag(closed K-12 schools, 21) & $-$0.232$^{**}$ & $-$0.250$^{**}$ & $-$0.178$^{*}$ & $-$0.198$^{**}$ \\ 
  & (0.102) & (0.098) & (0.103) & (0.100) \\ 
  lag(stay at home, 21) & $-$0.066 & $-$0.050 & $-$0.065 & $-$0.050 \\ 
  & (0.067) & (0.065) & (0.067) & (0.064) \\ 
  lag(business closure policies, 21) & 0.098 &  & 0.107 &  \\ 
  & (0.087) &  & (0.092) &  \\ 
  lag(closed movie theaters, 21) &  & 0.006 &  & 0.021 \\ 
  &  & (0.090) &  & (0.088) \\ 
  lag(closed restaurants, 21) &  & 0.087 &  & 0.083 \\ 
  &  & (0.072) &  & (0.070) \\ 
  lag(closed non-essent bus, 21) &  & $-$0.001 &  & $-$0.001 \\ 
  &  & (0.058) &  & (0.059) \\ 
  lag(workplaces, 21) & 1.297$^{**}$ & 1.279$^{**}$ & 0.896 & 0.889 \\ 
  & (0.515) & (0.510) & (0.554) & (0.558) \\ 
  lag(retail, 21) & 0.572 & 0.598 & 0.523 & 0.546 \\ 
  & (0.441) & (0.460) & (0.438) & (0.455) \\ 
  lag(grocery, 21) & $-$0.935$^{**}$ & $-$0.966$^{**}$ & $-$0.887$^{**}$ & $-$0.912$^{**}$ \\ 
  & (0.388) & (0.397) & (0.368) & (0.377) \\ 
  lag(transit, 21) & 0.348 & 0.368 & 0.384 & 0.396 \\ 
  & (0.284) & (0.273) & (0.283) & (0.272) \\ 
  lag($\Delta \log \Delta D_{it}$, 21) & 0.016 & 0.015 & 0.016 & 0.015 \\ 
  & (0.035) & (0.035) & (0.037) & (0.037) \\ 
  lag($\log \Delta D_{it}$, 21) & $-$0.055$^{**}$ & $-$0.052$^{**}$ & $-$0.053$^{**}$ & $-$0.050$^{**}$ \\ 
  & (0.024) & (0.025) & (0.024) & (0.024) \\ 
  lag($\Delta \log \Delta D_{it}$.national, 21) &  &  & $-$0.034 & $-$0.036 \\ 
  &  &  & (0.044) & (0.046) \\ 
  lag($\log \Delta D_{it}$.national, 21) &  &  & $-$0.047 & $-$0.046 \\ 
  &  &  & (0.039) & (0.038) \\ 
   &  &  &  &  \\ 
  &  &  &  &  \\ 
 \hline \\[-1.8ex] 
state variables & Yes & Yes & Yes & Yes \\ 
Month $\times$ state variables & Yes & Yes & Yes & Yes \\ 
\hline \\[-1.8ex] 
$\sum_j \mathrm{Policy}_j$ & -0.346$^{**}$ & -0.358$^{**}$ & -0.283$^{*}$ & -0.296$^{*}$ \\ 
 & (0.162) & (0.164) & (0.172) & (0.175) \\ 
$\sum_k w_k \mathrm{Behavior}_k$ & -0.837$^{***}$ & -0.845$^{***}$ & -0.661$^{***}$ & -0.670$^{***}$ \\ 
 & (0.164) & (0.170) & (0.176) & (0.179) \\ 
Observations & 3,468 & 3,468 & 3,468 & 3,468 \\ 
R$^{2}$ & 0.518 & 0.518 & 0.518 & 0.519 \\ 
Adjusted R$^{2}$ & 0.512 & 0.512 & 0.513 & 0.513 \\ 
\hline 
\hline \\[-1.8ex] 
\textit{Note:}  & \multicolumn{4}{r}{$^{*}$p$<$0.1; $^{**}$p$<$0.05; $^{***}$p$<$0.01} \\ 
\end{tabular} 
   \end{minipage}
   \end{tabular}
   }
 \begin{flushleft}
     \scriptsize Dependent variable is the weekly growth rate of
     confirmed cases (in the left panel) or deaths (in the right
     panel) as defined in equation (\ref{eq:y}). The covariates
     include lagged policy and behavior variables, which are
     constructed as 7 day moving averages between $t$ to $t-7$ of
     corresponding daily measures.  The row
     ``$\sum_j \mathrm{Policies}_j$'' reports the sum of all policy
     coefficients.  The row ``$\sum_k w_k \mathrm{Behavior}_k$''
     reports the sum of four coefficients of behavior variables
     weighted by the average of each behavioral variable from April
     1st-10th.   ``business closure policies'' is defined by the average of   closured movie theaters, closured restaurants, and closured non-essential businesses.  Standard errors are clustered at the state level.
   \end{flushleft}
 \end{minipage}
\end{table}
\end{landscape}
\clearpage
}

We more formally analyze the effect of policies by estimating
regressions. We first look at the direct effect of policies on case and death growth
conditional on behavior by estimating equation (\ref{eq:R1}):
\begin{align}
  {\ycolor  Y_{i,t+\ell}}
  & = {\bcolor \alpha ' B_{it}} + {\pcolor\pi 'P_{it}} + {\icolor \mu'I_{it}} +  {\wcolor\delta_Y 'W_{it}}  + \varepsilon^y_{it},
  \label{eq:reg-y}
\end{align}
where the outcome variable, $\ycolor Y_{i,t+\ell}$, is either case growth  or death growth.

For case growth as the outcome, we choose a lag length of $\ell=14$ days for behavior, policy, and information variables to reflect the delay between infection and
confirmation of case.\footnote{As we review in the Appendix \ref{sec:timing},
  a lag length of 14 days between exposure and case reporting, as well as a lag length of 21 days between exposure and deaths, is broadly consistent with currently
  available evidence. Section \ref{sec:sensitivity} provides a sensitivity analysis under different timing assumptions.  }   $\bcolor B_{it}=(B_{it}^1,...,B_{it}^4)'$ is a vector of four behavior
variables  in state $i$. $\pcolor P_{it}$ includes the Covid-related policies
in state $i$  that directly affect the spread of Covid-19
after controlling for behavior variables (e.g., masks for employees).
We include information variables,  $\icolor I_{it}$, that include the past cases and case growths
because  the past cases may be
correlated with (latent) government  policies or people's behaviors
that are not fully captured by our observed policy and behavior
variables. We also consider a specification that includes the past cases and case growth at the national level as additional information variables. $\wcolor W_{it}$ is a set of confounders that includes month dummies, state-level covariates,  and the interaction terms between month dummies and state-level covariates.\footnote{Month dummies also represent the latent information that is not fully captured by the past cases and growths.}
For case growth, $\wcolor W_{it}$ also includes the  test rate growth  $\Delta \log(T)_{it}$  to capture the effect of changing test rates on confirmed cases.
Equation (\ref{eq:reg-y}) corresponds to (\ref{eq:M})  derived from the SIR model.

For death growth as the outcome, we take a lag length of $\ell=21$ days. The information variables $\icolor I_{it}$ include  past deaths and death growth rates;  $\wcolor W_{it}$ is the same as that of the case growth equation except that the growth rate of test rates is excluded from $\wcolor W_{it}$ as implied by equation (\ref{eq:M-D}).

Table \ref{tab:BPItoY} shows
the results of estimating
(\ref{eq:reg-y}) for case and death growth rates. 

Column (1) represents our baseline specification while column (2) replaces business closure policies with closed movie theaters, closed restaurants, and closed non-essential businesses. Columns (3) and (4) include past cases/deaths and growth rates at national level as additional regressors.

The estimates indicate that mandatory face masks for employees
reduce the growth rate of infections and deaths by 9-15 percent, while holding behavior constant. This suggests that
requiring masks for employees in public-facing businesses may be an effective preventive measure.\footnote{Note that we are \textit{not} evaluating the effect of \textit{universal} mask-wearing for the public but that of mask-wearing for employees. The effect of \textit{universal} mask-wearing for the public could be larger if people comply with such a policy measure. \cite{tian2020calibrated} considered a  model in which mask wearing reduces  the reproduction number by a factor $(1-e \cdot pm)^2$, where $e$ is the efficacy of trapping viral particles inside the mask and $pm$ is the percentage of mask-wearing population. Given an estimate of $R_0=2.4$, \cite{howard2020} argue that  50\% mask usage and a 50\% mask efficacy level would reduce the reproduction number from 2.4 to 1.35, an order of magnitude
impact.} The estimated effect of masks on death growth is larger than the effect on case growth, but this difference between the two estimated effects is not statistically significant.

Closing schools has a large and statistically
significant coefficient in the death growth regressions. As discussed
above, however, there is little cross-state variation in the timing of school
closures, making estimates of its effect less reliable.

Neither the effect of stay-at-home orders nor that of business closure policies is estimated significantly different from zero, suggesting that these lockdown policies may  have little direct
effect on case or death growth when behavior is held constant.


The row ``$\sum_k w_k \mathrm{Behavior}_k$'' reports the sum of estimated coefficients weighted by the average of the behavioral
variables from April 1st-10th. The estimates of $-0.80$ and $-0.84$ for ``$\sum_k w_k \mathrm{Behavior}_k$'' in column (1)  imply that a reduction in mobility measures relative to the baseline in January and February have induced a decrease in case and death growth rates by 80 and 84 percent, respectively, suggesting an importance of social distancing for reducing the spread of Covid-19.
When including national cases and deaths in information, as shown in columns (3) and (4), the estimated aggregate impact of behavior is substantially smaller but remains large and statistically significant.

A useful practical implication of these results is that Google Mobility Reports and similar data might be useful as a leading indicator of potential case or death growth. This should be done with caution, however, because other changes in the environment might alter the relationship between behavior and infections. Preventative measures, including mandatory face masks, and changes in habit that are not captured in our data might alter the future relationship between Google Mobility Reports and case/death growth.

The negative coefficients of the log of past cases or deaths in
Table \ref{tab:BPItoY} is
consistent with a hypothesis that higher reported cases and deaths change people's behavior to reduce transmission risks. Such behavioral changes in response to new information   are  partly captured by Google mobility measures, but the negative estimated coefficient of past cases or deaths imply that other latent behavioral changes that are not fully captured by Google mobility measures (e.g., frequent hand-washing, wearing masks, and keeping 6ft/2m distancing) are also important for reducing future cases and deaths.

If policies are enacted and behavior changes, then future cases/deaths and information will change, which will induce further behavior changes.  However, since the model includes lags of cases/deaths as well as their growth rates,  computing a long-run effect is not completely straightforward. We investigate dynamic effects that incorporate feedback through information in section \ref{counterfactuals}.

  \afterpage{
 \begin{landscape}
\begin{table}[!htbp] \centering
 \caption{\label{tab:PtoY}
   The Total Effect of Policies on Case and Death Growth ($PI \to Y$)}\vspace{-0.3cm}
 \begin{minipage}{\linewidth}
   \resizebox{\textwidth}{!}{
   \centering
   \tiny
   \begin{tabular}{c|c}
   \begin{minipage}{0.51\linewidth}
     \centering
     \begin{tabular}{@{\extracolsep{1pt}}lcccc} 
\\[-1.8ex]\hline 
\hline \\[-1.8ex] 
 & \multicolumn{4}{c}{\textit{Dependent variable:}} \\ 
\cline{2-5} 
 & \multicolumn{4}{c}{$\Delta \log \Delta C_{it}$} \\ 
\\[-1.8ex] & (1) & (2) & (3) & (4)\\ 
\hline \\[-1.8ex] 
 lag(masks for employees, 14) & $-$0.083$^{**}$ & $-$0.081$^{**}$ & $-$0.103$^{***}$ & $-$0.102$^{***}$ \\ 
  & (0.038) & (0.040) & (0.033) & (0.035) \\ 
  lag(closed K-12 schools, 14) & $-$0.226$^{**}$ & $-$0.236$^{**}$ & 0.029 & 0.017 \\ 
  & (0.089) & (0.097) & (0.102) & (0.107) \\ 
  lag(stay at home, 14) & $-$0.127$^{**}$ & $-$0.121$^{**}$ & $-$0.115$^{**}$ & $-$0.103$^{**}$ \\ 
  & (0.057) & (0.054) & (0.054) & (0.052) \\ 
  lag(business closure policies, 14) & $-$0.076 &  & $-$0.001 &  \\ 
  & (0.068) &  & (0.061) &  \\ 
  lag(closed movie theaters, 14) &  & 0.027 &  & 0.062 \\ 
  &  & (0.051) &  & (0.046) \\ 
  lag(closed restaurants, 14) &  & $-$0.041 &  & $-$0.011 \\ 
  &  & (0.049) &  & (0.045) \\ 
  lag(closed non-essent bus, 14) &  & $-$0.051 &  & $-$0.038 \\ 
  &  & (0.050) &  & (0.043) \\ 
  lag($\Delta \log \Delta C_{it}$, 14) & 0.040 & 0.041$^{*}$ & 0.036 & 0.035 \\ 
  & (0.024) & (0.025) & (0.028) & (0.028) \\ 
  lag($\log \Delta C_{it}$, 14) & $-$0.137$^{***}$ & $-$0.137$^{***}$ & $-$0.091$^{***}$ & $-$0.090$^{***}$ \\ 
  & (0.022) & (0.022) & (0.026) & (0.026) \\ 
  lag($\Delta \log \Delta C_{it}$.national, 14) &  &  & $-$0.128$^{***}$ & $-$0.123$^{***}$ \\ 
  &  &  & (0.039) & (0.041) \\ 
  lag($\log \Delta C_{it}$.national, 14) &  &  & $-$0.243$^{***}$ & $-$0.245$^{***}$ \\ 
  &  &  & (0.045) & (0.045) \\ 
  $\Delta \log T_{it}$ & 0.156$^{***}$ & 0.157$^{***}$ & 0.158$^{***}$ & 0.160$^{***}$ \\ 
  & (0.044) & (0.044) & (0.042) & (0.041) \\ 
 \hline \\[-1.8ex] 
state variables & Yes & Yes & Yes & Yes \\ 
Month $\times$ state variables & Yes & Yes & Yes & Yes \\ 
\hline \\[-1.8ex] 
$\sum_j \mathrm{Policy}_j$ & -0.512$^{***}$ & -0.504$^{***}$ & -0.190 & -0.175 \\ 
 & (0.150) & (0.154) & (0.156) & (0.159) \\ 
Observations & 3,825 & 3,825 & 3,825 & 3,825 \\ 
R$^{2}$ & 0.749 & 0.750 & 0.763 & 0.763 \\ 
Adjusted R$^{2}$ & 0.747 & 0.747 & 0.760 & 0.761 \\ 
\hline 
\hline \\[-1.8ex] 
\textit{Note:}  & \multicolumn{4}{r}{$^{*}$p$<$0.1; $^{**}$p$<$0.05; $^{***}$p$<$0.01} \\ 
\end{tabular} 
   \end{minipage}
     &
   \begin{minipage}{0.51\linewidth}
     \centering
     \begin{tabular}{@{\extracolsep{1pt}}lcccc} 
\\[-1.8ex]\hline 
\hline \\[-1.8ex] 
 & \multicolumn{4}{c}{\textit{Dependent variable:}} \\ 
\cline{2-5} 
 & \multicolumn{4}{c}{$\Delta \log \Delta D_{it}$} \\ 
\\[-1.8ex] & (1) & (2) & (3) & (4)\\ 
\hline \\[-1.8ex] 
 lag(masks for employees, 21) & $-$0.134$^{***}$ & $-$0.133$^{**}$ & $-$0.156$^{***}$ & $-$0.155$^{***}$ \\ 
  & (0.051) & (0.053) & (0.050) & (0.052) \\ 
  lag(closed K-12 schools, 21) & $-$0.610$^{***}$ & $-$0.621$^{***}$ & $-$0.234$^{**}$ & $-$0.248$^{**}$ \\ 
  & (0.115) & (0.121) & (0.111) & (0.109) \\ 
  lag(stay at home, 21) & $-$0.082 & $-$0.075 & $-$0.068 & $-$0.057 \\ 
  & (0.066) & (0.064) & (0.066) & (0.062) \\ 
  lag(business closure policies, 21) & $-$0.059 &  & 0.059 &  \\ 
  & (0.086) &  & (0.086) &  \\ 
  lag(closed movie theaters, 21) &  & $-$0.006 &  & 0.050 \\ 
  &  & (0.089) &  & (0.082) \\ 
  lag(closed restaurants, 21) &  & $-$0.012 &  & 0.030 \\ 
  &  & (0.061) &  & (0.055) \\ 
  lag(closed non-essent bus, 21) &  & $-$0.040 &  & $-$0.016 \\ 
  &  & (0.066) &  & (0.063) \\ 
  lag($\Delta \log \Delta D_{it}$, 21) & $-$0.001 & $-$0.001 & 0.017 & 0.016 \\ 
  & (0.033) & (0.033) & (0.036) & (0.037) \\ 
  lag($\log \Delta D_{it}$, 21) & $-$0.078$^{***}$ & $-$0.078$^{***}$ & $-$0.064$^{**}$ & $-$0.063$^{**}$ \\ 
  & (0.026) & (0.027) & (0.027) & (0.027) \\ 
  lag($\Delta \log \Delta D_{it}$.national, 21) &  &  & $-$0.147$^{***}$ & $-$0.148$^{**}$ \\ 
  &  &  & (0.056) & (0.057) \\ 
  lag($\log \Delta D_{it}$.national, 21) &  &  & $-$0.116$^{***}$ & $-$0.117$^{***}$ \\ 
  &  &  & (0.032) & (0.032) \\ 
   &  &  &  &  \\ 
  &  &  &  &  \\ 
 \hline \\[-1.8ex] 
state variables & Yes & Yes & Yes & Yes \\ 
Month $\times$ state variables & Yes & Yes & Yes & Yes \\ 
\hline \\[-1.8ex] 
$\sum_j \mathrm{Policy}_j$ & -0.885$^{***}$ & -0.887$^{***}$ & -0.399$^{**}$ & -0.396$^{**}$ \\ 
 & (0.159) & (0.166) & (0.183) & (0.188) \\ 
Observations & 3,468 & 3,468 & 3,468 & 3,468 \\ 
R$^{2}$ & 0.502 & 0.502 & 0.512 & 0.512 \\ 
Adjusted R$^{2}$ & 0.497 & 0.497 & 0.507 & 0.507 \\ 
\hline 
\hline \\[-1.8ex] 
\textit{Note:}  & \multicolumn{4}{r}{$^{*}$p$<$0.1; $^{**}$p$<$0.05; $^{***}$p$<$0.01} \\ 
\end{tabular} 
   \end{minipage}

   \end{tabular}
   }
   \begin{flushleft}
     \scriptsize Dependent variable is the weekly growth rate of
     confirmed cases (in the left panel) or deaths (in the right
     panel) as defined in equation (\ref{eq:y}). The covariates
     include lagged policy variables, which are
     constructed as 7 day moving averages between $t$ to $t-7$ of
     corresponding daily measures.  The row
     ``$\sum_j \mathrm{Policies}_j$'' reports the sum of all policy
     coefficients.   ``business closure policies'' is defined by the average of   closed movie theaters, closed restaurants, and closed non-essential businesses.  Standard errors are clustered at the state level.    \end{flushleft}
 \end{minipage}
\end{table}
\end{landscape}
}

\afterpage{
\begin{table}[!h]
  \caption{\label{tab:dieff-si}Direct and Indirect Policy Effects
    without national case/death variables}
\begin{minipage}{\linewidth}
  \centering
    \scriptsize
  \begin{tabular}{c}
    \textbf{Cases}
    \\
    
\begin{tabular}{lccccc|>{}c}
\toprule
  & Direct & Indirect & Total & PI$\to$Y Coef. & Average & Difference\\
\midrule
masks for employees & -0.090$^{***}$ & 0.043 & -0.047 & -0.083$^{**}$ & -0.065 & 0.036$^{**}$\\
 & (0.031) & (0.028) & (0.043) & (0.038) & (0.040) & (0.015)\\
closed K-12 schools & -0.074 & -0.374$^{***}$ & -0.448$^{***}$ & -0.226$^{***}$ & -0.337$^{***}$ & -0.223$^{***}$\\
 & (0.080) & (0.095) & (0.116) & (0.086) & (0.098) & (0.055)\\
stay at home & -0.063 & -0.034 & -0.096$^{*}$ & -0.127$^{**}$ & -0.112$^{**}$ & 0.031$^{**}$\\
 & (0.049) & (0.029) & (0.055) & (0.058) & (0.056) & (0.015)\\
business closure policies & 0.051 & -0.153$^{***}$ & -0.101 & -0.076 & -0.089 & -0.025\\
 & (0.062) & (0.045) & (0.068) & (0.067) & (0.067) & (0.021)\\
$\sum_j \mathrm{Policy}_j$ & -0.176 & -0.517$^{***}$ & -0.693$^{***}$ & -0.512$^{***}$ & -0.603$^{***}$ & -0.181$^{***}$\\
 & (0.129) & (0.145) & (0.182) & (0.147) & (0.162) & (0.061)\\
$\Delta \log \Delta C_{it}$ & 0.015 & 0.012 & 0.027 & 0.040 & 0.033 & -0.013$^{**}$\\
 & (0.026) & (0.009) & (0.025) & (0.025) & (0.025) & (0.007)\\
$\log \Delta C_{it}$ & -0.105$^{***}$ & -0.045$^{***}$ & -0.150$^{***}$ & -0.137$^{***}$ & -0.144$^{***}$ & -0.013$^{*}$\\
 & (0.018) & (0.015) & (0.025) & (0.021) & (0.023) & (0.008)\\
\bottomrule
\end{tabular}
    \\
    \textbf{Deaths}
    \\
    
\begin{tabular}{lccccc|>{}c}
\toprule
  & Direct & Indirect & Total & PI$\to$Y Coef. & Average & Difference\\
\midrule
masks for employees & -0.146$^{***}$ & 0.035$^{***}$ & -0.111$^{**}$ & -0.134$^{***}$ & -0.122$^{**}$ & 0.022\\
 & (0.049) & (0.000) & (0.049) & (0.051) & (0.049) & (0.021)\\
closed K-12 schools & -0.232$^{**}$ & -0.420$^{***}$ & -0.653$^{***}$ & -0.610$^{***}$ & -0.632$^{***}$ & -0.042$^{*}$\\
 & (0.095) & (0.084) & (0.111) & (0.109) & (0.109) & (0.023)\\
stay at home & -0.066 & -0.008 & -0.073 & -0.082 & -0.078 & 0.008\\
 & (0.068) & (0.031) & (0.067) & (0.067) & (0.067) & (0.016)\\
business closure policies & 0.098$^{***}$ & -0.163$^{**}$ & -0.065 & -0.059 & -0.062 & -0.006\\
 & (0.000) & (0.065) & (0.065) & (0.086) & (0.061) & (0.089)\\
$\sum_j \mathrm{Policy}_j$ & -0.346$^{***}$ & -0.557$^{***}$ & -0.902$^{***}$ & -0.885$^{***}$ & -0.894$^{***}$ & -0.018\\
 & (0.122) & (0.137) & (0.156) & (0.157) & (0.147) & (0.106)\\
$\Delta \log \Delta D_{it}$ & 0.016 & -0.040$^{***}$ & -0.024 & -0.001 & -0.012 & -0.023$^{***}$\\
 & (0.034) & (0.011) & (0.030) & (0.032) & (0.031) & (0.005)\\
$\log \Delta D_{it}$ & -0.055$^{**}$ & -0.015 & -0.070$^{**}$ & -0.078$^{***}$ & -0.074$^{***}$ & 0.009\\
 & (0.024) & (0.009) & (0.028) & (0.026) & (0.027) & (0.005)\\
\bottomrule
\end{tabular}
  \end{tabular}
  \smallskip
  \begin{flushleft}
    {\tiny Direct effects capture the effect of policy on case
      growth holding behavior, information, and confounders
      constant. Direct effects are given by ${\pcolor \pi}$ in
      equation (\ref{eq:R1}). Indirect effects capture how policy
      changes behavior and behavior shift case growth. They are given
      by ${\bcolor \alpha}$ from (\ref{eq:R1}) times ${\pcolor \beta}$
      from (\ref{eq:R2}). The total effect is
      ${\pcolor \pi} + {\pcolor \beta} {\bcolor \alpha}$. Column
      ``PI$\to$Y Coefficients'' shows the coefficient estimates from
      \ref{eq:R4}. Columns ``Difference''    are  the differences between
      the estimates from (\ref{eq:R4}) and the combination of
      (\ref{eq:R1}) and (\ref{eq:R2}) while column ``Average'' are their averages.
      Standard errors are computed by
      bootstrap and clustered on state.}
    \end{flushleft}
  \end{minipage}
\end{table}
}

\afterpage{
\begin{table}[!h]
  \caption{\label{tab:dieff-con-si}Direct and Indirect Policy Effects
    without national case/death variables, \textbf{restrictions imposed}}
\begin{minipage}{\linewidth}
  \centering
    \scriptsize
  \begin{tabular}{c}
    \textbf{Cases}
    \\
    
\begin{tabular}{lccccc|>{}c}
\toprule
  & Direct & Indirect & Total & PI$\to$Y Coef. & Average & Difference\\
\midrule
masks for employees & -0.096$^{***}$ & 0.000 & -0.096$^{***}$ & -0.083$^{**}$ & -0.089$^{***}$ & -0.013\\
 & (0.030) & (0.000) & (0.030) & (0.039) & (0.032) & (0.025)\\
closed K-12 schools & -0.073 & -0.364$^{***}$ & -0.436$^{***}$ & -0.226$^{**}$ & -0.331$^{***}$ & -0.210$^{***}$\\
 & (0.078) & (0.094) & (0.119) & (0.092) & (0.102) & (0.056)\\
stay at home & -0.053 & -0.032 & -0.085 & -0.127$^{**}$ & -0.106$^{*}$ & 0.042$^{**}$\\
 & (0.052) & (0.028) & (0.058) & (0.057) & (0.057) & (0.020)\\
business closure policies & 0.000 & -0.157$^{***}$ & -0.157$^{***}$ & -0.076 & -0.117$^{**}$ & -0.081\\
 & (0.000) & (0.042) & (0.042) & (0.066) & (0.048) & (0.054)\\
$\sum_j \mathrm{Policy}_j$ & -0.221$^{**}$ & -0.553$^{***}$ & -0.774$^{***}$ & -0.512$^{***}$ & -0.643$^{***}$ & -0.262$^{***}$\\
 & (0.108) & (0.124) & (0.166) & (0.151) & (0.156) & (0.061)\\
$\Delta \log \Delta C_{it}$ & 0.013 & 0.009 & 0.021 & 0.040$^{*}$ & 0.031 & -0.019$^{***}$\\
 & (0.026) & (0.009) & (0.024) & (0.024) & (0.024) & (0.007)\\
$\log \Delta C_{it}$ & -0.104$^{***}$ & -0.042$^{***}$ & -0.147$^{***}$ & -0.137$^{***}$ & -0.142$^{***}$ & -0.010\\
 & (0.017) & (0.015) & (0.024) & (0.021) & (0.023) & (0.007)\\
\bottomrule
\end{tabular}
    \\
    \textbf{Deaths}
    \\
    
\begin{tabular}{lccccc|>{}c}
\toprule
  & Direct & Indirect & Total & PI$\to$Y Coef. & Average & Difference\\
\midrule
masks for employees & -0.152$^{***}$ & 0.000 & -0.152$^{***}$ & -0.134$^{***}$ & -0.143$^{***}$ & -0.019\\
 & (0.048) & (0.000) & (0.048) & (0.051) & (0.049) & (0.020)\\
closed K-12 schools & -0.228$^{**}$ & -0.391$^{***}$ & -0.619$^{***}$ & -0.610$^{***}$ & -0.615$^{***}$ & -0.009\\
 & (0.101) & (0.079) & (0.117) & (0.112) & (0.114) & (0.031)\\
stay at home & -0.047 & -0.006 & -0.054 & -0.082 & -0.068 & 0.028\\
 & (0.063) & (0.031) & (0.063) & (0.067) & (0.064) & (0.023)\\
business closure policies & 0.000 & -0.154$^{***}$ & -0.154$^{***}$ & -0.059 & -0.106$^{*}$ & -0.095\\
 & (0.000) & (0.059) & (0.059) & (0.084) & (0.061) & (0.078)\\
$\sum_j \mathrm{Policy}_j$ & -0.427$^{***}$ & -0.551$^{***}$ & -0.978$^{***}$ & -0.885$^{***}$ & -0.931$^{***}$ & -0.093\\
 & (0.123) & (0.124) & (0.149) & (0.158) & (0.150) & (0.060)\\
$\Delta \log \Delta D_{it}$ & 0.014 & -0.040$^{***}$ & -0.025 & -0.001 & -0.013 & -0.025$^{***}$\\
 & (0.035) & (0.010) & (0.032) & (0.033) & (0.032) & (0.005)\\
$\log \Delta D_{it}$ & -0.052$^{**}$ & -0.012 & -0.064$^{**}$ & -0.078$^{***}$ & -0.071$^{***}$ & 0.014$^{**}$\\
 & (0.023) & (0.010) & (0.026) & (0.026) & (0.026) & (0.007)\\
\bottomrule
\end{tabular}
  \end{tabular}
  \smallskip
  \begin{flushleft}
    {\tiny Direct effects capture the effect of policy on case
      growth holding behavior, information, and confounders
      constant. Direct effects are given by ${\pcolor \pi}$ in
      equation (\ref{eq:R1}). Indirect effects capture how policy
      changes behavior and behavior shift case growth. They are given
      by ${\bcolor \alpha}$ from (\ref{eq:R1}) times ${\pcolor \beta}$
      from (\ref{eq:R2}). The total effect is
      ${\pcolor \pi} + {\pcolor \beta} {\bcolor \alpha}$. Column
      ``PI$\to$Y Coefficients'' shows the coefficient estimates from
      \ref{eq:R4}. Columns ``Difference''    are  the differences between
      the estimates from (\ref{eq:R4}) and the combination of
      (\ref{eq:R1}) and (\ref{eq:R2}) while column ``Average'' are their averages.
      Standard errors are computed by
      bootstrap and clustered on state.}
    \end{flushleft}
  \end{minipage}
\end{table}
}

\afterpage{
\begin{table}
  \caption{\label{tab:dieff}Direct and Indirect Policy Effects with national case/death variables}
\begin{minipage}{\linewidth}
  \centering
    \scriptsize
  \begin{tabular}{c}
    \textbf{Cases}
    \\
    
\begin{tabular}{lccccc|>{}c}
\toprule
  & Direct & Indirect & Total & PI$\to$Y Coef. & Average & Difference\\
\midrule
masks for employees & -0.100$^{***}$ & -0.002 & -0.102$^{***}$ & -0.103$^{***}$ & -0.103$^{***}$ & 0.001\\
 & (0.028) & (0.018) & (0.035) & (0.032) & (0.033) & (0.011)\\
closed K-12 schools & 0.043 & -0.024 & 0.019 & 0.029 & 0.024 & -0.011\\
 & (0.092) & (0.036) & (0.101) & (0.099) & (0.100) & (0.015)\\
stay at home & -0.079 & -0.036$^{*}$ & -0.115$^{**}$ & -0.115$^{**}$ & -0.115$^{**}$ & 0.000\\
 & (0.051) & (0.019) & (0.052) & (0.054) & (0.053) & (0.011)\\
business closure policies & 0.045 & -0.045$^{*}$ & 0.000 & -0.001 & -0.000 & 0.001\\
 & (0.057) & (0.026) & (0.059) & (0.057) & (0.058) & (0.012)\\
$\sum_j \mathrm{Policy}_j$ & -0.091 & -0.107$^{*}$ & -0.198 & -0.190 & -0.194 & -0.008\\
 & (0.151) & (0.061) & (0.159) & (0.154) & (0.156) & (0.022)\\
$\Delta \log \Delta C_{it}$ & 0.024 & 0.014$^{*}$ & 0.038 & 0.036 & 0.037 & 0.002\\
 & (0.028) & (0.007) & (0.027) & (0.028) & (0.027) & (0.004)\\
$\log \Delta C_{it}$ & -0.088$^{***}$ & -0.003 & -0.091$^{***}$ & -0.091$^{***}$ & -0.091$^{***}$ & -0.000\\
 & (0.020) & (0.010) & (0.026) & (0.025) & (0.025) & (0.005)\\
$\Delta \log \Delta C_{it}$.national & -0.095$^{**}$ & -0.054$^{***}$ & -0.149$^{***}$ & -0.128$^{***}$ & -0.138$^{***}$ & -0.022\\
 & (0.040) & (0.016) & (0.040) & (0.037) & (0.038) & (0.013)\\
$\log \Delta C_{it}$.national & -0.177$^{***}$ & -0.084$^{***}$ & -0.261$^{***}$ & -0.243$^{***}$ & -0.252$^{***}$ & -0.018$^{*}$\\
 & (0.048) & (0.023) & (0.044) & (0.044) & (0.043) & (0.010)\\
\bottomrule
\end{tabular}
    \\ \\
    \textbf{Deaths}
    \\
    
\begin{tabular}{lccccc|>{}c}
\toprule
  & Direct & Indirect & Total & PI$\to$Y Coef. & Average & Difference\\
\midrule
masks for employees & -0.147$^{***}$ & 0.009$^{***}$ & -0.138$^{***}$ & -0.156$^{***}$ & -0.147$^{***}$ & 0.018\\
 & (0.050) & (0.000) & (0.050) & (0.052) & (0.050) & (0.017)\\
closed K-12 schools & -0.178$^{*}$ & -0.055 & -0.234$^{**}$ & -0.234$^{**}$ & -0.234$^{**}$ & 0.000\\
 & (0.100) & (0.038) & (0.112) & (0.108) & (0.110) & (0.019)\\
stay at home & -0.065 & 0.001 & -0.063 & -0.068 & -0.066 & 0.005\\
 & (0.067) & (0.029) & (0.066) & (0.067) & (0.066) & (0.015)\\
business closure policies & 0.107$^{***}$ & -0.048 & 0.059 & 0.059 & 0.059 & -0.001\\
 & (0.000) & (0.036) & (0.036) & (0.082) & (0.044) & (0.091)\\
$\sum_j \mathrm{Policy}_j$ & -0.283$^{**}$ & -0.094 & -0.377$^{**}$ & -0.399$^{**}$ & -0.388$^{**}$ & 0.023\\
 & (0.130) & (0.069) & (0.151) & (0.186) & (0.163) & (0.092)\\
$\Delta \log \Delta D_{it}$ & 0.016 & -0.010$^{**}$ & 0.006 & 0.017 & 0.012 & -0.010$^{***}$\\
 & (0.037) & (0.005) & (0.036) & (0.036) & (0.036) & (0.004)\\
$\log \Delta D_{it}$ & -0.053$^{**}$ & -0.006 & -0.059$^{**}$ & -0.064$^{**}$ & -0.062$^{**}$ & 0.005\\
 & (0.023) & (0.009) & (0.026) & (0.026) & (0.026) & (0.005)\\
$\Delta \log \Delta D_{it}$.national & -0.034 & -0.120$^{***}$ & -0.154$^{***}$ & -0.147$^{***}$ & -0.151$^{***}$ & -0.007\\
 & (0.043) & (0.022) & (0.048) & (0.054) & (0.051) & (0.013)\\
$\log \Delta D_{it}$.national & -0.047 & -0.077$^{**}$ & -0.124$^{***}$ & -0.116$^{***}$ & -0.120$^{***}$ & -0.008\\
 & (0.038) & (0.030) & (0.035) & (0.032) & (0.033) & (0.013)\\
\bottomrule
\end{tabular}
    \\
  \end{tabular}
  \smallskip
  \begin{flushleft}
      \scriptsize Direct effects capture the effect of policy on case
      growth holding behavior, information, and confounders
      constant. Direct effects are given by ${\pcolor \pi}$ in
      equation (\ref{eq:R1}). Indirect effects capture how policy
      changes behavior and behavior shift case growth. They are given
      by ${\bcolor \alpha}$ from (\ref{eq:R1}) times ${\pcolor \beta}$
      from (\ref{eq:R2}). The total effect is
      ${\pcolor \pi} + {\pcolor \beta} {\bcolor \alpha}$. Column
      ``PI$\to$Y Coefficients'' shows the coefficient estimates from
      \ref{eq:R4}. Columns ``Difference''    are  the differences between
      the estimates from (\ref{eq:R4}) and the combination of
      (\ref{eq:R1}) and (\ref{eq:R2}) while column ``Average'' are their averages.
      Standard errors are computed by
      bootstrap and clustered on state.
    \end{flushleft}
  \end{minipage}
\end{table}
\clearpage}

\afterpage{
\begin{table}
  \caption{\label{tab:dieff-con}Direct and Indirect Policy Effects with national case/death variables, \textbf{restrictions imposed}}
\begin{minipage}{\linewidth}
  \centering
    \scriptsize
  \begin{tabular}{c}
    \textbf{Cases}
    \\
    
\begin{tabular}{lccccc|>{}c}
\toprule
  & Direct & Indirect & Total & PI$\to$Y Coef. & Average & Difference\\
\midrule
masks for employees & -0.105$^{***}$ & 0.000 & -0.105$^{***}$ & -0.103$^{***}$ & -0.104$^{***}$ & -0.001\\
 & (0.027) & (0.000) & (0.027) & (0.031) & (0.028) & (0.016)\\
closed K-12 schools & 0.045 & -0.022 & 0.023 & 0.029 & 0.026 & -0.007\\
 & (0.092) & (0.034) & (0.101) & (0.099) & (0.100) & (0.015)\\
stay at home & -0.071 & -0.033$^{*}$ & -0.104$^{*}$ & -0.115$^{**}$ & -0.110$^{**}$ & 0.011\\
 & (0.052) & (0.019) & (0.056) & (0.052) & (0.053) & (0.017)\\
business closure policies & 0.000 & -0.038 & -0.038 & -0.001 & -0.019 & -0.038\\
 & (0.000) & (0.024) & (0.024) & (0.061) & (0.038) & (0.054)\\
$\sum_j \mathrm{Policy}_j$ & -0.131 & -0.094$^{*}$ & -0.225$^{*}$ & -0.190 & -0.207 & -0.035\\
 & (0.123) & (0.049) & (0.134) & (0.155) & (0.143) & (0.047)\\
$\Delta \log \Delta C_{it}$ & 0.023 & 0.013$^{**}$ & 0.036 & 0.036 & 0.036 & 0.000\\
 & (0.028) & (0.007) & (0.027) & (0.027) & (0.027) & (0.004)\\
$\log \Delta C_{it}$ & -0.087$^{***}$ & -0.003 & -0.090$^{***}$ & -0.091$^{***}$ & -0.091$^{***}$ & 0.000\\
 & (0.021) & (0.011) & (0.027) & (0.025) & (0.026) & (0.005)\\
$\Delta \log \Delta C_{it}$.national & -0.098$^{**}$ & -0.050$^{***}$ & -0.147$^{***}$ & -0.128$^{***}$ & -0.137$^{***}$ & -0.020\\
 & (0.040) & (0.016) & (0.040) & (0.037) & (0.038) & (0.014)\\
$\log \Delta C_{it}$.national & -0.178$^{***}$ & -0.078$^{***}$ & -0.256$^{***}$ & -0.243$^{***}$ & -0.249$^{***}$ & -0.013\\
 & (0.048) & (0.024) & (0.042) & (0.044) & (0.043) & (0.011)\\
\bottomrule
\end{tabular}
    \\ \\
    \textbf{Deaths}
    \\
    
\begin{tabular}{lccccc|>{}c}
\toprule
  & Direct & Indirect & Total & PI$\to$Y Coef. & Average & Difference\\
\midrule
masks for employees & -0.154$^{***}$ & 0.000 & -0.154$^{***}$ & -0.156$^{***}$ & -0.155$^{***}$ & 0.002\\
 & (0.046) & (0.000) & (0.046) & (0.048) & (0.047) & (0.018)\\
closed K-12 schools & -0.177$^{*}$ & -0.051 & -0.227$^{**}$ & -0.234$^{**}$ & -0.231$^{**}$ & 0.006\\
 & (0.102) & (0.036) & (0.113) & (0.109) & (0.111) & (0.025)\\
stay at home & -0.046 & 0.004 & -0.043 & -0.068 & -0.055 & 0.026\\
 & (0.063) & (0.027) & (0.062) & (0.065) & (0.063) & (0.020)\\
business closure policies & 0.000 & -0.040 & -0.040 & 0.059 & 0.009 & -0.099\\
 & (0.000) & (0.032) & (0.032) & (0.084) & (0.048) & (0.084)\\
$\sum_j \mathrm{Policy}_j$ & -0.377$^{***}$ & -0.087 & -0.464$^{***}$ & -0.399$^{**}$ & -0.432$^{***}$ & -0.065\\
 & (0.124) & (0.063) & (0.144) & (0.178) & (0.157) & (0.074)\\
$\Delta \log \Delta D_{it}$ & 0.015 & -0.010$^{**}$ & 0.006 & 0.017 & 0.011 & -0.011$^{***}$\\
 & (0.037) & (0.005) & (0.036) & (0.036) & (0.036) & (0.004)\\
$\log \Delta D_{it}$ & -0.050$^{**}$ & -0.006 & -0.056$^{**}$ & -0.064$^{**}$ & -0.060$^{**}$ & 0.008\\
 & (0.023) & (0.008) & (0.026) & (0.026) & (0.026) & (0.006)\\
$\Delta \log \Delta D_{it}$.national & -0.039 & -0.112$^{***}$ & -0.151$^{***}$ & -0.147$^{***}$ & -0.149$^{***}$ & -0.004\\
 & (0.044) & (0.020) & (0.048) & (0.054) & (0.051) & (0.013)\\
$\log \Delta D_{it}$.national & -0.043 & -0.070$^{**}$ & -0.113$^{***}$ & -0.116$^{***}$ & -0.114$^{***}$ & 0.002\\
 & (0.036) & (0.029) & (0.030) & (0.031) & (0.030) & (0.013)\\
\bottomrule
\end{tabular}
    \\
  \end{tabular}
  \smallskip
  \begin{flushleft}
      \scriptsize Direct effects capture the effect of policy on case
      growth holding behavior, information, and confounders
      constant. Direct effects are given by ${\pcolor \pi}$ in
      equation (\ref{eq:R1}). Indirect effects capture how policy
      changes behavior and behavior shift case growth. They are given
      by ${\bcolor \alpha}$ from (\ref{eq:R1}) times ${\pcolor \beta}$
      from (\ref{eq:R2}). The total effect is
      ${\pcolor \pi} + {\pcolor \beta} {\bcolor \alpha}$. Column
      ``PI$\to$Y Coefficients'' shows the coefficient estimates from
      \ref{eq:R4}. Columns ``Difference''    are  the differences between
      the estimates from (\ref{eq:R4}) and the combination of
      (\ref{eq:R1}) and (\ref{eq:R2}) while column ``Average'' are their averages.
      Standard errors are computed by
      bootstrap and clustered on state.
    \end{flushleft}
  \end{minipage}
\end{table}
\clearpage}

\subsection{The Total Effect of Policies on Case Growth\label{total-policy-effect}}

In this section, we focus our analysis on policy effects when we hold
information constant. The estimated effect of policy on
behavior in Table \ref{tab:PItoB} and those of policies and
behavior on case/death growth in Table  \ref{tab:BPItoY}  can be
combined to calculate the total effect of policy as well as its
decomposition into  direct and indirect effects.

The first three columns of Table \ref{tab:dieff-si} show the direct
(holding behavior constant) and indirect (through behavior changes)
effects of policy under a specification that excludes national information variables. These are computed from the specification with
national cases or deaths included as information (columns (1)-(4) of
Table \ref{tab:PItoB} and column (1) of Table \ref{tab:BPItoY}). The estimates imply
that all policies combined would reduce the growth rate of cases  and deaths by
0.69 and 0.90, respectively,   out of which more than one-half to
two-third is attributable to the indirect effect through their impact on behavior.  The estimate also indicates that the
effect of mandatory masks for employees is mostly direct.

We can also examine the total effect of policies and information on
case  or death growth, by estimating (\ref{eq:R4}). The coefficients on policy
in this regression combine both the direct and indirect effects.

Table \ref{tab:PtoY} shows
the full set of coefficient estimates for
(\ref{eq:R4}). The results are broadly consistent with what we found
above.  As in Table \ref{tab:PItoB}, the effect of school closures is
sensitive to the inclusion of national information variables.  Also as
above, mask mandates have a significant negative effect on growth
rates.

Similarly to Table  \ref{tab:dieff-si}, the first three columns of  Table \ref{tab:dieff-con-si} report  the estimated direct and indirect effects of policy but impose that masks for employees only affect cases/deaths directly without affecting behavior and that business closure policies only affect  cases/deaths indirectly through their effects on behavior.\footnote{These are computed from estimating the specification in columns (1)-(4) of
Table \ref{tab:PItoB} and column (1) of Table \ref{tab:BPItoY} but imposing that the coefficient of masks for employees is zero in (\ref{eq:R1}) and that the coefficient of business closure policies is zero in (\ref{eq:R2}).} The estimated total effect of masks for employees in the third column of Table \ref{tab:dieff-con-si} is higher than that of Table  \ref{tab:dieff-si}. Similarly, the total effect of business closure policies is estimated to be larger in Table \ref{tab:dieff-con-si} than in  Table  \ref{tab:dieff-si}.

Column ``Difference''  in Tables   \ref{tab:dieff-si} and \ref{tab:dieff-con-si}  show the
difference between the estimate of (\ref{eq:R4}) in column  ``PI$\to$Y Coefficient''   and the implied estimate from
(\ref{eq:R1})-(\ref{eq:R2}) in   column ``Total.''   Differences  are generally small except for the coefficient of closed K-12 schools and  the sum of all policies in Table   \ref{tab:dieff-si}. The large differences in school closures  may be due to the  difficulty in identifying the effect of school closures given a lack of cross-sectional variation. Imposing the coefficients of masks for employees and business closure policies to be zero in (\ref{eq:R1}) and  (\ref{eq:R2}), respectively, does not increase differences between  ``Total" and  ``PI$\to$Y Coefficient''  as reported in the last column of  Table \ref{tab:dieff-con-si}.

 Tables  \ref{tab:dieff} and \ref{tab:dieff-con}  present the results for a specification that includes national information variables, where the estimates on masks for employees are similar to those in  Tables  \ref{tab:dieff-si} and \ref{tab:dieff-con-si}. Column ``Difference''  of Table \ref{tab:dieff-con} indicates that the restrictions of zero coefficients for masks for employees in (\ref{eq:R1}) and for business closure policies in  (\ref{eq:R2}) are not statistically rejected.



Column ``Average'' of  Tables \ref{tab:dieff-si} and \ref{tab:dieff}   reports the average of
``Total'' and ``PI$\to$Y Coefficient'' columns.  The average is an appealing and simple way
to combine the two estimates of the total effect: one relying on the causal structure and another inferred from a direct estimation of equation (PI $\to$ Y).
We shall be using the average estimate in generating the counterfactuals in the next section. Turning to the results, the estimates of Tables \ref{tab:dieff-si} and \ref{tab:dieff}  imply that all policies combined would reduce
\(\Delta \log \Delta D\) by  0.93 and  0.39, respectively. For
comparison, the median of \(\Delta \log \Delta D_{it}\) reached its
peak in mid-March of about 1.3 (see Figure \ref{fig:growthq} in the
appendix). Since then it has declined to near 0. Therefore,  -0.93 and -0.39 imply
that policy changes can account for roughly one-third to two-third of the observed
decrease in death growth.  The remainder of the decline is likely due
to changes in behavior from information.

\section{Sensitivity Analysis}\label{sec:sensitivity}

\subsection{Specifications, Timings, and Flexible Controls via Machine Learning}
\begin{figure}[ht]
  \caption{Estimated Coefficients for  Policy Variables: Sensitivity Analysis \label{fig:whisker}}\bigskip
  \begin{minipage}{\linewidth}
    \centering
   {\textbf{(A)  masks for employees}}\\
    \medskip
    \begin{tabular}{cc}
 $\quad$  (i) baseline timing$^\dagger$ &$\quad$ (ii) alternative timing$^\ddagger$\\
      \includegraphics[width=0.5\textwidth]{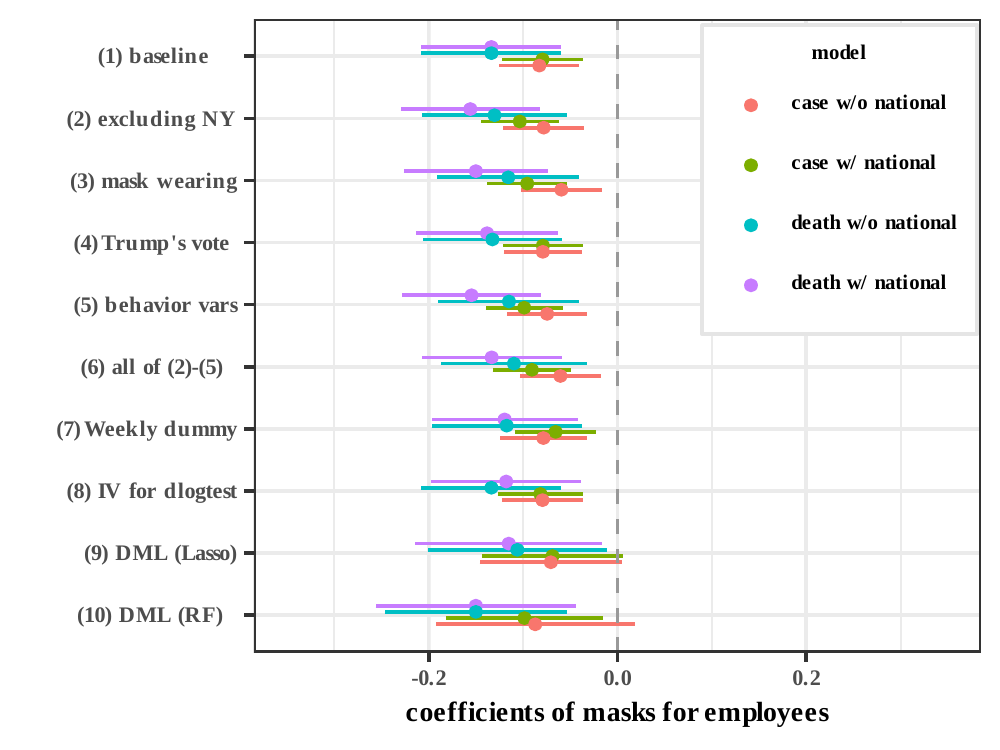}
      &
      \includegraphics[width=0.5\textwidth]{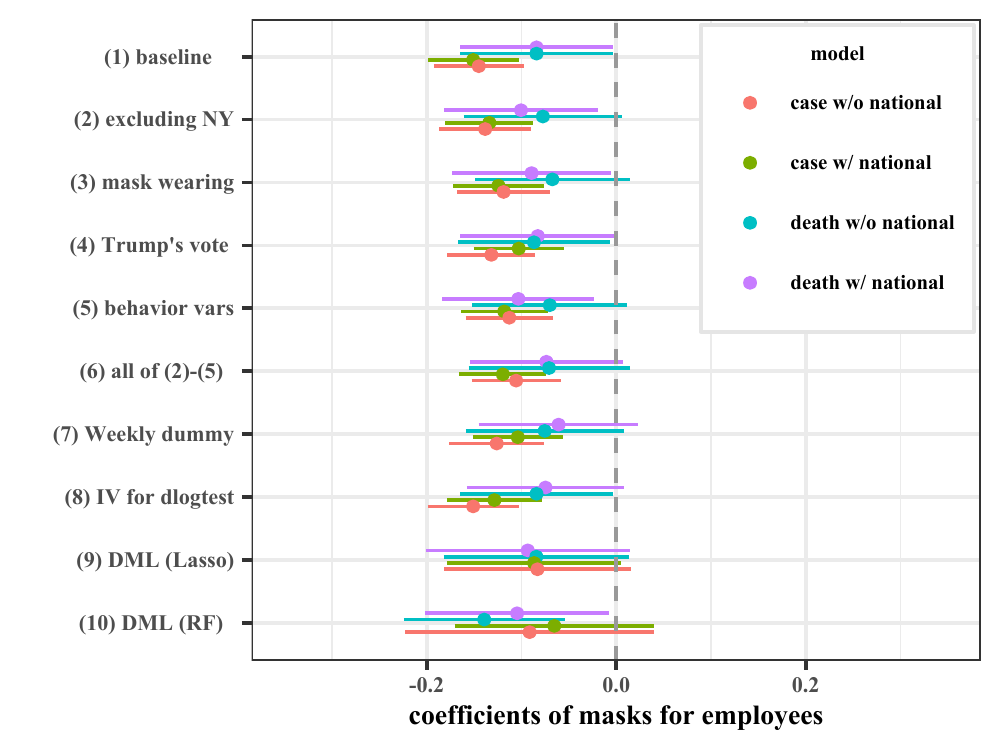}
    \end{tabular}
  \end{minipage} \\\smallskip
    \begin{minipage}{\linewidth}
    \centering
     {\textbf{(B)  closed K-12 Schools}}\\
    \medskip
    \begin{tabular}{cc}
 $\quad$  (i) baseline timing$^\dagger$ &$\quad$ (ii) alternative timing$^\ddagger$\\
      \includegraphics[width=0.5\textwidth]{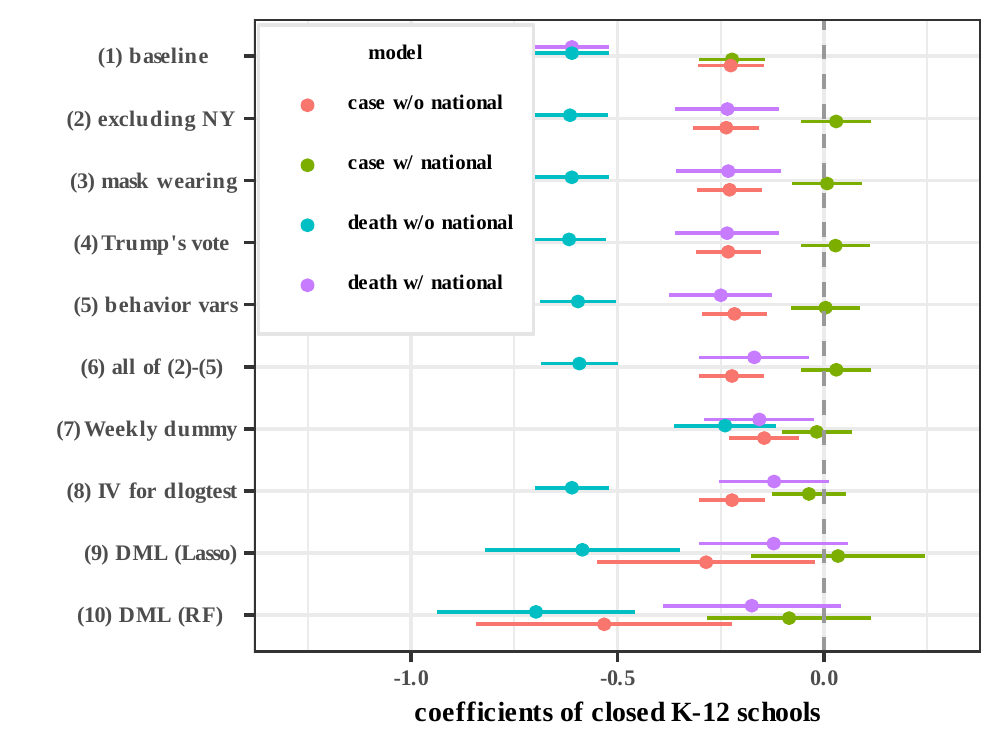}
      &
      \includegraphics[width=0.5\textwidth]{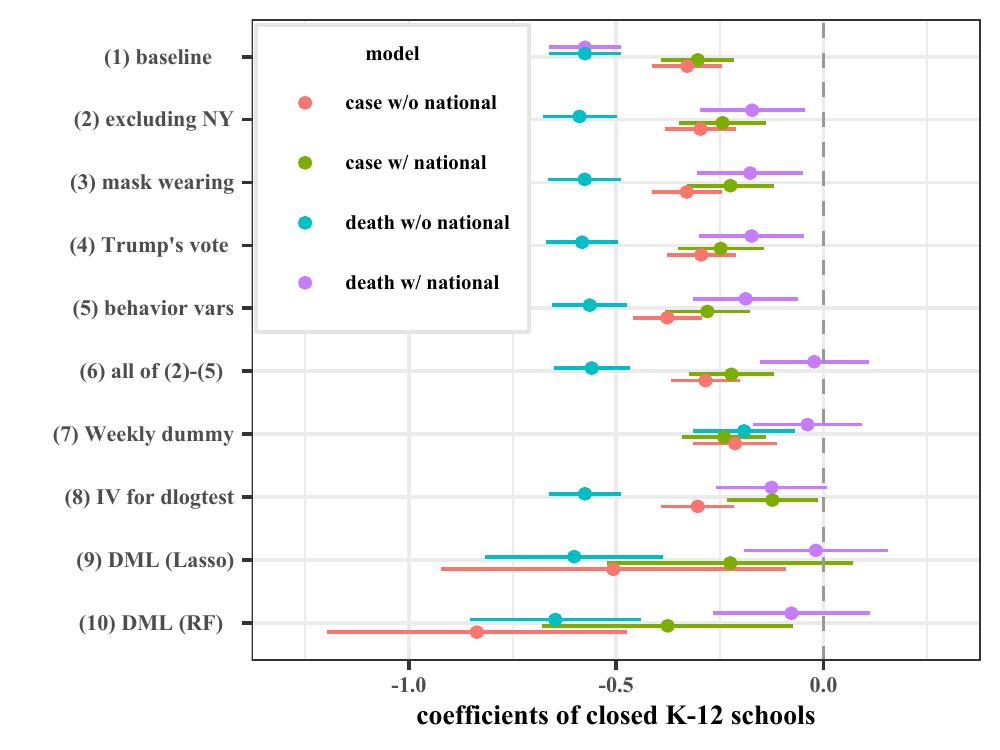}
          \end{tabular}
  \end{minipage}
    \begin{flushleft}
      \footnotesize
      $^\dagger$The times from exposure to case confirmation and death reporting  are assumed to be 14 and 21 days, respectively. $^\ddagger$The times from exposure to case confirmation and death reporting  are assumed to be 7 and 24 days, respectively.
    \end{flushleft}
\end{figure}

In this section, we provide sensitivity analysis by estimating (\ref{eq:R4}) with alternative specifications and methods. Figure \ref{fig:whisker} shows  the 90\% confidence intervals of   coefficients of (A) masks for employees, (B)  closed K-12 school, (C) stay-at-home, and (D) the average variable of  closed movie theaters, closed restaurants, and closed non-essential businesses for the following specifications and estimation methods:
  \begin{itemize}
  \item[(1)] Baseline specification  in columns (1) and (3) of Table  \ref{tab:PtoY}.
  \item[(2)]  Exclude  the state of New York from the sample because it may be viewed as an outlier in the early pandemic period.
  \item[(3)]   Add  the percentage of people who wear masks to protect themselves in March and April as a confounder for unobserved  personal risk-aversion and initial attitude toward mask wearing.\footnote{ The survey is conducted online by YouGov and is based on the interviews of 89,347 US adults aged 18 and over between March 26-April 29, 2020.  The survey question is ``Which, if any, of the following measures have you taken in the past 2 weeks to protect yourself from the Coronavirus (COVID-19)?''.}
   \item[(4)]   Add the log of Trump's vote share  in the 2016 presidential election as a confounder for unobserved private behavioral response.
   \item[(5)]   Add  past behavior variables  as information used to set policies. Under this specification, our causal interpretation is valid when policy variables are sufficiently random conditional on past behavior variables.
   \item[(6)]   Include all additional controls in (3)-(5) with the sample that excludes New York as in (2).
   \item[(7)] Add weekly dummies  to the baseline specification.
   \item[(8)] Baseline specification estimated by instrumenting $\Delta \log T_{it}$ with one week lagged  logarithm  value of the number of tests per 1000 people.
   \item[(9)]   Estimated by Double Machine Learning (DML) \citep[e.g.,][]{chernozhukov18} with Lasso to reduce dimensionality while including all additional controls in (3)-(5).     \item[(10)]  Estimated by DML with Random Forest to reduce dimensionality  and capture some nonlinearities while including all additional controls in (3)-(5).   
  \end{itemize}
In Figure \ref{fig:whisker}, ``red'', ``green'', ``blue'', and ``purple" indicate the regression models for case growth without national information variables, case growth with national information variables, death growth without national information variables, and death growth with national information variables, respectively. The left panel of ``(i) baseline timing'' assume that the times from exposure to case confirmation and death reporting are 14 and 21 days, respectively,  while they are 7 and 24 days, respectively, in the right panel of ``(ii) alternative timing.''\footnote{This alternative timing assumption is motivated by the lower bound estimate of median times from exposure to case confirmation or death reporting in Table 2 of \url{https://www.cdc.gov/coronavirus/2019-ncov/hcp/planning-scenarios.html}, which are based on data received by CDC through June 29, 2020.  We thank C. Jessica E. Metcalf for recommending us to do a sensitivity analysis on the timing assumption while suggesting this reference to us.}

Panel (A) of Figure  \ref{fig:whisker} illustrates that the estimated coefficients of mask mandates are negative and significant in most specifications, methods, and timing assumptions, confirming the importance of mask policy on reducing case and death growths.  

In Panel (B) of  Figure  \ref{fig:whisker}, many estimates of closures of K-12 schools  suggest that  the effect of school closures is large. The visual evidence on growth rates for states with and without school closures in Figure \ref{fig:growthpolicies-school} also suggests that there may be a potentially large effect, though the history is very short.  This evidence is consistent
with the emerging evidence of prevalence of Covid-19 among children  \citep{Lee2020jama,Szablewski2020cdc}. \cite{children:nature} find that although children's
transmission and susceptibility rates are half that of ages 20-30,
children's contact rates are much higher.
This type of evidence, as well as, evidence that children carry viral
loads similar to older people (\cite{children:germany}), led Germany to make the early decision of closing schools.

Our estimates of school closures substantially vary  across  specifications, however. In particular, the estimated effects of school closures on case/death growth become notably smaller once national cases/deaths or weekly dummies are controlled for.
In the US state-level data,  there is little variation across states in the timing of school closures.
Consequently, its estimate is particularly sensitive to an inclusion of some aggregate variables such as national cases or weekly dummies.  Given this sensitivity,  there still exists a lot of uncertainty as to the magnitude of the effect of school closures. Any analyses of re-opening plans need to be aware of this uncertainty.  An important research question is how to resolve this uncertainty using additional data sources.

Panel (C) of Figure  \ref{fig:whisker} indicates that the estimated coefficients of stay-at-home orders  are generally negative and often significant, providing evidence that stay-at-home orders reduce the spread of COVID-19, although the estimates are sometimes sensitive to timing assumptions.  Panel (D) shows that the estimated coefficients of business closure policies substantially vary across specifications, providing a mixed evidence for the effect of closures of businesses on case/death growth.

   \addtocounter{figure}{-1}
\begin{figure}[ht]
  \caption{Estimated Coefficients for Policy Variables:  Sensitivity Analysis (cont.) \label{fig:whisker-2}}\bigskip
  \begin{minipage}{\linewidth}
    \centering
   {\textbf{(C)  stay-at-home orders}}\\
    \medskip
    \begin{tabular}{cc}
 $\quad$  (i) baseline timing$^\dagger$ &$\quad$ (ii) alternative timing$^\ddagger$\\
      \includegraphics[width=0.5\textwidth]{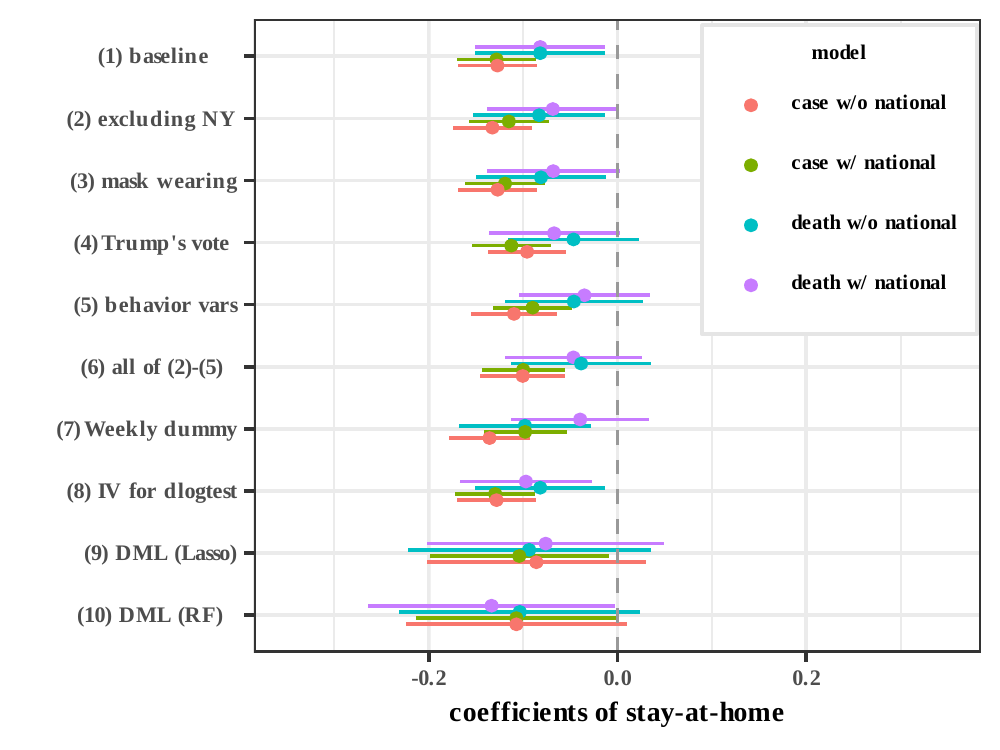}
      &
      \includegraphics[width=0.5\textwidth]{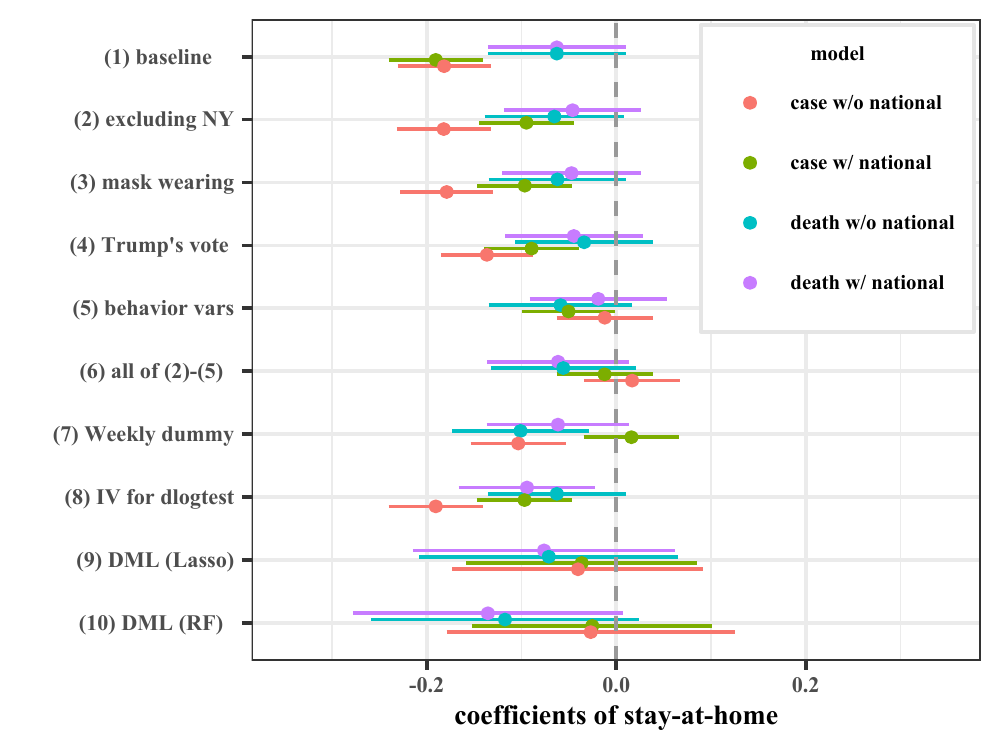}
    \end{tabular}
  \end{minipage} \\ \smallskip
    \begin{minipage}{\linewidth}
    \centering
     {\textbf{(D)  Average of  closed movie theaters, closed restaurants, and closed  businesses}}\\
    \medskip
    \begin{tabular}{cc}
 $\quad$  (i) baseline timing$^\dagger$ &$\quad$ (ii) alternative timing$^\ddagger$\\
      \includegraphics[width=0.5\textwidth]{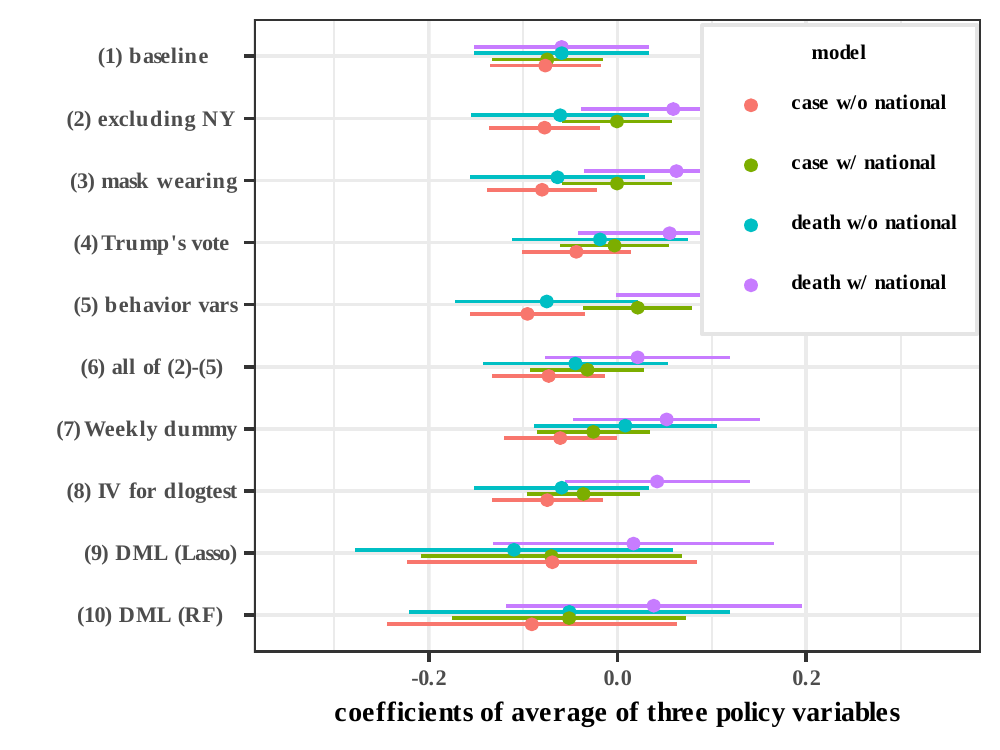}
      &
      \includegraphics[width=0.5\textwidth]{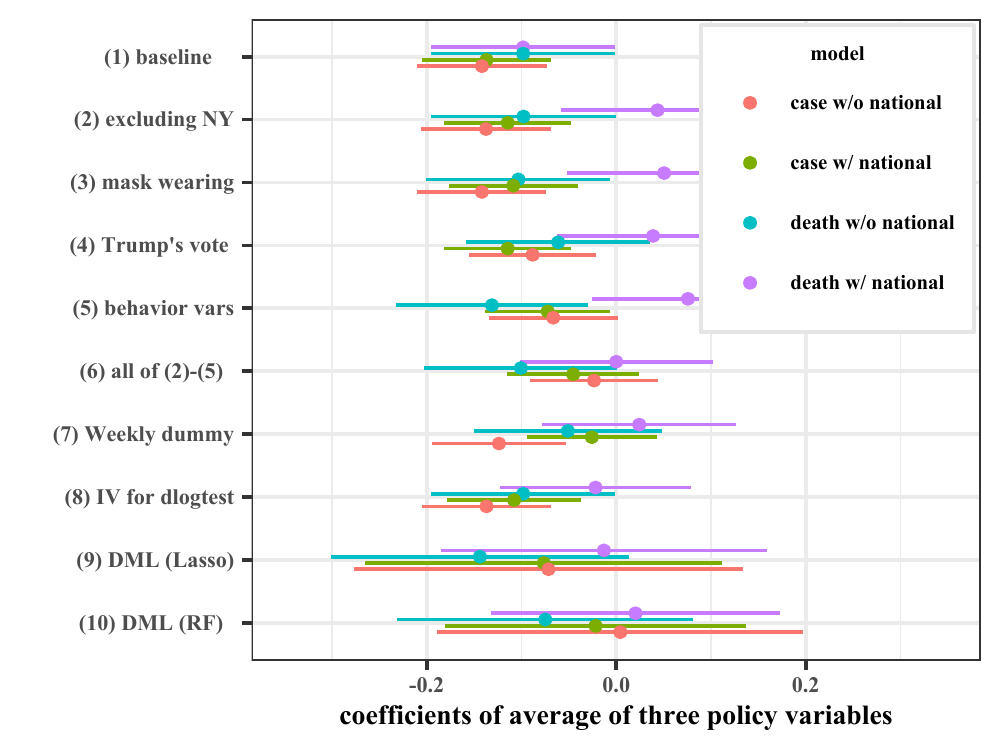}
          \end{tabular}
  \end{minipage}
    \begin{flushleft}
      \footnotesize
      $^\dagger$The times from exposure to case confirmation and death reporting  are assumed to be 14 and 21 days, respectively. $^\ddagger$The times from exposure to case confirmation and death reporting  are assumed to be 7 and 24 days, respectively.
    \end{flushleft}
\end{figure}

 \begin{figure}
  \caption{Case and death growth conditional on school closures \label{fig:growthpolicies-school}}
  \begin{minipage}{\linewidth}
    \centering
    \begin{tabular}{cc}
       \includegraphics[width=0.483\textwidth]{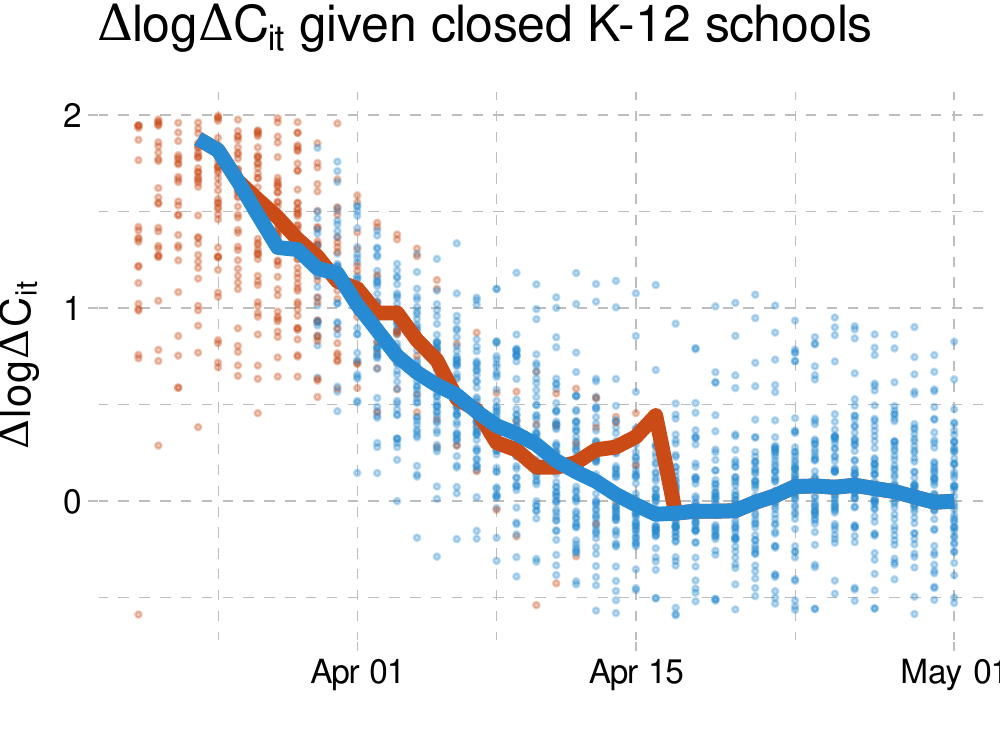}
      &
        \includegraphics[width=0.483\textwidth]{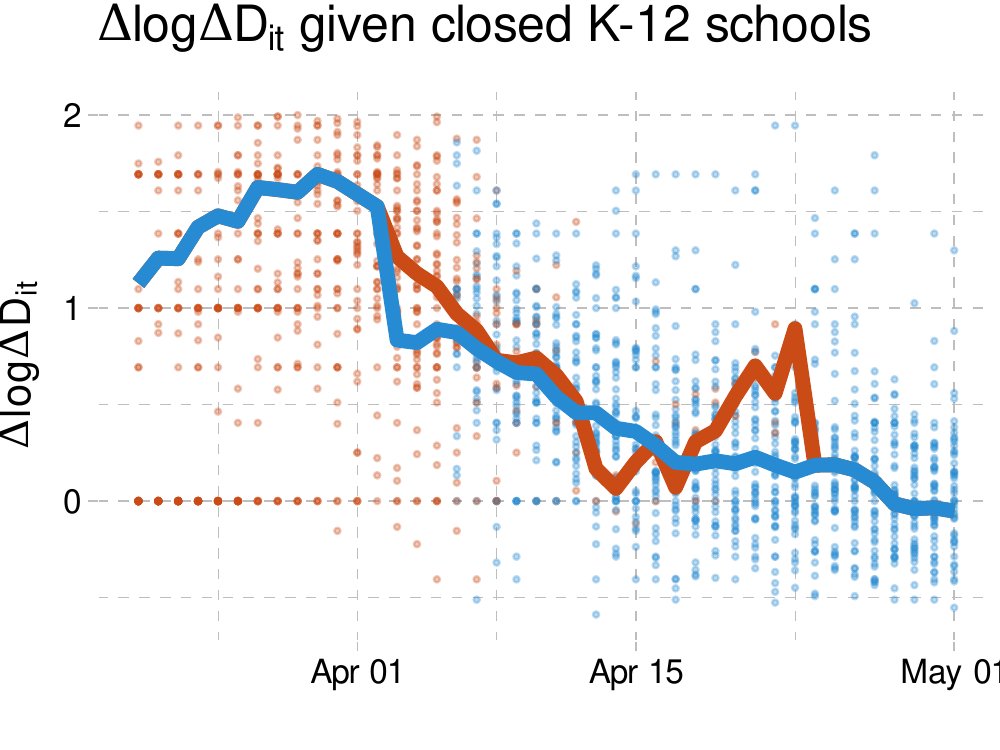}
    \end{tabular}
      \footnotesize In these figures, red points are the case or death
      growth rate in states without each policy 14 (or 21 for deaths)
      days earlier. Blue points are states with each policy 14 (or 21
      for deaths) days earlier. The red line is the average across
      states without each policy. The blue line is the average across
      states with each policy.
  \end{minipage}
\end{figure}

\subsection{Fixed Effects Specification}

Table \ref{tab:PtoY-fe} presents the results of estimating (\ref{eq:R4})  with  state fixed effects and weekly dummies.  Because fixed effects estimator could be substantially biased when the time dimension is relatively short, we report not only the standard fixed effects estimator in columns (1) and (3) but also the debiased fixed effects estimator \citep[e.g.,][]{chen2019mastering} in columns (2) and (4), where the state-level clustered bootstrap standard errors are reported in all columns.

The estimated coefficients of masks for employees largely confirm our finding that mandatory mask policies reduce the case and death growths. The estimated coefficients of stay-at-home orders and business closures are negative in columns (1) and (2) but their magnitudes as well as statistical significance are somewhat sensitive to whether bias corrections are applied or not. 

The results in Table \ref{tab:PtoY-fe} should be interpreted with caution. The fixed effects approach may not be preferred to random effects approach here because the former relies on long time and cross-sectional histories but, in our data, the effective time-dimension is short and the number of states is not large. Furthermore, the fixed effects approach could suffer more from  measurement errors, which could be a concern for our behavior and policy variables.

  \afterpage{
 \begin{landscape}
\begin{table}[!htbp] \centering
 \caption{\label{tab:PtoY-fe}
Fixed Effects Specification: the Total Effect of Policies on Case and Death Growth ($PI \to Y$)}\vspace{-0.3cm}
 \begin{minipage}{\linewidth}
   \resizebox{\textwidth}{!}{
   \centering
   \tiny
   \begin{tabular}{c|c}
   \begin{minipage}{0.51\linewidth}
     \centering
     \begin{tabular}{@{\extracolsep{1pt}}lcccc} 
\\[-1.8ex]\hline 
\hline \\[-1.8ex] 
 & \multicolumn{4}{c}{\textit{Dependent variable:}} \\ 
\cline{2-5} 
 & \multicolumn{4}{c}{$\Delta \log \Delta C_{it}$} \\ 
\\[-1.8ex] & (1) & (2) & (3) & (4)\\ 
\hline \\[-1.8ex] 
 lag(masks for employees, 14) & $-$0.103$^{**}$ & $-$0.271$^{***}$ & $-$0.097$^{**}$ & $-$0.255$^{***}$ \\ 
  & (0.044) & (0.075) & (0.044) & (0.071) \\ 
  lag(closed K-12 schools, 14) & $-$0.023 & 0.085 & $-$0.050 & 0.042 \\ 
  & (0.060) & (0.098) & (0.066) & (0.094) \\ 
  lag(stay at home, 14) & $-$0.123$^{**}$ & $-$0.088 & $-$0.099$^{**}$ & $-$0.055 \\ 
  & (0.051) & (0.068) & (0.049) & (0.075) \\ 
  lag(business closure policies, 14) & $-$0.080 & $-$0.162$^{*}$ &  &  \\ 
  & (0.076) & (0.086) &  &  \\ 
  lag(closed movie theaters, 14) &  &  & 0.049 & 0.077 \\ 
  &  &  & (0.068) & (0.084) \\ 
  lag(closed restaurants, 14) &  &  & $-$0.031 & $-$0.068 \\ 
  &  &  & (0.053) & (0.065) \\ 
  lag(closed non-essent bus, 14) &  &  & $-$0.099$^{**}$ & $-$0.166$^{***}$ \\ 
  &  &  & (0.047) & (0.063) \\ 
  lag($\Delta \log \Delta C_{it}$, 14) & 0.063$^{**}$ & 0.079$^{***}$ & 0.061$^{**}$ & 0.076$^{***}$ \\ 
  & (0.029) & (0.026) & (0.029) & (0.028) \\ 
  lag($\log \Delta C_{it}$, 14) & $-$0.216$^{***}$ & $-$0.185$^{***}$ & $-$0.214$^{***}$ & $-$0.181$^{***}$ \\ 
  & (0.020) & (0.032) & (0.021) & (0.033) \\ 
  $\Delta \log T_{it}$ & 0.116$^{***}$ & 0.116$^{***}$ & 0.118$^{***}$ & 0.147$^{***}$ \\ 
  & (0.042) & (0.042) & (0.042) & (0.047) \\ 
 \hline \\[-1.8ex] 
Observations & 3,825 & 3,825 & 3,825 & 3,825 \\ 
R$^{2}$ & 0.782 & 0.782 & 0.782 & 0.782 \\ 
Adjusted R$^{2}$ & 0.778 & 0.778 & 0.778 & 0.778 \\ 
\hline 
\hline \\[-1.8ex] 
\textit{Note:}  & \multicolumn{4}{r}{$^{*}$p$<$0.1; $^{**}$p$<$0.05; $^{***}$p$<$0.01} \\ 
\end{tabular} 
   \end{minipage}
     &
   \begin{minipage}{0.51\linewidth}
     \centering
     \begin{tabular}{@{\extracolsep{1pt}}lcccc} 
\\[-1.8ex]\hline 
\hline \\[-1.8ex] 
 & \multicolumn{4}{c}{\textit{Dependent variable:}} \\ 
\cline{2-5} 
 & \multicolumn{4}{c}{$\Delta \log \Delta D_{it}$} \\ 
\\[-1.8ex] & (1) & (2) & (3) & (4)\\ 
\hline \\[-1.8ex] 
 lag(masks for employees, 21) & $-$0.136$^{***}$ & $-$0.144$^{*}$ & $-$0.129$^{***}$ & $-$0.109 \\ 
  & (0.047) & (0.074) & (0.047) & (0.075) \\ 
  lag(closed K-12 schools, 21) & $-$0.115 & $-$0.007 & $-$0.161 & $-$0.026 \\ 
  & (0.108) & (0.091) & (0.123) & (0.103) \\ 
  lag(stay at home, 21) & $-$0.047 & $-$0.097 & $-$0.010 & $-$0.151 \\ 
  & (0.077) & (0.099) & (0.072) & (0.112) \\ 
  lag(business closure policies, 21) & $-$0.081 & $-$0.166 &  &  \\ 
  & (0.099) & (0.137) &  &  \\ 
  lag(closed movie theaters, 21) &  &  & 0.041 & $-$0.035 \\ 
  &  &  & (0.099) & (0.128) \\ 
  lag(closed restaurants, 21) &  &  & 0.015 & $-$0.029 \\ 
  &  &  & (0.086) & (0.097) \\ 
  lag(closed non-essent bus, 21) &  &  & $-$0.143$^{**}$ & $-$0.024 \\ 
  &  &  & (0.067) & (0.084) \\ 
  lag($\Delta \log \Delta D_{it}$, 21) & 0.009 & 0.083$^{*}$ & 0.006 & 0.084 \\ 
  & (0.031) & (0.045) & (0.032) & (0.055) \\ 
  lag($\log \Delta D_{it}$, 21) & $-$0.113$^{***}$ & $-$0.113$^{***}$ & $-$0.108$^{***}$ & $-$0.232$^{***}$ \\ 
  & (0.018) & (0.018) & (0.018) & (0.040) \\ 
 \hline \\[-1.8ex] 
Observations & 3,468 & 3,468 & 3,468 & 3,468 \\ 
R$^{2}$ & 0.536 & 0.536 & 0.537 & 0.537 \\ 
Adjusted R$^{2}$ & 0.527 & 0.527 & 0.528 & 0.528 \\ 
\hline 
\hline \\[-1.8ex] 
\textit{Note:}  & \multicolumn{4}{r}{$^{*}$p$<$0.1; $^{**}$p$<$0.05; $^{***}$p$<$0.01} \\ 
\end{tabular} 
   \end{minipage}

   \end{tabular}
   }
   \begin{flushleft}
     \scriptsize   All specifications include weekly dummies and state fixed effects. Dependent variable is the weekly growth rate of
     confirmed cases (in the left panel) or deaths (in the right
     panel) as defined in equation (\ref{eq:y}).   Columns (1) and (3) report the standard fixed effects estimator  while Columns (2) and (4) report the fixed effects estimator with bias corrections.    Standard errors are computed by multiplier bootstrap clustered at the state level.

   \end{flushleft}
 \end{minipage}
\end{table}
\end{landscape}
}

\section{Empirical Evaluation of Counterfactual Policies}\label{counterfactuals}

We now turn our focus to dynamic feedback effects. Policy and behavior
changes that reduce case and death growth today can lead to a more
optimistic, riskier behavior in the future, attenuating longer run
effects.  We perform the main counterfactual experiments using the
average of two estimated coefficients as reported in column
``Average'' of Table \ref{tab:dieff-con-si} under a specification that
excludes the number of past national cases and deaths from information
variables and constrains masks to have only direct effects and
business policies only indirect effects.\footnote{Results (not shown)
  using the unconstrained estimates in Table \ref{tab:dieff-si} are
  very similar.} Details of the counterfactual computations are in
appendix \ref{app:counterfactuals}.

\subsection{Business Mask Mandate}

\begin{figure}[ht]
  \caption{Effect of mandating masks on April 1st in Washington State \label{fig:WA-mask}}
  \begin{minipage}{\linewidth}
    \centering
    \begin{tabular}{ccc}
      \multicolumn{3}{c}{\textbf{Cases}} \\
      \includegraphics[width=0.31\textwidth]{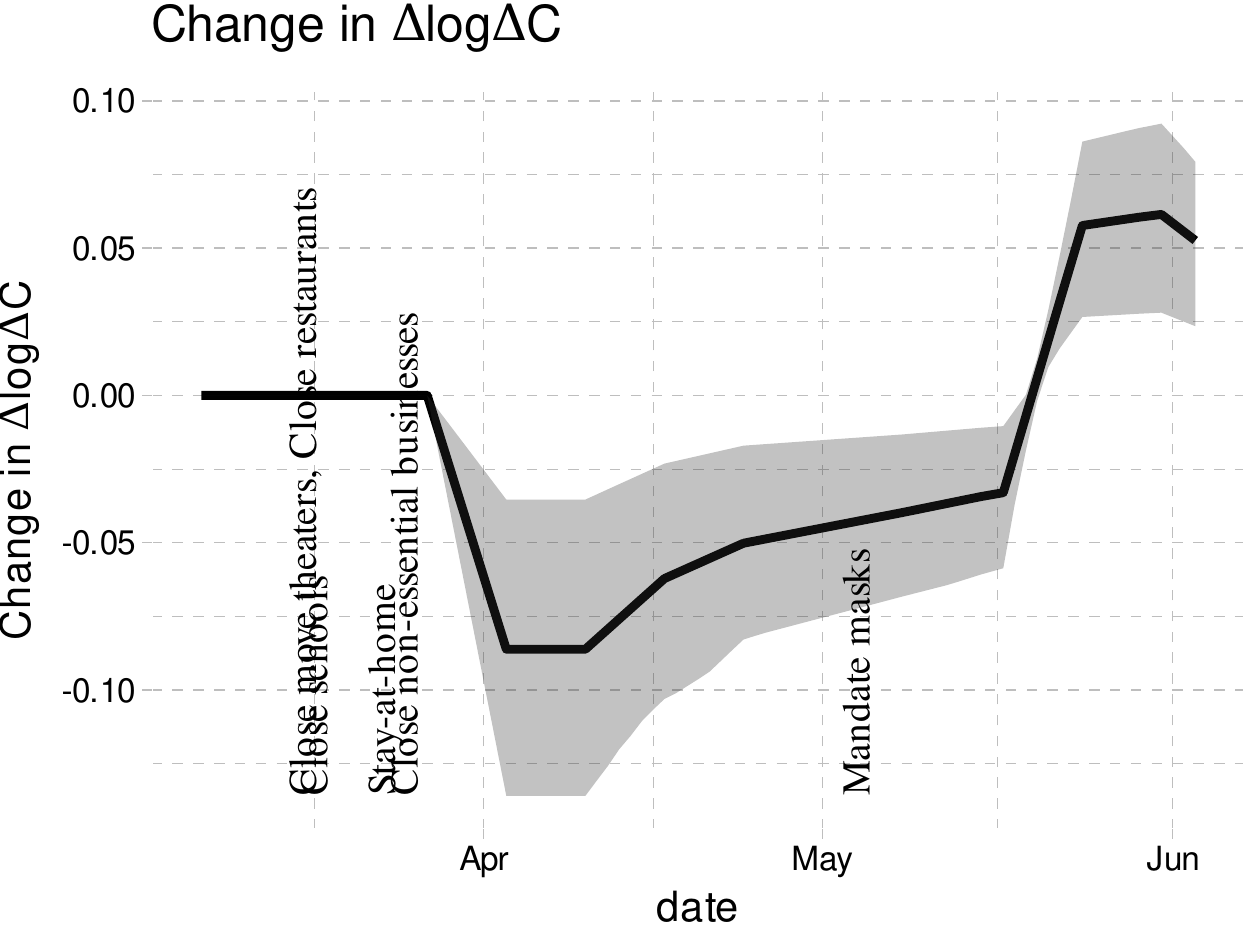}
      &
        \includegraphics[width=0.31\textwidth]{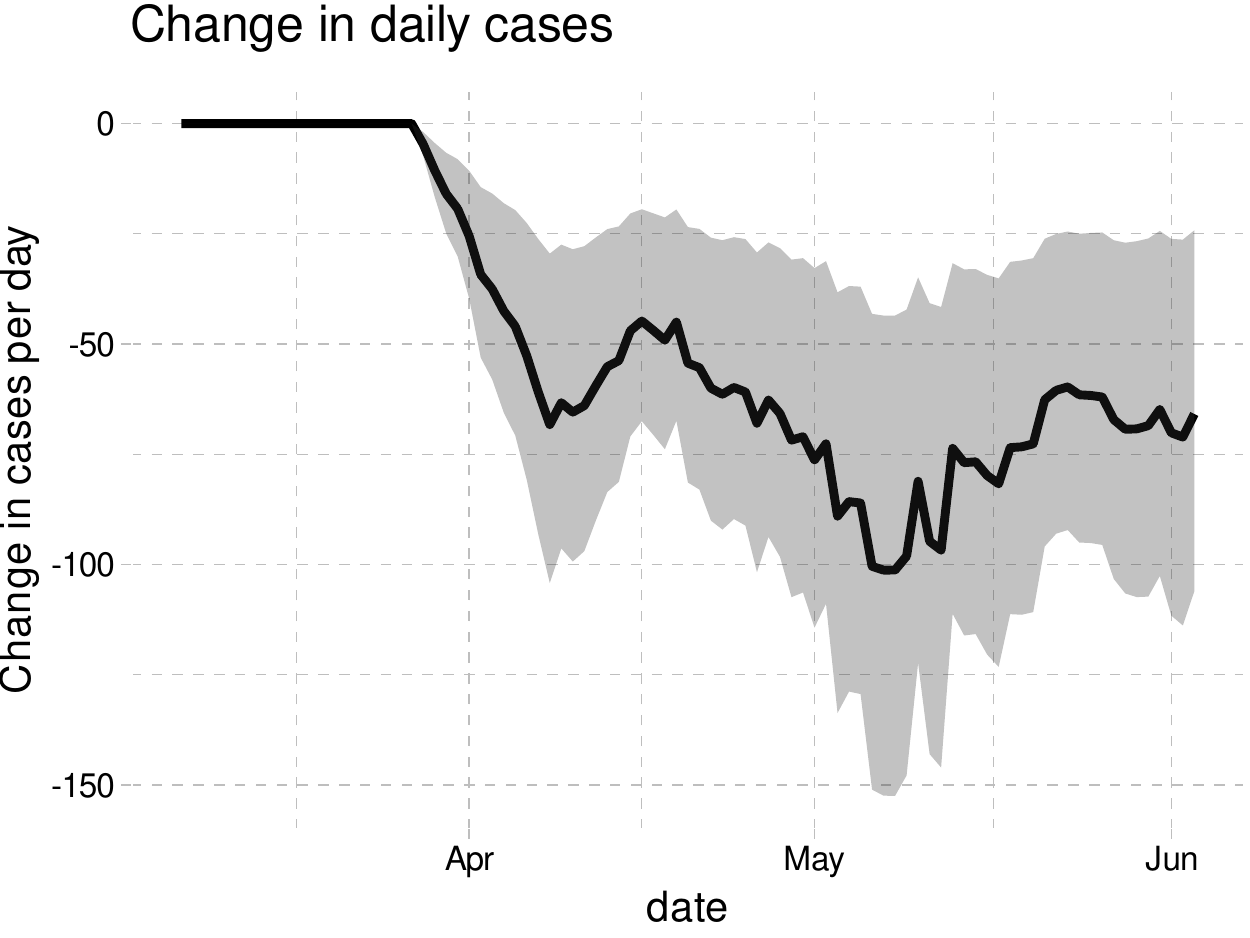}
      &
        \includegraphics[width=0.31\textwidth]{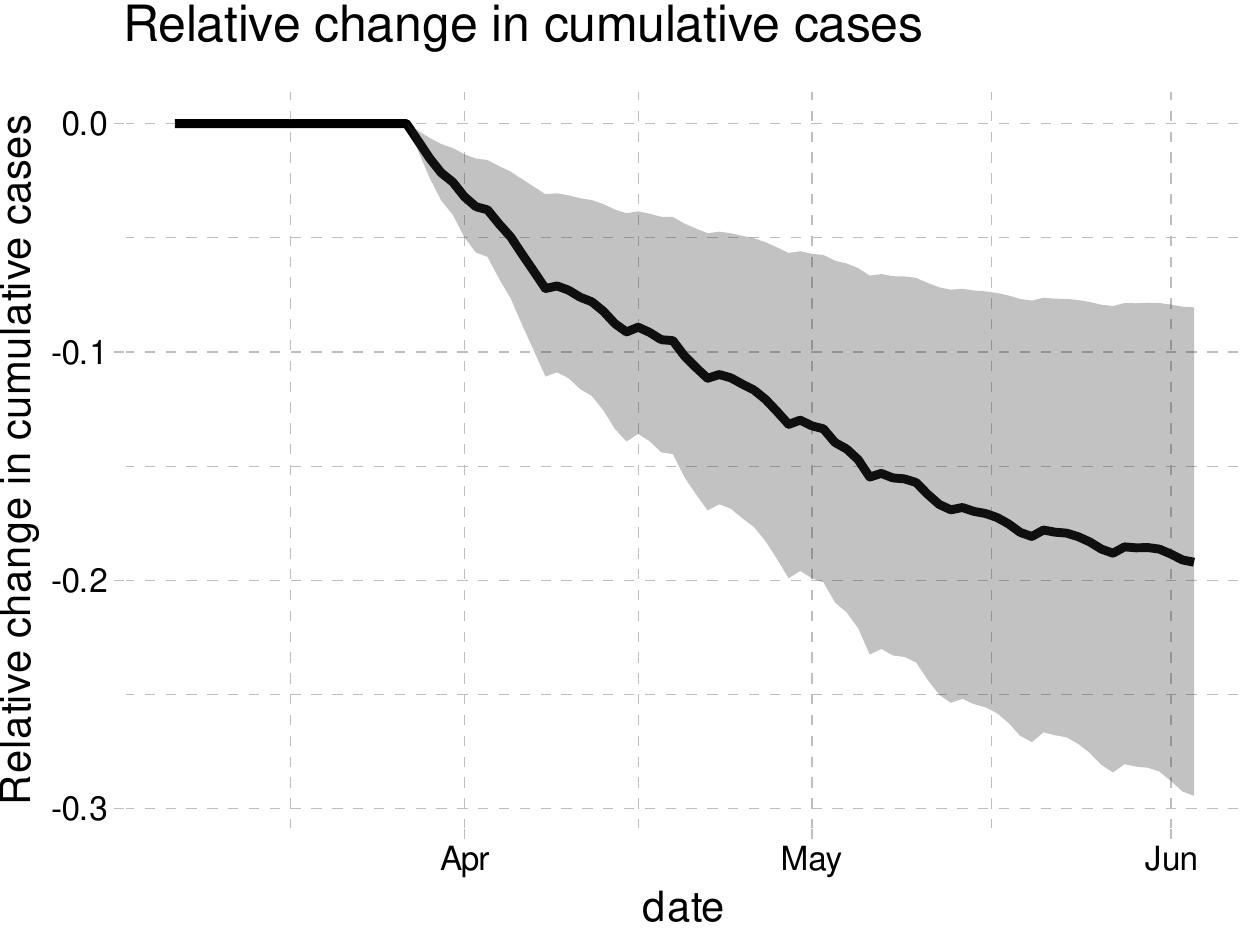}
      \\
      \\
      \multicolumn{3}{c}{\textbf{Deaths}}
      \\
      \includegraphics[width=0.31\textwidth]{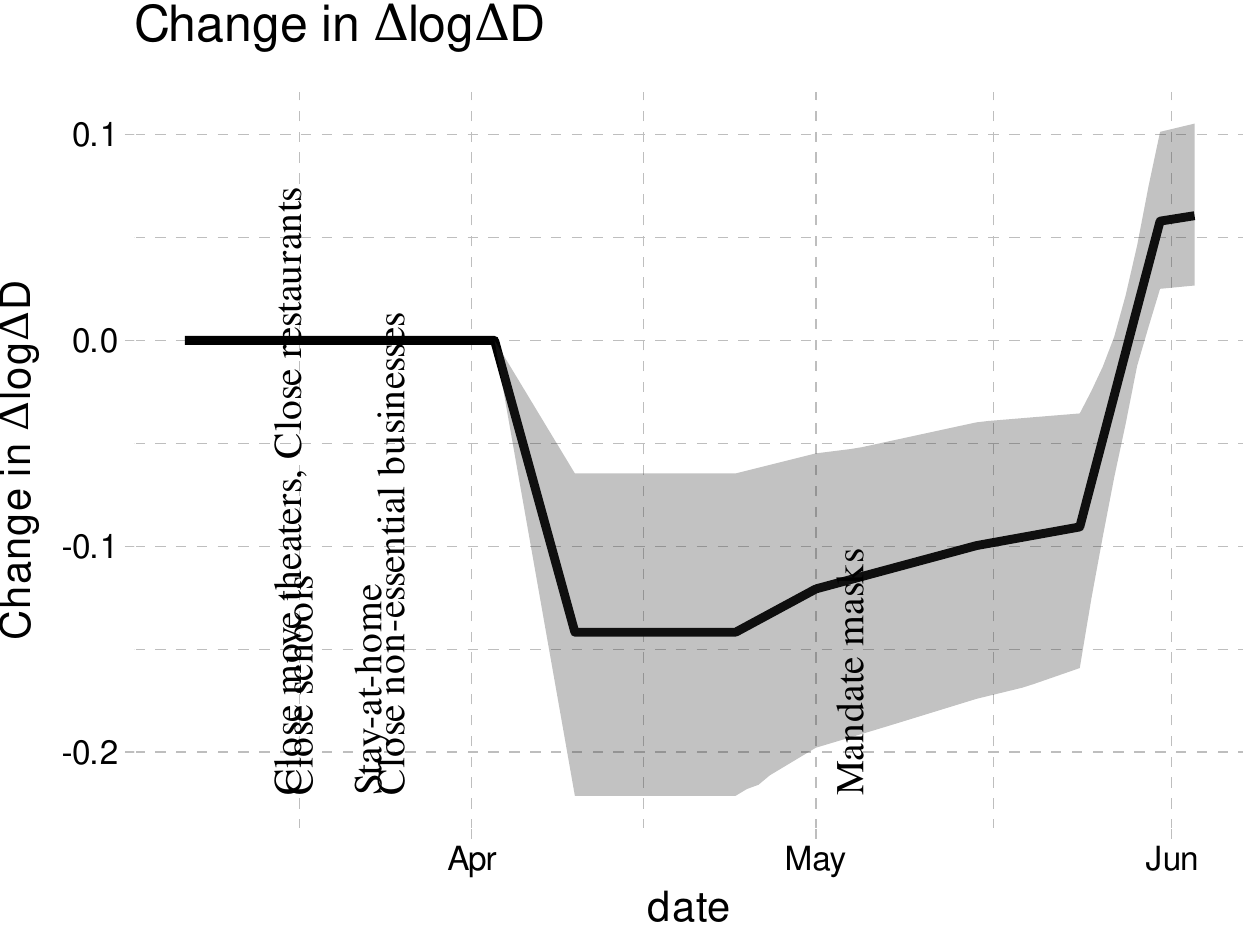}
      &
        \includegraphics[width=0.31\textwidth]{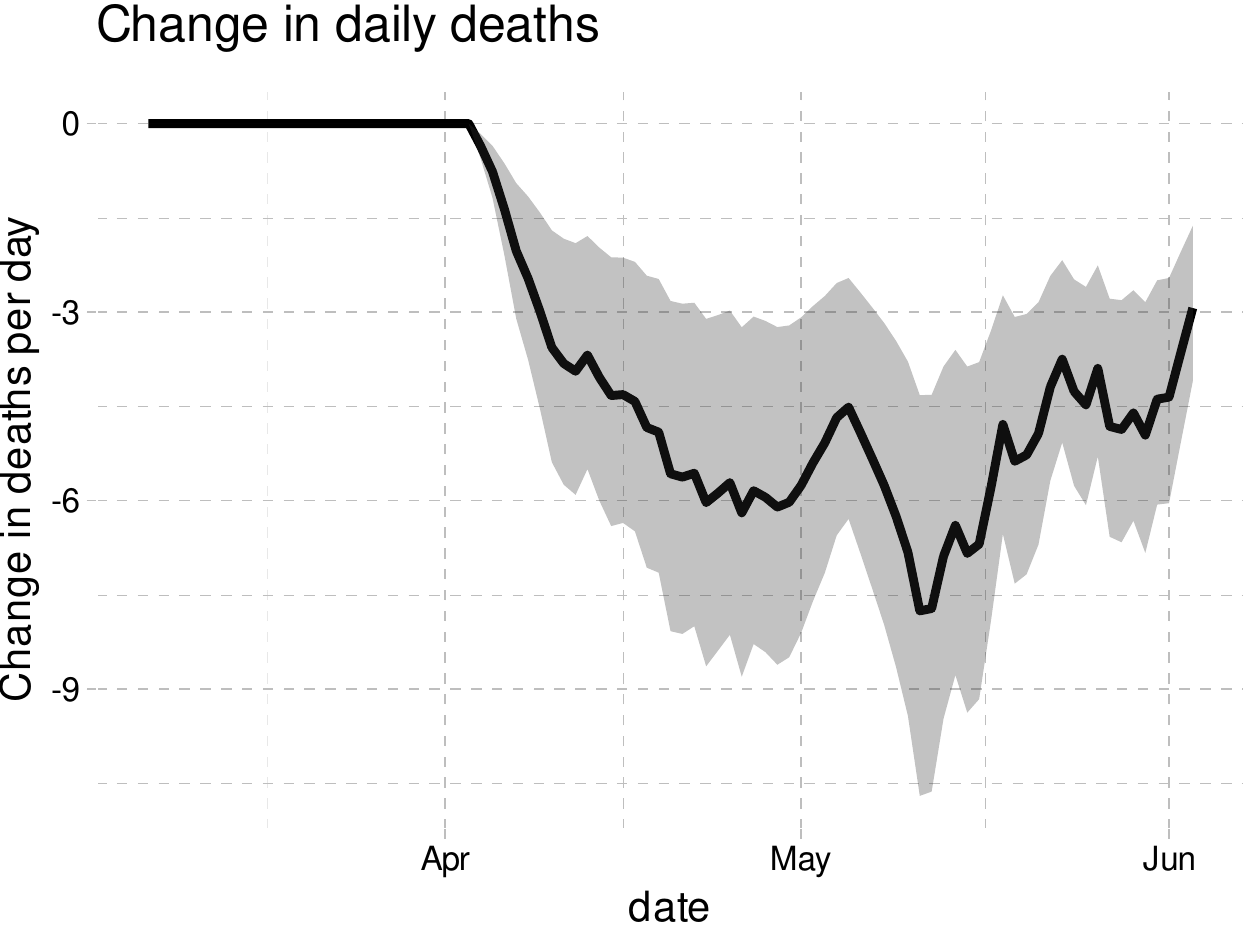}
      & \includegraphics[width=0.31\textwidth]{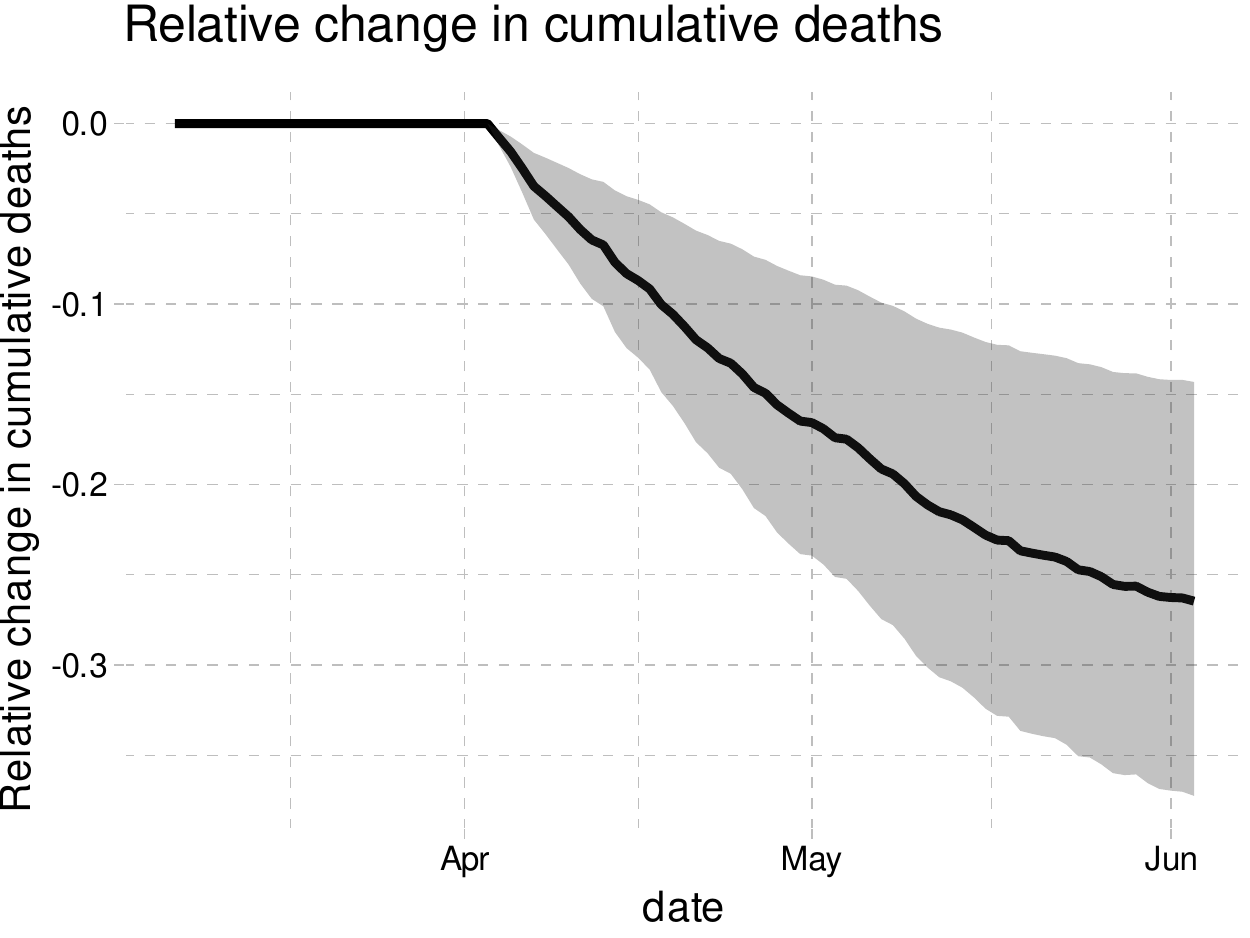}
    \end{tabular}

    \begin{flushleft}
      \footnotesize The vertical text denotes the dates that
      Washington imposed each policy.  To compute the estimated and
      counterfactual paths we use the average of two estimated
      coefficients as reported in column ``Average'' of Table
      \ref{tab:dieff-con-si}. We set initial $\Delta \log \Delta C$
      and $\log \Delta C$ to their values first observed in the state
      we are simulating. We hold all other regressors at their
      observed values. The shaded region is a point-wise 90\%
      confidence interval. See appendix \ref{app:counterfactuals} for
      more details.
    \end{flushleft}
  \end{minipage}
\end{figure}

We first consider the impact of a nationwide mask mandate for
employees beginning on March 14th. As discussed earlier, we find that
mask mandates reduce case and death growth even when holding behavior
constant. In other words, mask mandates may reduce infections with
relatively little economic disruption. This makes mask mandates a
particularly attractive policy instrument. In this section we examine
what would have happened to the number of cases if all states had
imposed a mask mandate on March 14th.\footnote{We feel this is a
  plausible counterfactual policy. Many states began implement
  business restrictions and school closures in mid March. In a paper
  made publicly available on April 1st, \cite{abaluck2020} argued for
  mask usage based on comparisons between countries with and without
  pre-existing norms of widespread mask usage.}

For illustrative purpose, we begin by focusing on Washington State.
The left column of Figure \ref{fig:WA-mask} shows the change in
growth rate from mandating masks on March 14th. The shaded region is a
90\% pointwise confidence interval. As shown, mandating masks on March 14th lowers the growth of cases or deaths 14 or 21 days later by 0.07 or
0.14. This effect then gradually declines due to information
feedback. Mandatory masks reduce past cases or deaths, which leads to
less cautious behavior, attenuating the impact of the policy. The
reversal of the decrease in growth in late May is due to our
comparison of a mask mandate on March 14th with Washington's actual
mask mandate in early May. By late May, the counterfactual mask
effect has decayed through information feedback, and we are comparing
it to the undecayed impact of Washington's actual, later mask mandate.

The middle column of Figure \ref{fig:WA-mask} shows how the changes in
case and death growth translate into changes in daily cases and
deaths. The estimates imply that mandating masks on March 14th would
have led to about 70 fewer cases and 5 fewer deaths per day throughout
April and May. Cumulatively, this implies 19\% $[9,29]$\% fewer cases and 26\%
$[14,37]$\% fewer deaths in Washington by the start of June.

\begin{figure}[ht]
  \caption{Effect of nationally mandating masks for employees on March
    14th in the US\label{fig:US-mask}}
  \begin{minipage}{\linewidth}
    \centering
    \medskip
    \begin{tabular}{ccc}
      \multicolumn{3}{c}{\textbf{Cases}} \\
      \includegraphics[width=0.31\textwidth]{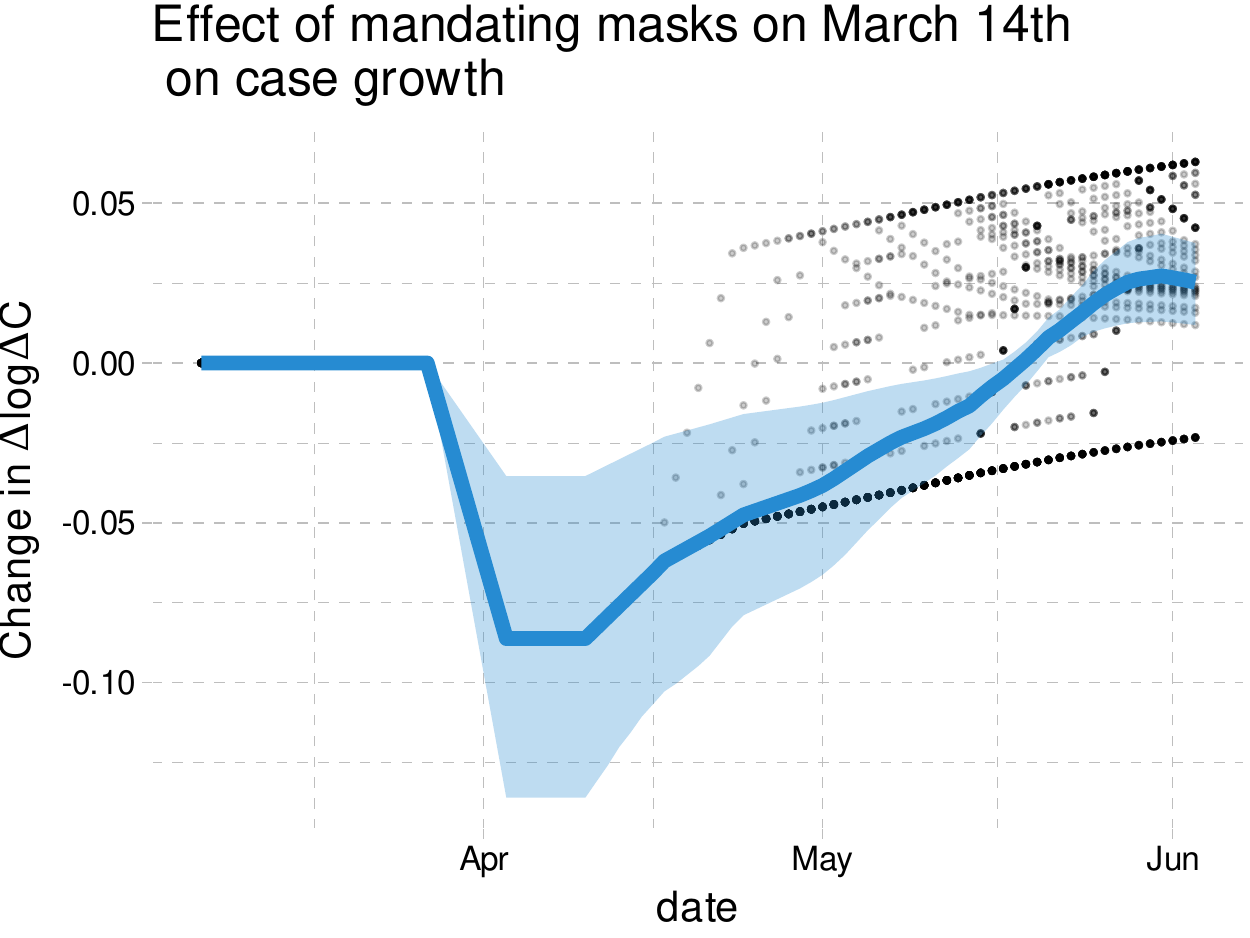}
      &
        \includegraphics[width=0.31\textwidth]{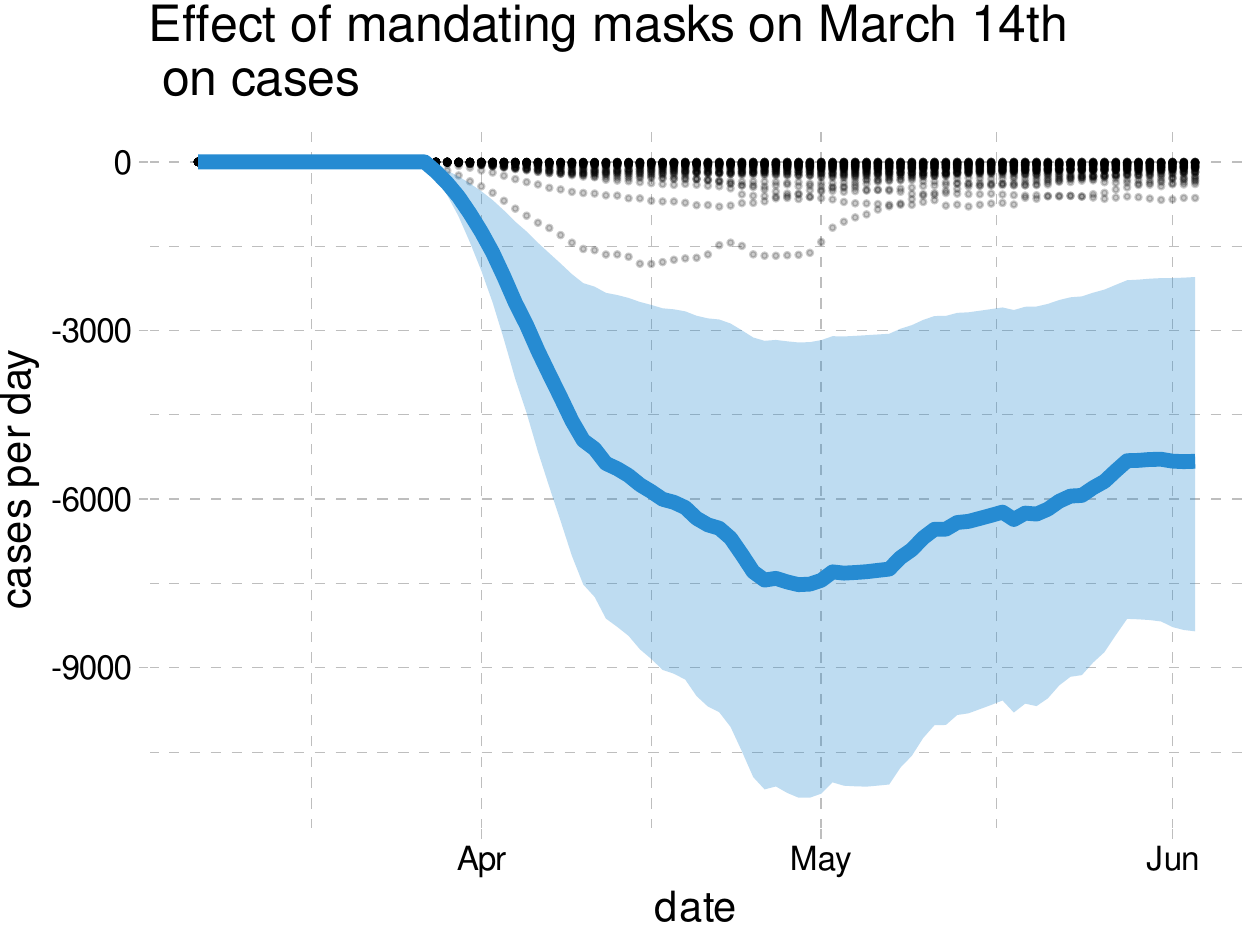}
      &
        \includegraphics[width=0.31\textwidth]{tables_and_figures/us-mask-rcumu_idx}
      \\
      \\
      \multicolumn{3}{c}{\textbf{Deaths}} \\
      \includegraphics[width=0.31\textwidth]{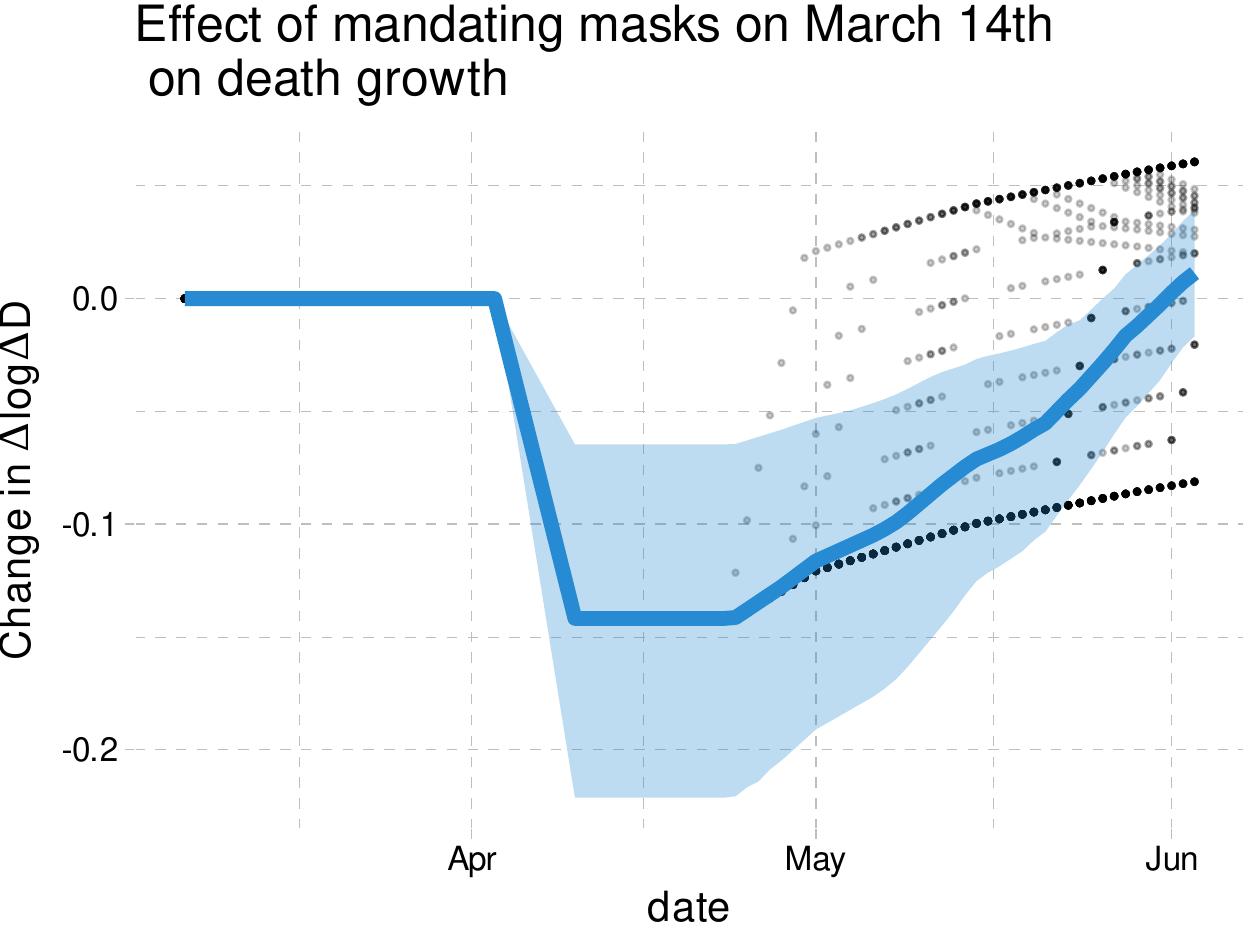}
      &
        \includegraphics[width=0.31\textwidth]{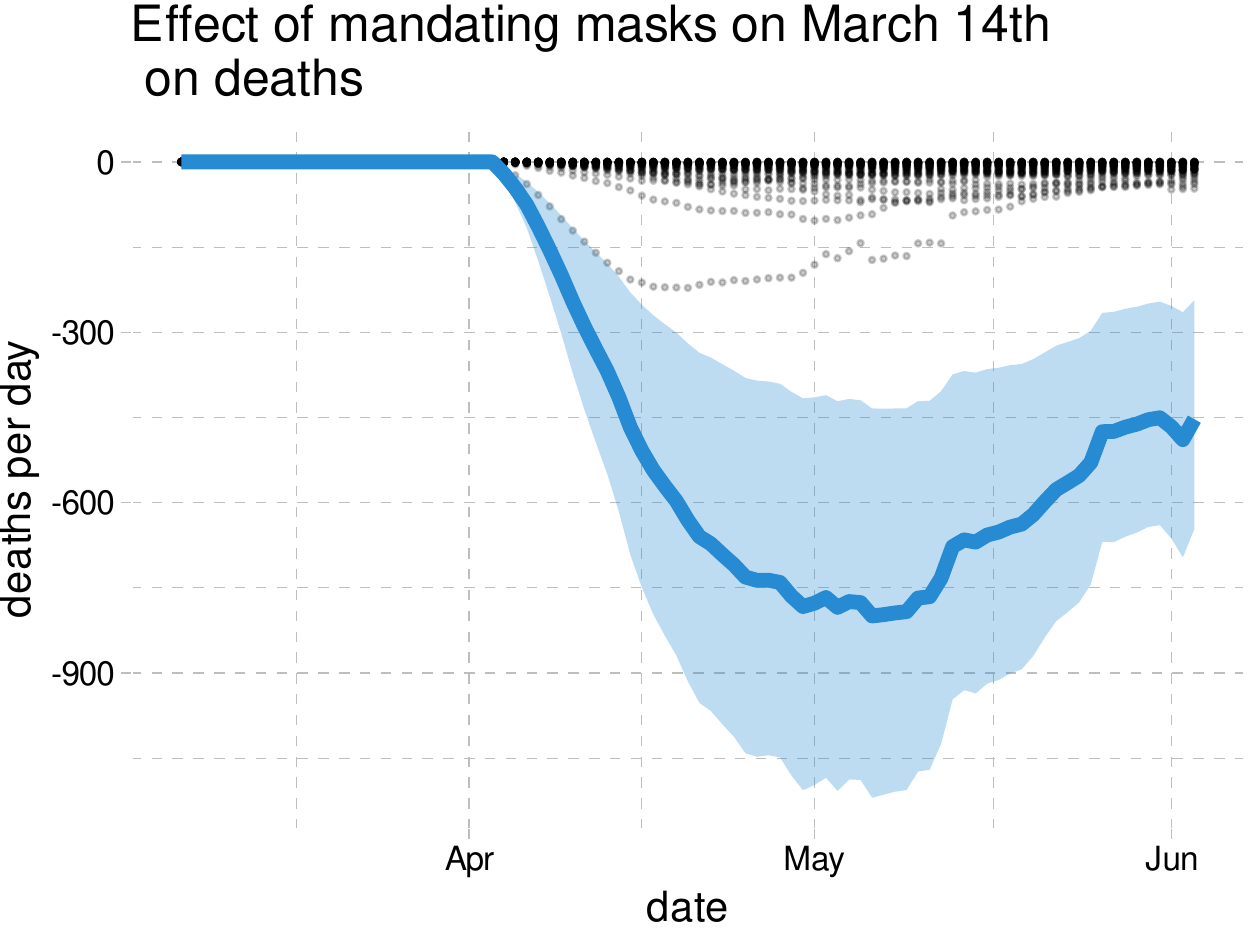}
     &
        \includegraphics[width=0.31\textwidth]{tables_and_figures/us-mask-rcumu_deaths_idx}
    \end{tabular}

    \begin{flushleft}
      \footnotesize In the left column, the dots are the average
      change in growth in each state. The blue line is the average
      across states of the change in growth. The shaded region is a
      point-wise 90\% confidence interval for the national average
      change in growth.  The middle column shows the total national
      change in daily cases and deaths. Black dots are the changes in each
      state.  The right column shows the change in cumulative cases or
      deaths relative to the baseline of actual policies.
    \end{flushleft}
  \end{minipage}
\end{figure}

The results for other states are similar to those for Washington.
Figure \ref{fig:US-mask} shows the national effect of mask mandate on
March 14th. The top panel shows the effect on cases and the bottom
panel shows the effect on deaths. The left column shows the change in
growth rates. Dots are the average change in growth rate in each state
(i.e. the dots are identical to the solid line in the left panel of
Figure \ref{fig:WA-mask}, except for each state instead of just
Washington). The solid blue line is the national average change in growth
rate. The shaded region is a 90\% pointwise confidence band for the
national average change in growth rate.

The middle column shows the changes in cases and deaths per day. Dots
are the expected change in each state. The thick line is the national
total change in daily cases or deaths. The shaded region is a
pointwise 90\% confidence band for the national total change. The
estimates that a mask mandate in mid March would have decreased cases
and deaths by about 6000 and 600 respectively throughout April and
May.

The right column of Figure \ref{fig:US-mask} shows how these daily
changes compare to the total cumulative cases and deaths. The right
column shows the national relative change in cumulative cases
or deaths. The point estimates indicate that mandating masks on March
14th could have led to 21\% fewer cumulative cases and 34\% fewer
deaths by the end of May with their 90 percent intervals
given by $[9,32]$\% and $[19,47]$\%, respectively.
The result roughly translates into $19$ to $47$ thousand saved lives.

\subsection{Business Closure Policies}

\begin{figure}[ht]
  \caption{Effect of having no business closure policies in the US\label{fig:US-business}}
  \begin{minipage}{\linewidth}
    \centering
    \begin{tabular}{ccc}
      \multicolumn{3}{c}{\textbf{Cases}} \\
      \includegraphics[width=0.31\textwidth]{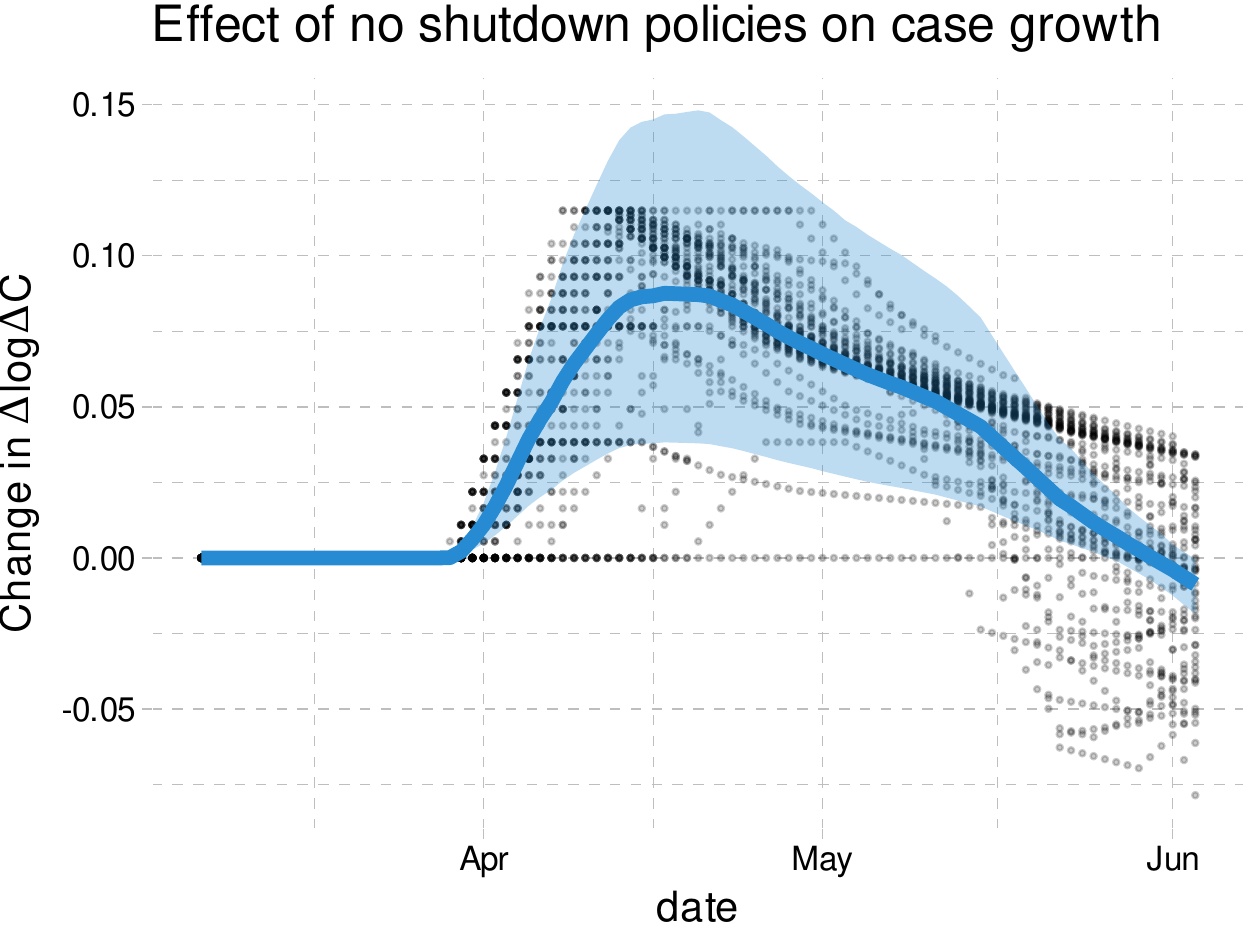}
      &
      \includegraphics[width=0.31\textwidth]{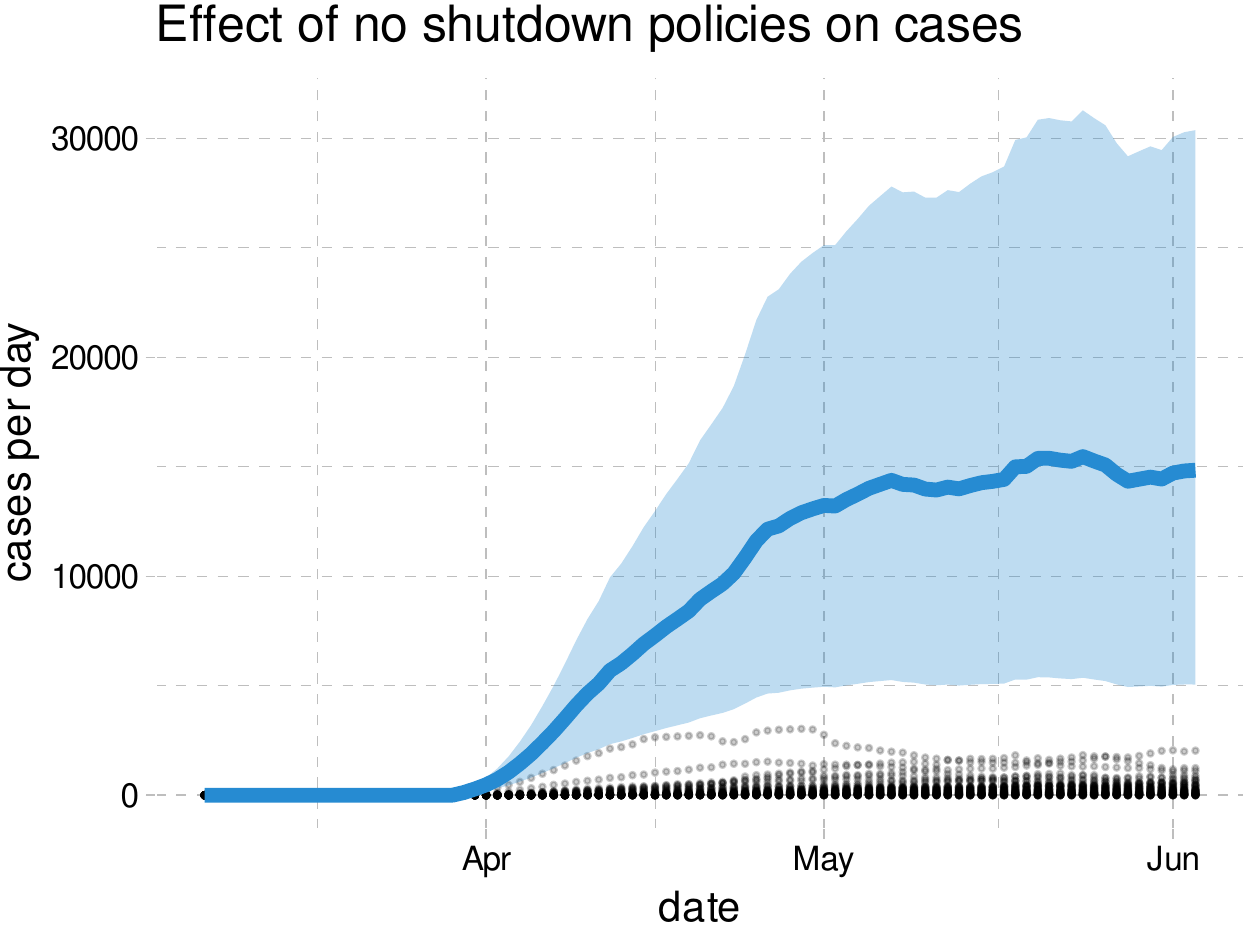}
      &
        \includegraphics[width=0.31\textwidth]{tables_and_figures/us-index-rcumu_idx}
      \\
      \\
      \multicolumn{3}{c}{\textbf{Deaths}} \\
      \includegraphics[width=0.31\textwidth]{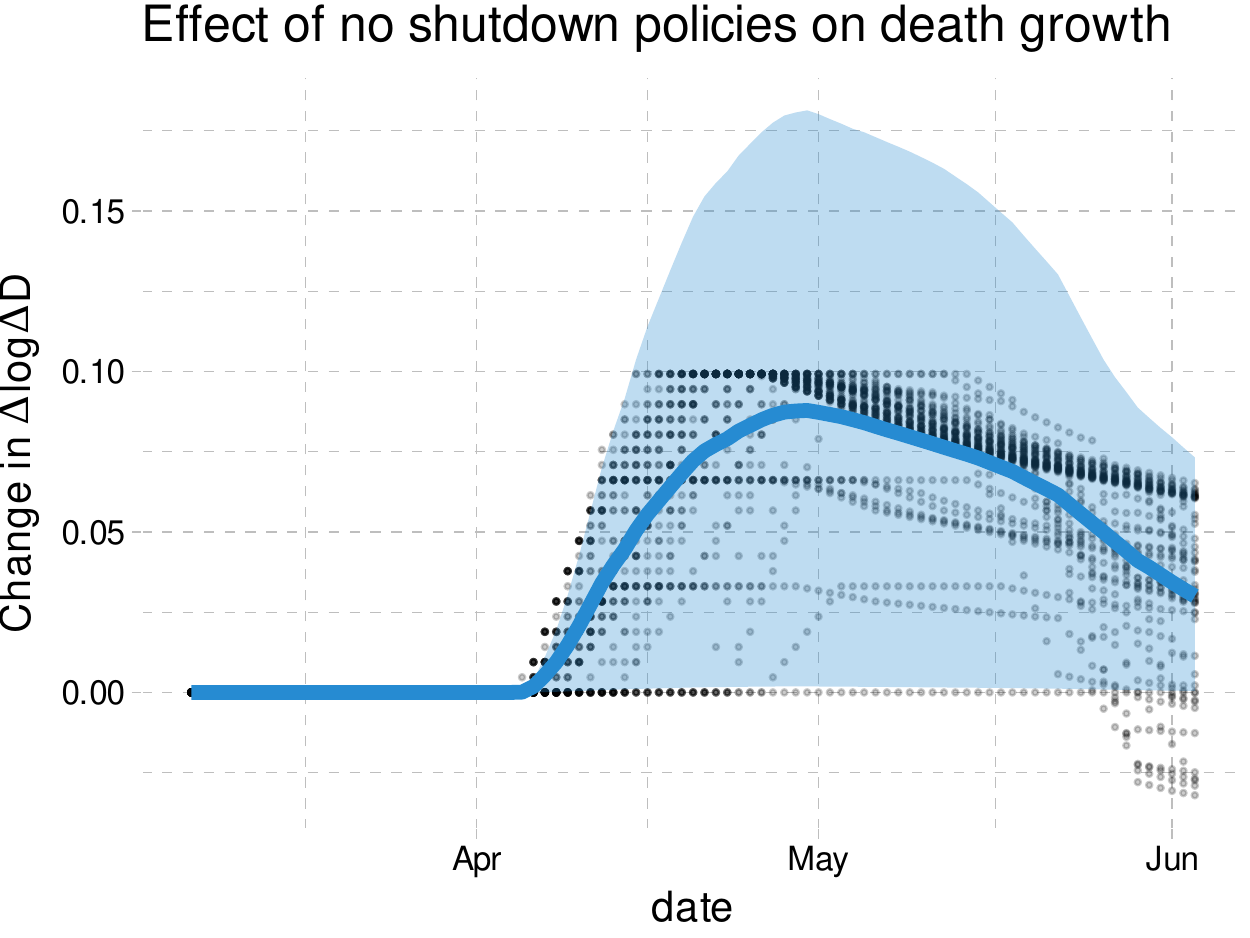}
      &
        \includegraphics[width=0.31\textwidth]{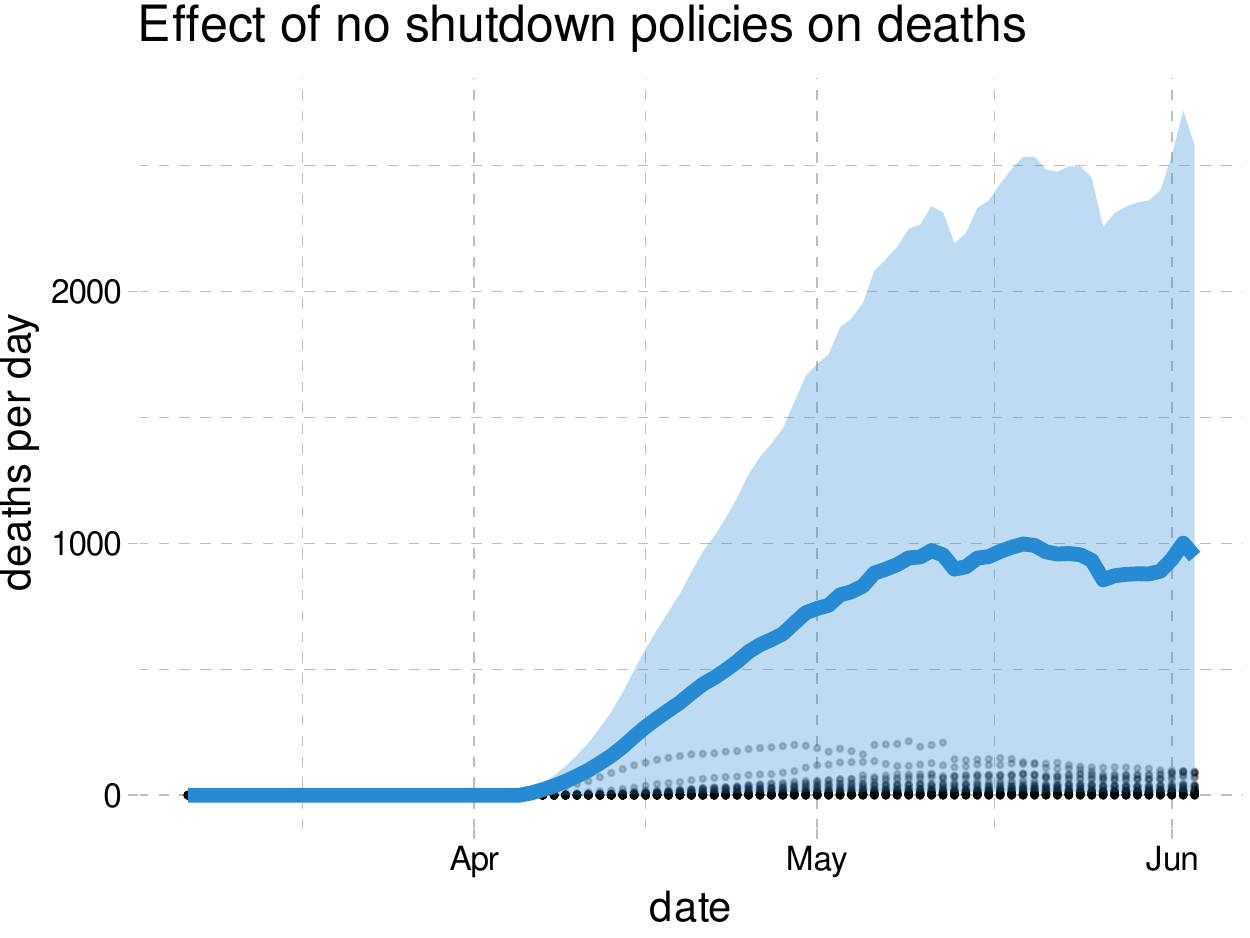}
      &
        \includegraphics[width=0.31\textwidth]{tables_and_figures/us-index-rcumu_deaths_idx}
    \end{tabular}

    \begin{flushleft}
      \footnotesize In the left column, the dots are the average
      change in growth in each state. The blue line is the average
      across states of the change in growth. The shaded region is a
      point-wise 90\% confidence interval for the national average
      change in growth.  The middle column shows the total national
      change in daily cases and deaths. Black dots are the changes in each
      state.  The right column shows the change in cumulative cases or
      deaths relative to the baseline of actual policies.
    \end{flushleft}
  \end{minipage}
\end{figure}

Business closures are particularly controversial. We now examine a
counterfactual where there are no business closure policies (no
restaurant, movie theater, or non-essential business closure).  Figure
\ref{fig:US-business} shows the national effect of never restricting
businesses on cases and deaths. The point estimates imply that,
without business closures, cases and deaths would be about 40\% higher
at the end of May. The confidence intervals are wide. A 90 percent
confidence interval for the increase in cases at the end of May is
$[17,78]$\%.
The confidence interval for deaths is even wider, $[1, 97]$\%.

\subsection{Stay-at-home orders}

\begin{figure}[ht]
  \caption{Effect of having no stay-at-home orders in the US\label{fig:US-shelter}}
  \begin{minipage}{\linewidth}
    \centering
    \begin{tabular}{ccc}
      \multicolumn{3}{c}{\textbf{Cases}} \\
      \includegraphics[width=0.31\textwidth]{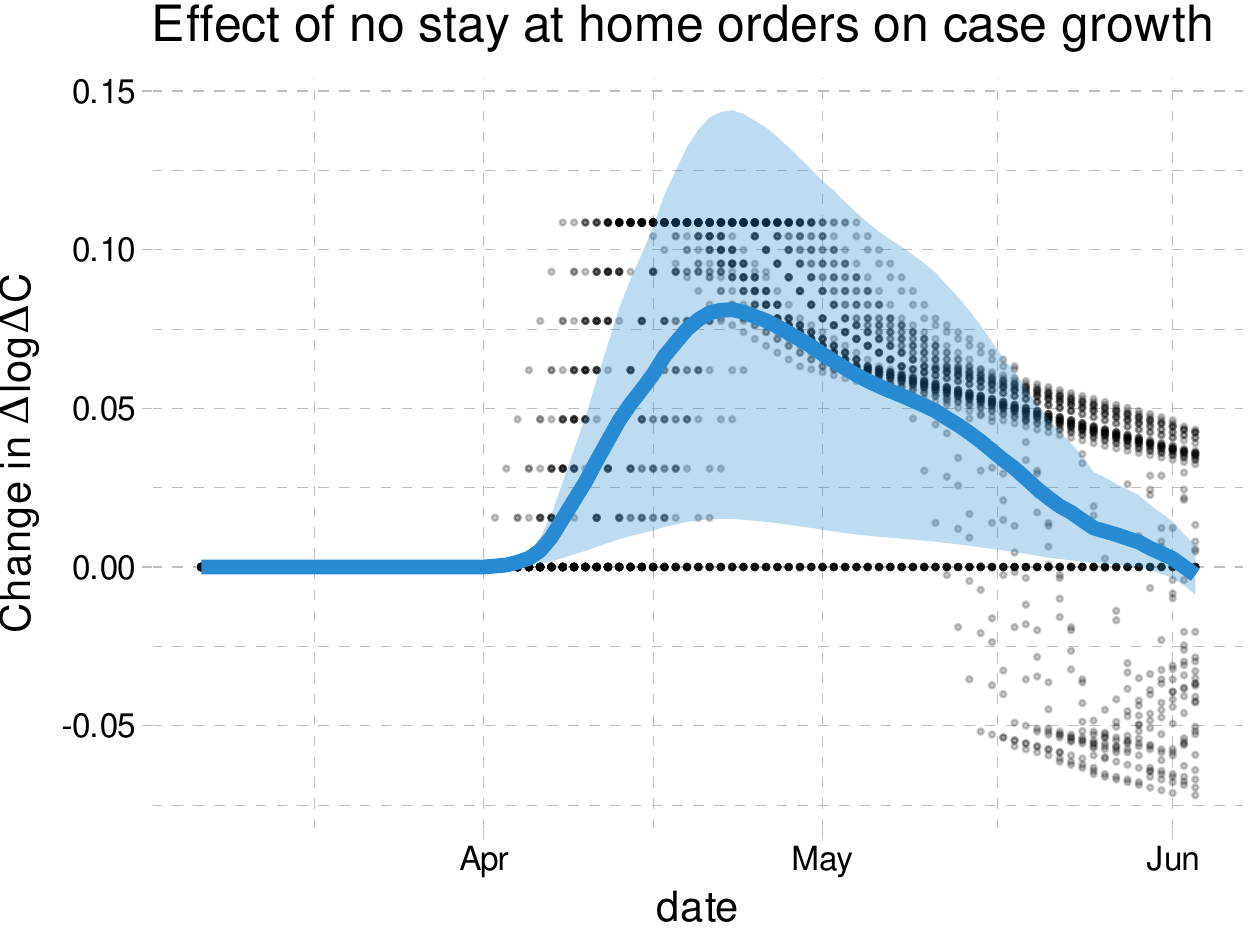}
      &
        \includegraphics[width=0.31\textwidth]{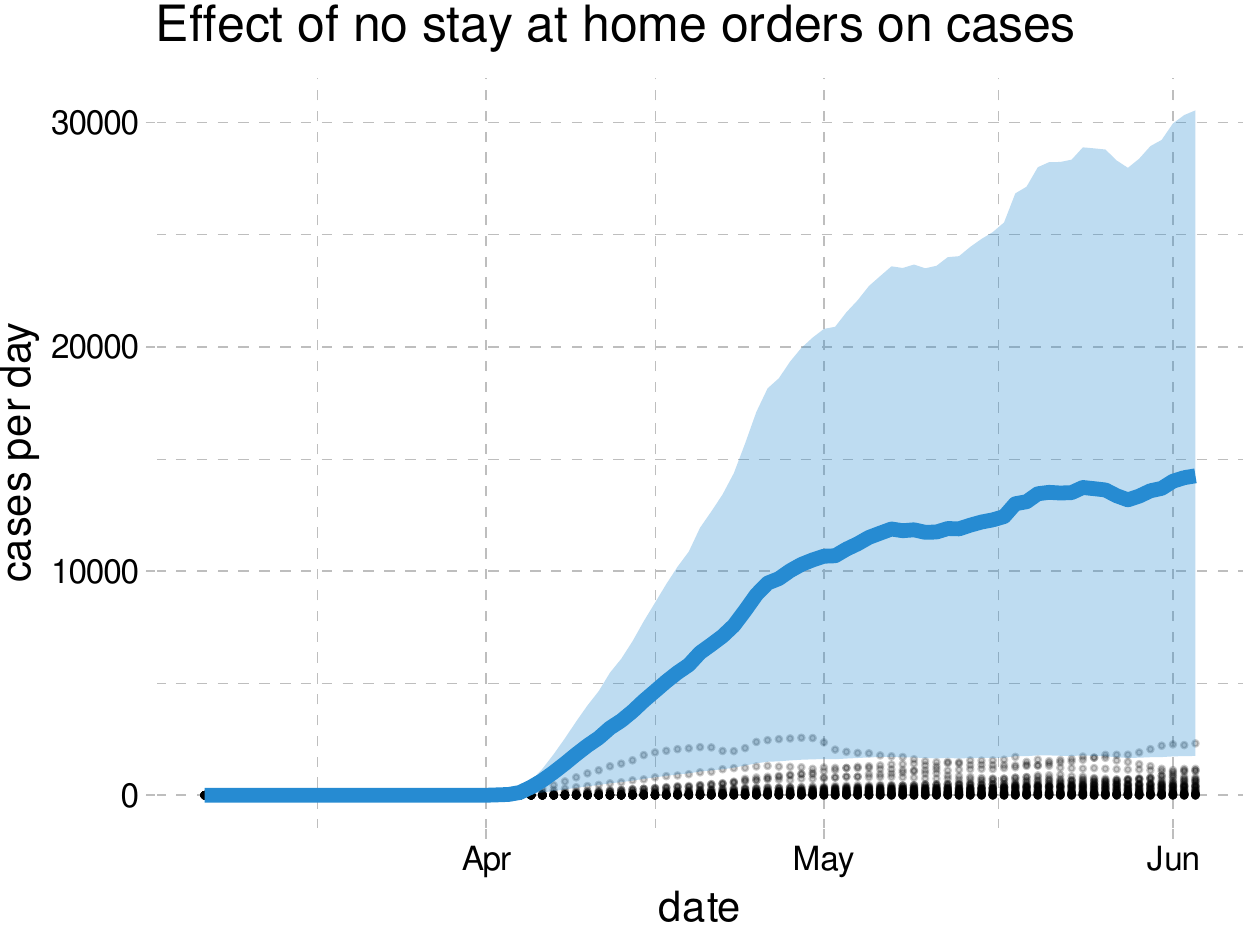}
      &

        \includegraphics[width=0.31\textwidth]{tables_and_figures/us-shelter-rcumu_idx}
      \\
      \\
      \multicolumn{3}{c}{\textbf{Deaths}} \\
      \includegraphics[width=0.31\textwidth]{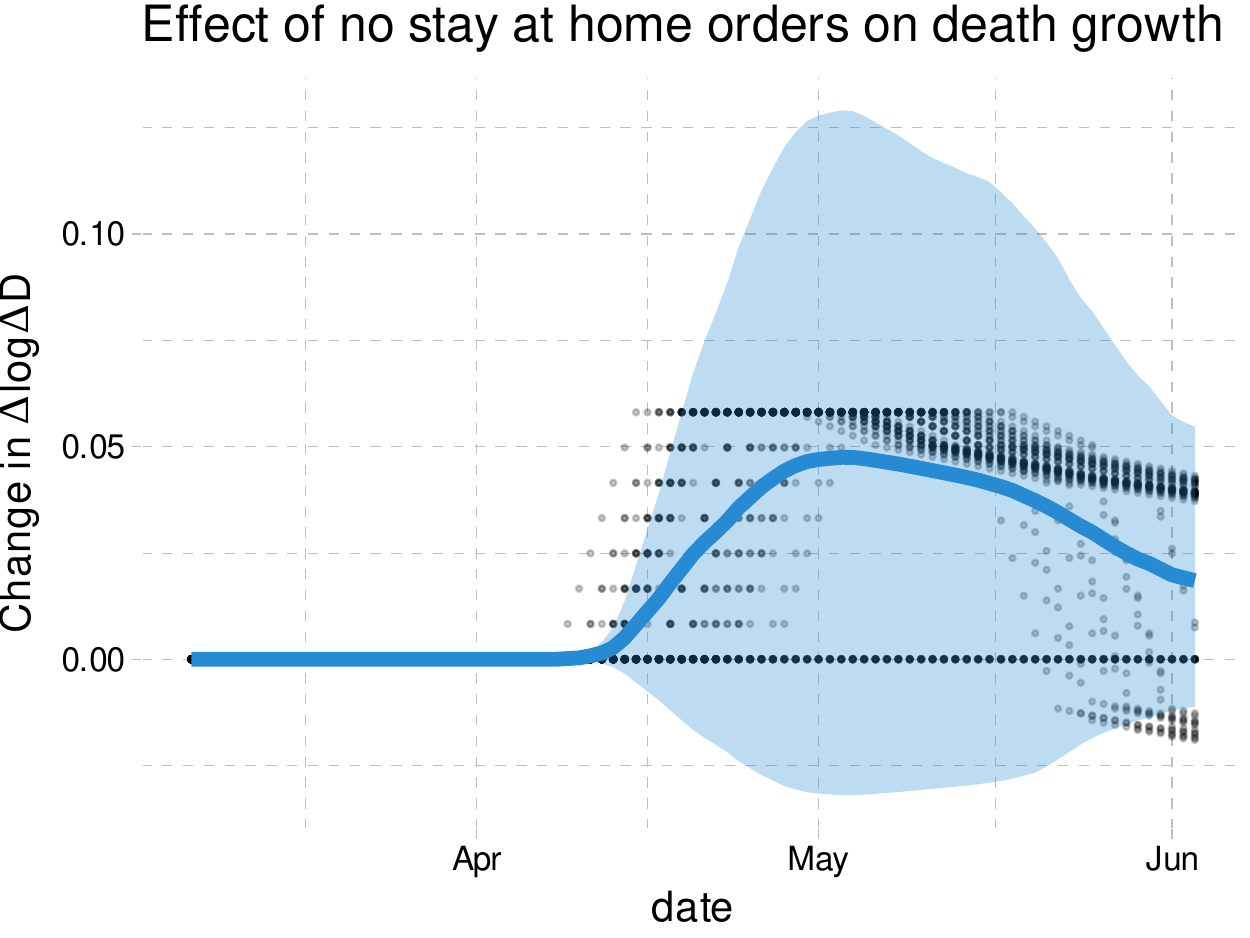}
      &
      \includegraphics[width=0.31\textwidth]{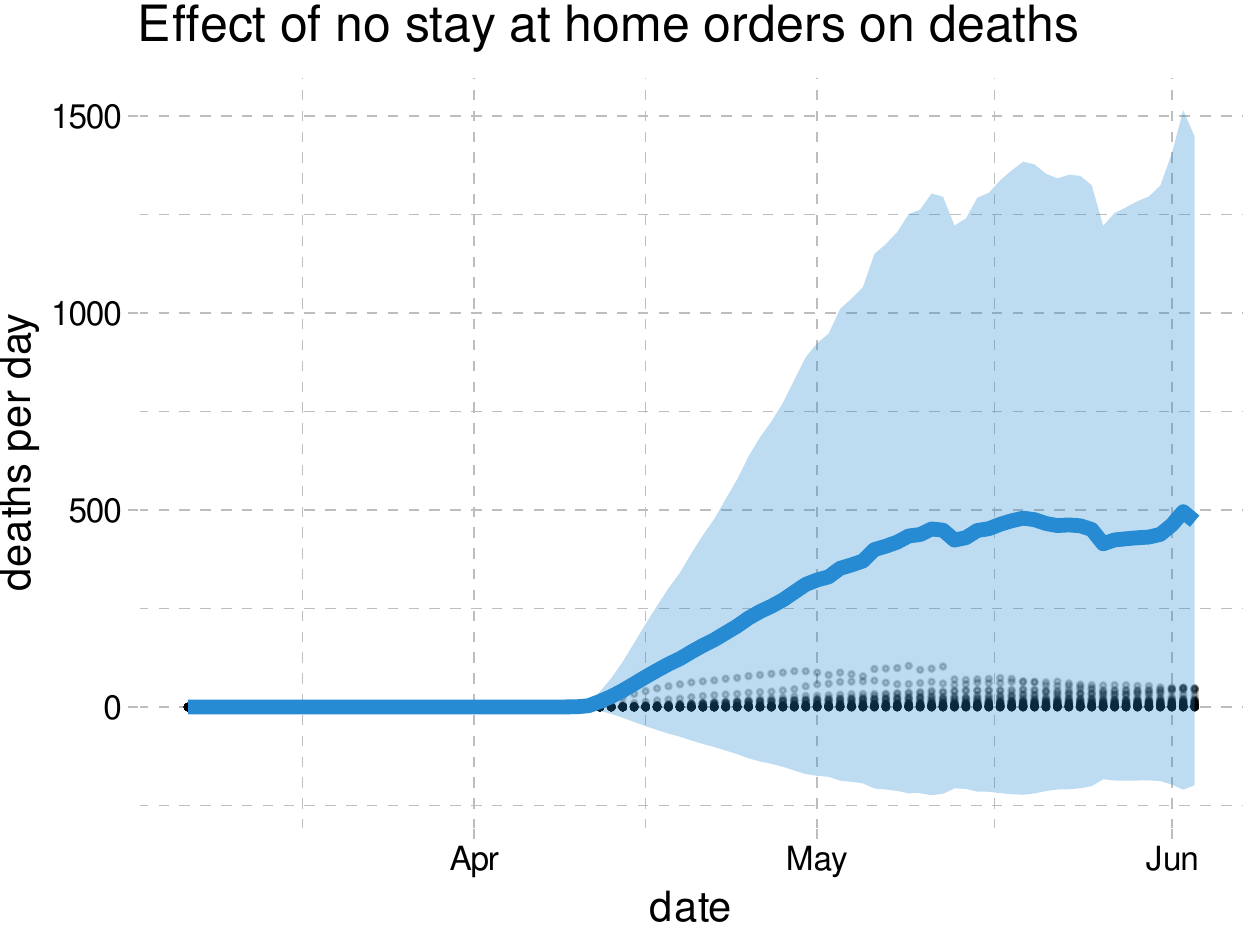}
      &
        \includegraphics[width=0.31\textwidth]{tables_and_figures/us-shelter-rcumu_deaths_idx}
    \end{tabular}

    \begin{flushleft}
      \footnotesize In the left column, the dots are the average
      change in growth in each state. The blue line is the average
      across states of the change in growth. The shaded region is a
      point-wise 90\% confidence interval for the national average
      change in growth.  The middle column shows the total national
      change in daily cases and deaths. Black dots are the changes in each
      state.  The right column shows the change in cumulative cases or
      deaths relative to the baseline of actual policies.
    \end{flushleft}
  \end{minipage}
\end{figure}

We next examine a counterfactual where stay-at-home orders had been never
issued.  Figure \ref{fig:US-shelter} shows the national effect of no
stay-at-home orders. On average, without stay-at-home orders, case
growth rate would have been nearly 0.075 higher in late April. This
translates to 37\% $[6,63]$\% more cases per week by the start of June. The
estimated effect for deaths is a bit smaller, with no
increase included in a 90 percent confidence interval, $[-7,50]$\%.

\afterpage{\clearpage}

\section{Conclusion}

This paper assesses the effects of policies on the spread of Covid-19 in the US using state-level data on cases, tests, policies, and social distancing behavior measures from Google Mobility Reports. Our findings are summarized as follows.

First, our empirical analysis robustly indicates that mandating face masks has
reduced the spread of Covid-19 without affecting people's social
distancing behavior measured by Google Mobility Reports. Our
counterfactual experiment suggests that
if all states had  adopted mandatory face mask policies on March 14th of 2020, then the cumulative number of deaths by the end of May would have been smaller by as much as  $19$ to $47$\%, which roughly translates into $19$ to $47$  thousand saved lives.
  
Second, our baseline counterfactual analysis suggests that keeping all businesses open would have led to 17 to 78\% more cases  while not implementing stay-at-home orders would have increased cases by 6 to 63\% by the end of May although we should interpret these numbers with some caution given that the estimated effect of business closures and stay-at-home orders vary across specifications in our sensitivity analysis. 

Third, we find considerable uncertainty over  the impact of school closures on case or death growth because it is difficult to identify the effect of school closures from the US state-level data due to the lack of variation in the timing of school closures across states.

Fourth, our analysis shows that people voluntarily reduce their visits to workplace, retail stores, grocery stores, and people limit their use of public transit when they receive  information on a higher number of new cases and deaths. This suggests that individuals make decisions to voluntarily limit their contact with others in response to greater transmission risks, leading to an important feedback mechanism that affects future cases and deaths. Model simulations that ignore this voluntary behavioral response to information on transmission risks would over-predict the future number of cases and deaths.

Beyond these findings, our paper presents a useful conceptual framework to investigate the relative roles of policies and information on determining the spread of Covid-19 through their impact on people's behavior. Our causal model allows us  to explicitly define counterfactual scenarios to properly evaluate the effect of alternative policies on the spread of Covid-19. 
 More broadly, our  causal framework can be useful for quantitatively analyzing not only health outcomes but also various economic outcomes \citep{bartik2020, chetty2020real}.

\FloatBarrier

\begin{footnotesize}

\bibliographystyle{jpe}
\bibliography{covid}

\end{footnotesize}

\newpage

\appendix

\section{Data Construction}

\subsection{Measuring $\Delta C$ and $\Delta\log \Delta C$}

We have three data sets with information on daily cumulative confirmed
cases in each state. As shown in Table \ref{tab:casecor}, these
cumulative case numbers are very highly correlated. However, Table
\ref{tab:casediff} shows that the numbers are different more often
than not.

\begin{table}[ht]
\centering
\begin{tabular}{rrrr}
  \hline
 & NYT & JHU & CTP \\ 
  \hline
NYT & 1.00000 & 0.99985 & 0.99970 \\ 
  JHU & 0.99985 & 1.00000 & 0.99985 \\ 
  CTP & 0.99970 & 0.99985 & 1.00000 \\ 
   \hline
\end{tabular}
\caption{Correlation of cumulative cases \label{tab:casecor}} 
\end{table}

\begin{table}[ht]
\centering
\begin{tabular}{rrrr}
  \hline
 & 1 & 2 & 3 \\ 
  \hline
NYT & 1.00 & 0.38 & 0.32 \\ 
  JHU & 0.38 & 1.00 & 0.51 \\ 
  CTP & 0.32 & 0.51 & 1.00 \\ 
   \hline
\end{tabular}
\caption{Portion of cumulative cases that are equal between data sets\label{tab:casediff}} 
\end{table}

\begin{figure}[!ht]\caption{Daily Cases, Weekly Cases,  Weekly Deaths from NYT Data, and Weekly Tests from JHU\label{fig:casedeathtest}}\smallskip
  \centering
  \begin{minipage}{\textwidth}
    \centering
    \includegraphics[width=0.8\textwidth,height=0.7\textwidth]{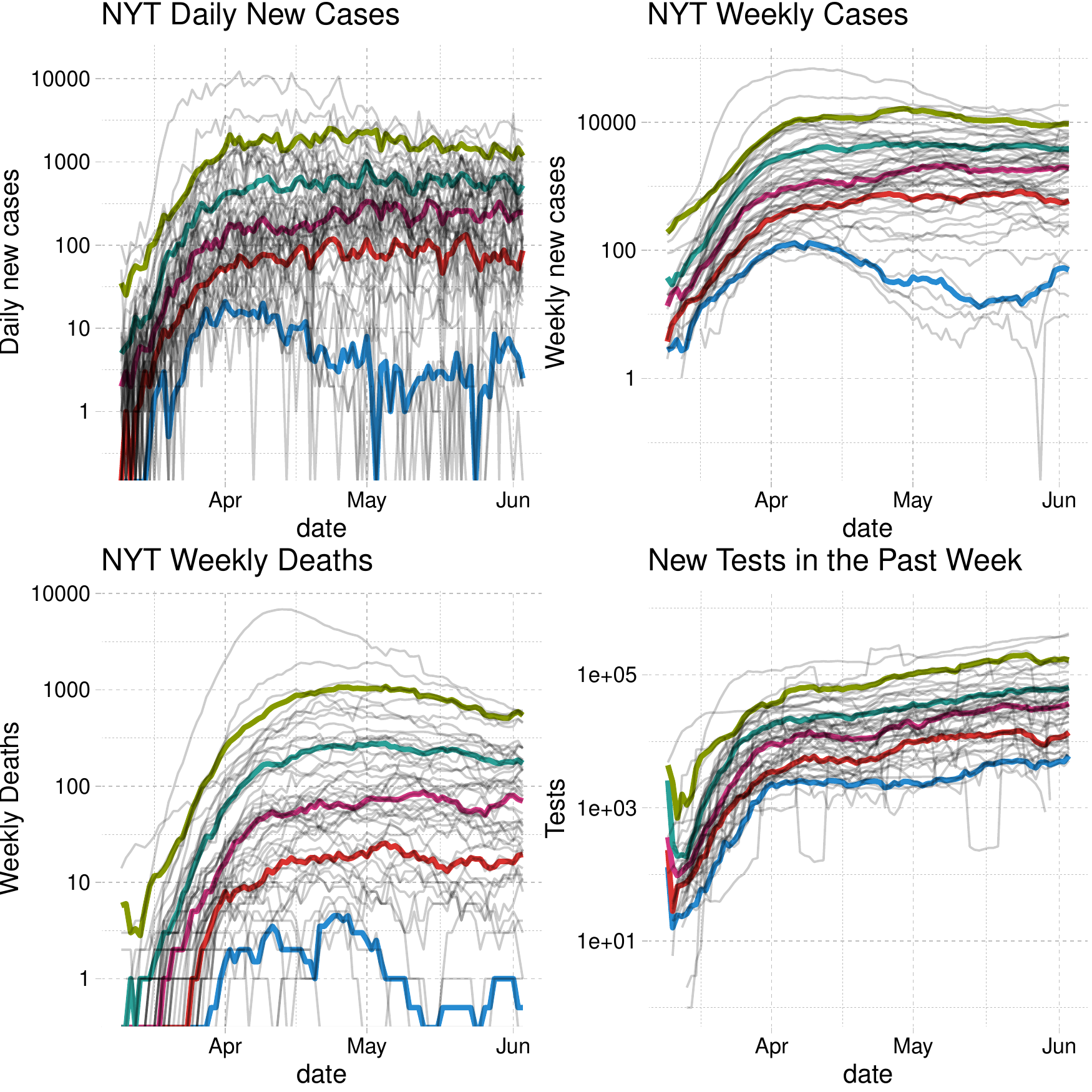}
  \end{minipage}
     \begin{flushleft}
      \footnotesize Thin gray lines are the log of cases, death, and tests  in each
      state and date. Thicker colored lines are their quantiles conditional on date.
      \end{flushleft}
\end{figure}

The upper left panel of Figure \ref{fig:casedeathtest} shows the evolution of new cases in the NYT data, where daily changes in cumulative
cases  exhibit some excessive volatility. This is likely due to delays
and bunching in testing and reporting of results.
Table
\ref{tab:newcasecor} shows the variance of log new cases in each data
set, as well as their correlations. As shown, the correlations are
approximately $0.9$. The NYT new case
numbers have the lowest variance.\footnote{This comparison is somewhat
  sensitive to how you handle negative and zero cases when taking
  logs. Here, we replaced $\log(0)$ with $-1$. In our main
  results, we work with weekly new cases, which are very rarely zero.}
In our subsequent results, we will primarily use the case numbers from
The New York Times.

\begin{table}[ht]
\centering
\begin{tabular}{rrrr}
  \hline
 & NYT & JHU & CTP \\ 
  \hline
NYT & 1.00 & 0.93 & 0.86 \\ 
  JHU & 0.93 & 1.00 & 0.82 \\ 
  CTP & 0.86 & 0.82 & 1.00 \\ 
  Variance & 5.05 & 5.81 & 6.11 \\ 
   \hline
\end{tabular}
\caption{Correlation and variance of log daily new cases\label{tab:newcasecor}} 
\end{table}

For most of our results, we focus on new cases in a week instead of in
a day. We do this for two reasons as discussed in the main text. First, a decline in new cases over
two weeks has become a key metric for decision makers. Secondly, aggregating to weekly new cases smooths out the noise associated with  the timing of
reporting and testing.

Table \ref{tab:weekcasecor} reports the correlation and variance of
weekly log new cases across the three data sets. The upper right panel of Figure
\ref{fig:casedeathtest} shows the evolution of weekly new cases in each
state over time  from the NYT data. The upper panel of Figure \ref{fig:growthq} shows the evolution of the weekly growth rate of new cases in each
state over time.

\begin{table}[ht]
\centering
\begin{tabular}{rrrr}
  \hline
 & NYT & JHU & CTP \\ 
  \hline
NYT & 1.00 & 1.00 & 0.98 \\ 
  JHU & 1.00 & 1.00 & 0.98 \\ 
  CTP & 0.98 & 0.98 & 1.00 \\ 
  Variance & 3.68 & 3.76 & 3.47 \\ 
   \hline
\end{tabular}
\caption{Correlation and variance of log weekly new cases\label{tab:weekcasecor}} 
\end{table}


\subsection{Deaths}

\begin{figure}[!ht]\caption{Case and death growth \label{fig:growthq}}
  \centering
  \begin{minipage}{\textwidth}
    \centering
    \begin{tabular}{c}
      \includegraphics[width=0.75\textwidth]{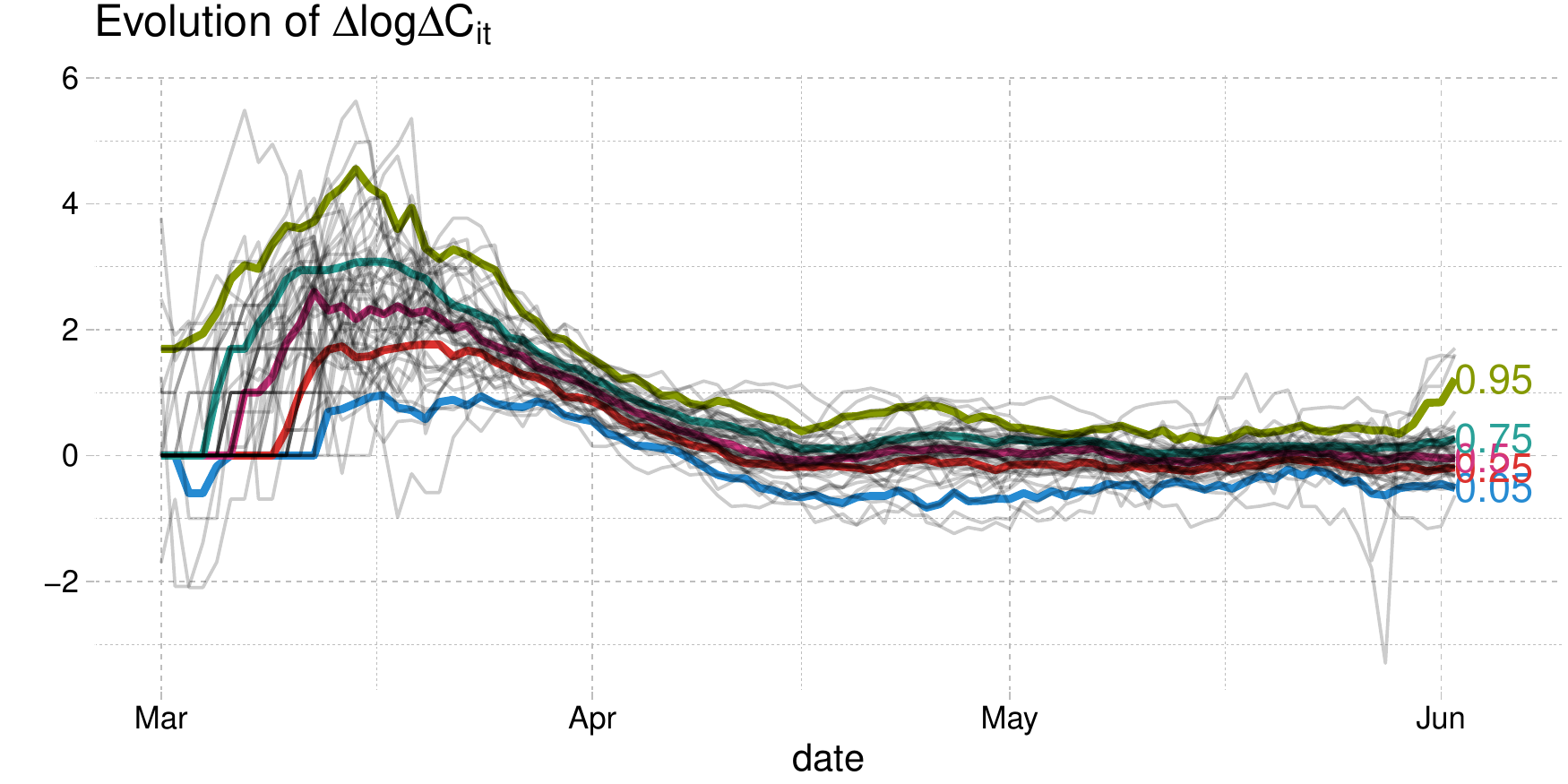}
      \\
      \includegraphics[width=0.75\textwidth]{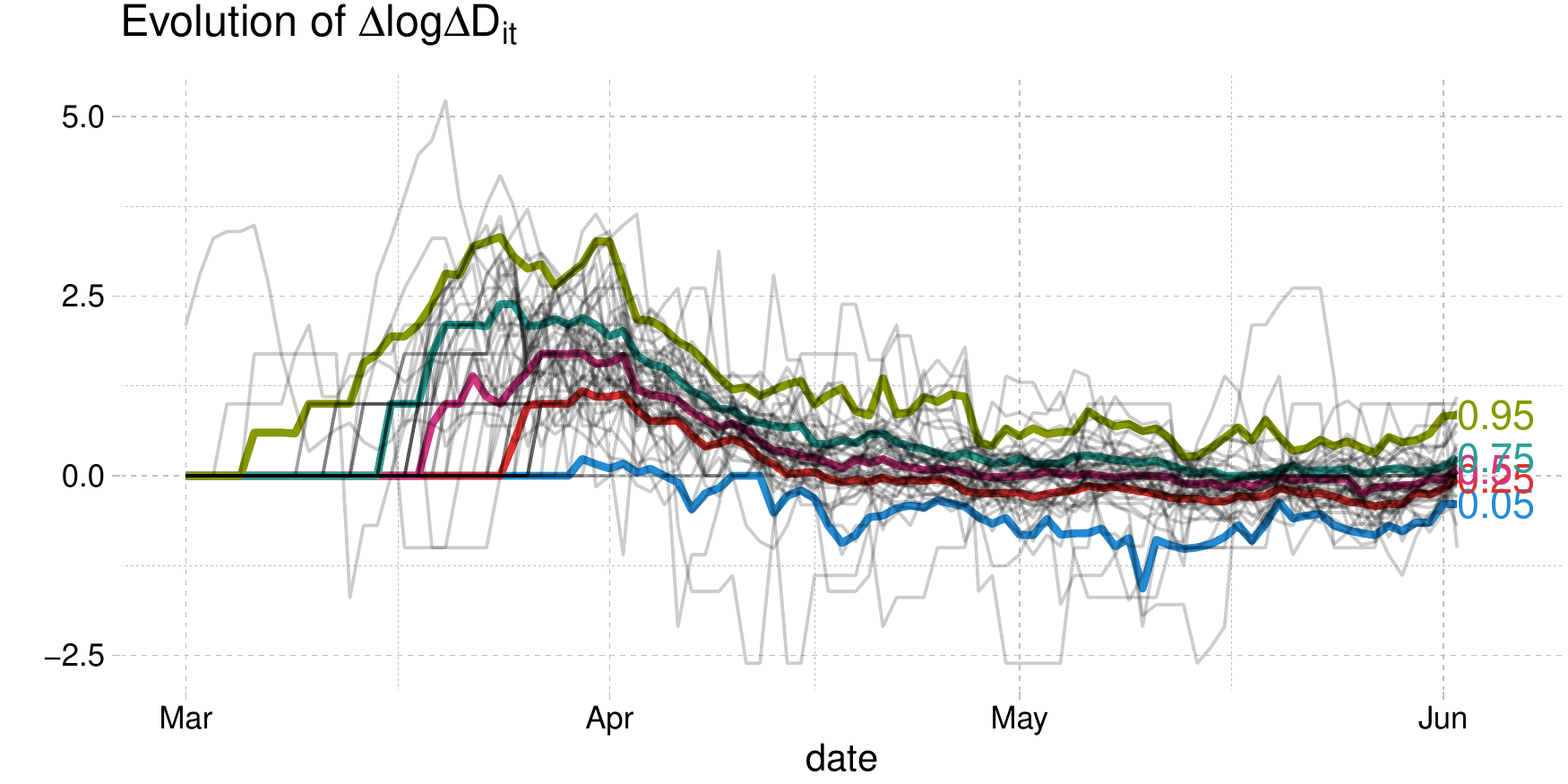}
    \end{tabular}
  \end{minipage}
     \begin{flushleft}
      \footnotesize Thin gray lines are case or death growth in each
      state and date. Thicker colored lines are quantiles of case or
      death growth conditional on date.      \end{flushleft}
\end{figure}

Table \ref{tab:weekdeathcor} reports the correlation and variance of
weekly deaths in the three data sets. The lower left panel of Figure \ref{fig:casedeathtest}
shows the evolution of weekly deaths in each state from the NYT data. As with cases, we
use death data from The New York Times in our main results. The lower panel of Figure \ref{fig:growthq} shows the evolution of the weekly growth rate of new deaths in each
state over time.

\begin{table}[ht]
\centering
\begin{tabular}{rrrr}
  \hline
 & NYT & JHU & CTP \\ 
  \hline
NYT & 1.00 & 0.98 & 0.97 \\ 
  JHU & 0.98 & 1.00 & 0.98 \\ 
  CTP & 0.97 & 0.98 & 1.00 \\ 
  Variance & 172292.15 & 174310.07 & 126952.94 \\ 
   \hline
\end{tabular}
\caption{Correlation and variance of weekly deaths\label{tab:weekdeathcor}} 
\end{table}


\subsection{Tests}

Our test data comes from The Covid Tracking Project. The lower right panel of Figure
\ref{fig:casedeathtest} shows the evolution of the weekly number of tests over time.


\FloatBarrier

\subsection{Social Distancing Measures}

\begin{figure}[ht]\caption{Evolution of Google Mobility Reports \label{fig:gmr}}
  \begin{minipage}{\linewidth}
    \begin{tabular}{cc}
      \includegraphics[width=0.5\linewidth]{tables_and_figures/workplaces}
      &
      \includegraphics[width=0.5\linewidth]{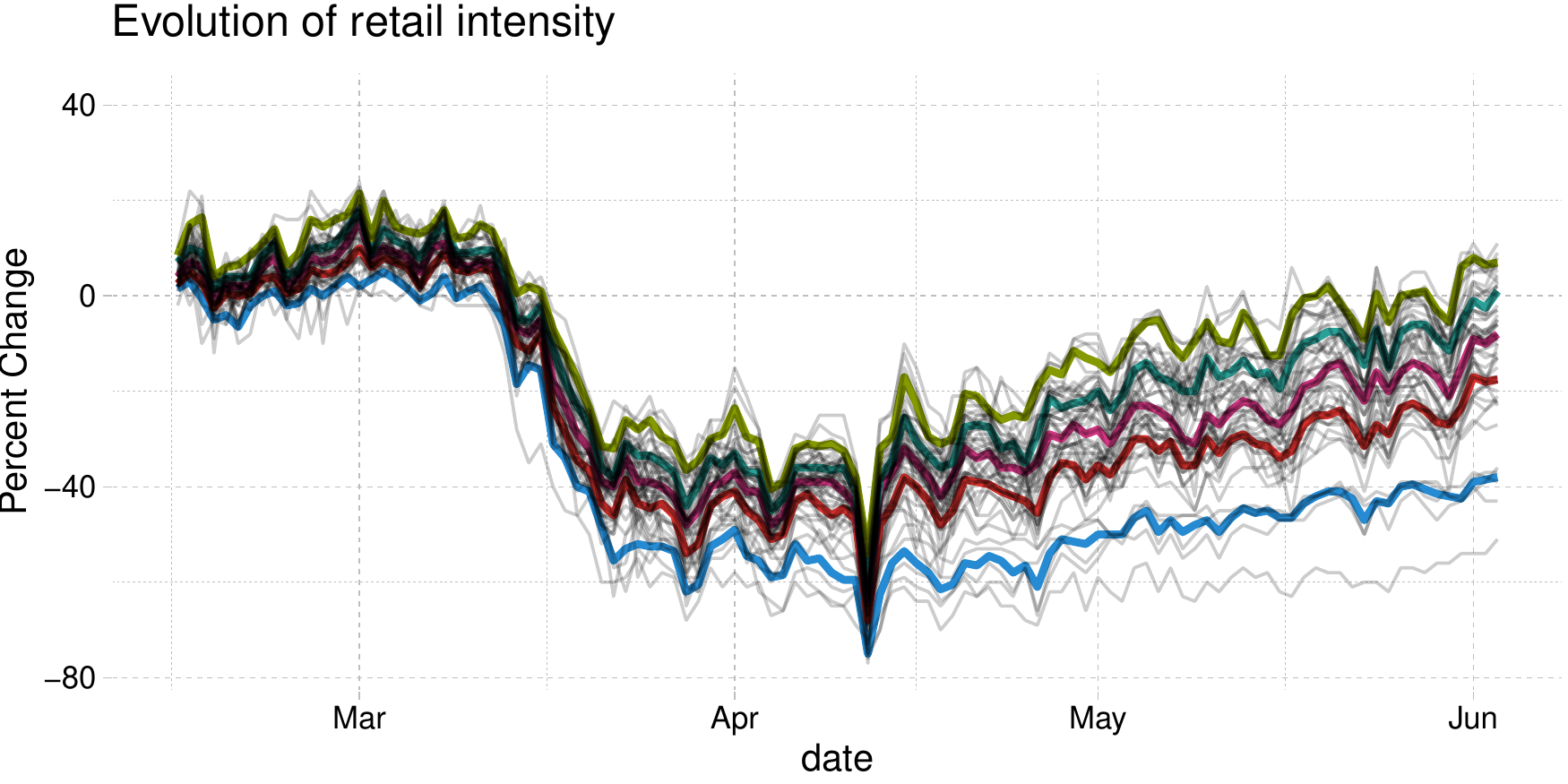}
      \\
      \includegraphics[width=0.5\linewidth]{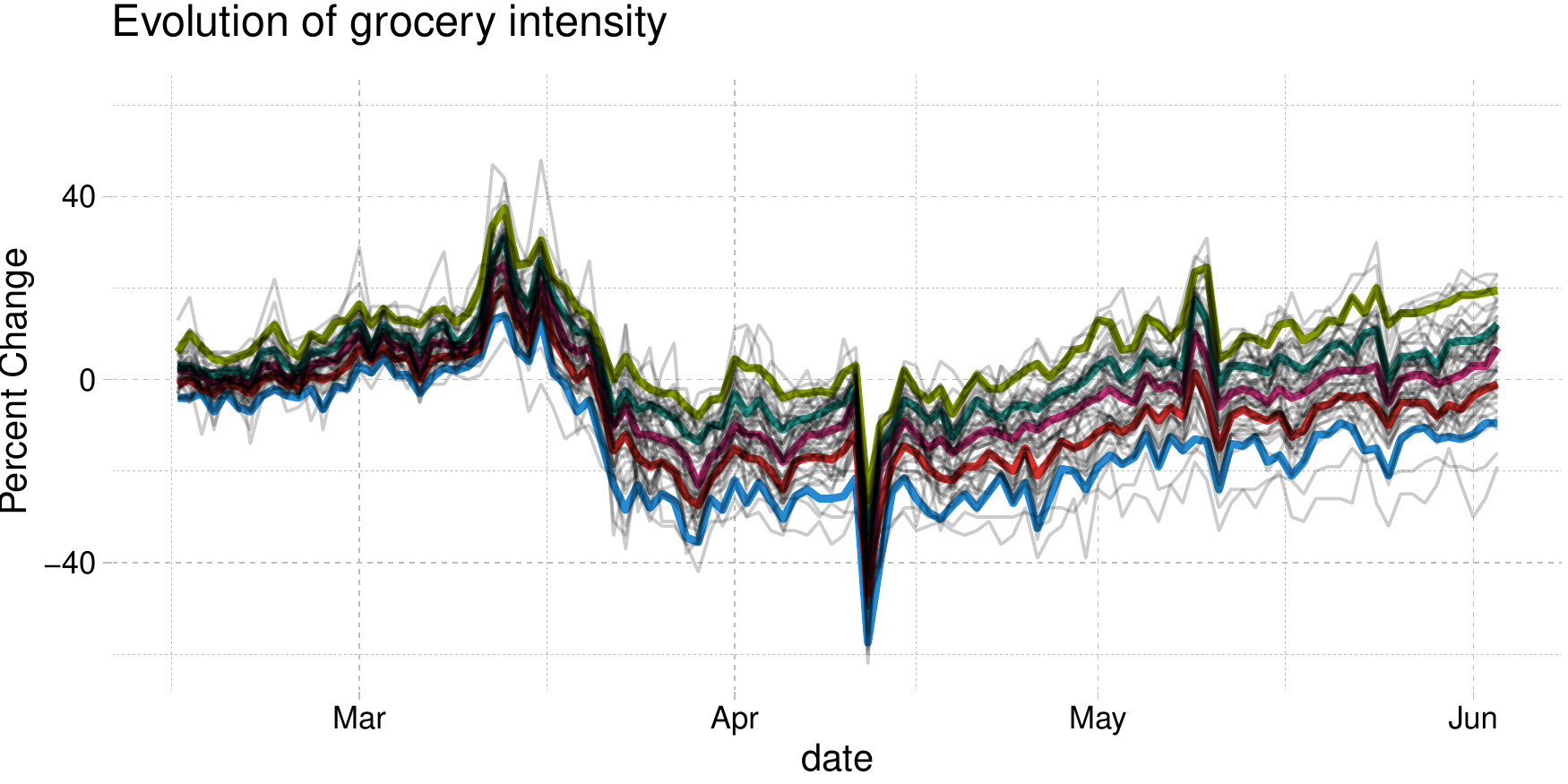}
      &
      \includegraphics[width=0.5\linewidth]{tables_and_figures/transit}
    \end{tabular}
  \end{minipage}
     \begin{flushleft}
      \footnotesize This figure shows the evolution of Google Mobility
      Reports over time. Thin gray lines are the value of the
      variables in each state and date. Thicker colored lines are
      quantiles of the variables conditional on date.     \end{flushleft}
\end{figure}

In measuring social distancing, we focus on Google Mobility
Reports. This data has international coverage and is publicly
available. Figure \ref{fig:gmr} shows the evolution of the four Google
Mobility Reports variables that we use in our analysis.

\subsection{Policy Variables}

We use the database on US state policies created by
\cite{raifman2020}. 
As discussed in the main text, our analysis focuses on six policies. For stay-at-home orders, closed nonessential
businesses, closed K-12 schools, closed restaurants except takeout,
and closed movie theaters, we double-checked any
state for which \cite{raifman2020} does not record a date. We filled
in a few missing dates. Our modified data is available
\href{"https://docs.google.com/spreadsheets/d/1E6HRkgbdSnZ9ZxrneydU6q4hhOCCt9oTl_5fa3OFVZE/edit?usp=sharing}{here}. Our
modifications fill in 1 value for school closures, 2 for stay-at-home orders, 3 for movie theater closure, and 4 for
non-essential business closures. Table \ref{tab:policies} displays all 25 dated policy variables in
\cite{raifman2020}'s database with our modifications described above.

\begin{figure}
  \caption{Case and death growth conditional on policies \label{fig:growthpolicies1}}
  \begin{minipage}{\linewidth}
    \centering
    \begin{tabular}{cc}
      \includegraphics[width=0.483\textwidth]{tables_and_figures/pmaskbus-cases-14}
      &
        \includegraphics[width=0.483\textwidth]{tables_and_figures/pmaskbus-deaths-21}
      \\
      \includegraphics[width=0.483\textwidth]{tables_and_figures/pk12-cases-14}
      &
        \includegraphics[width=0.483\textwidth]{tables_and_figures/pk12-deaths-21}
      \\
      \includegraphics[width=0.483\textwidth]{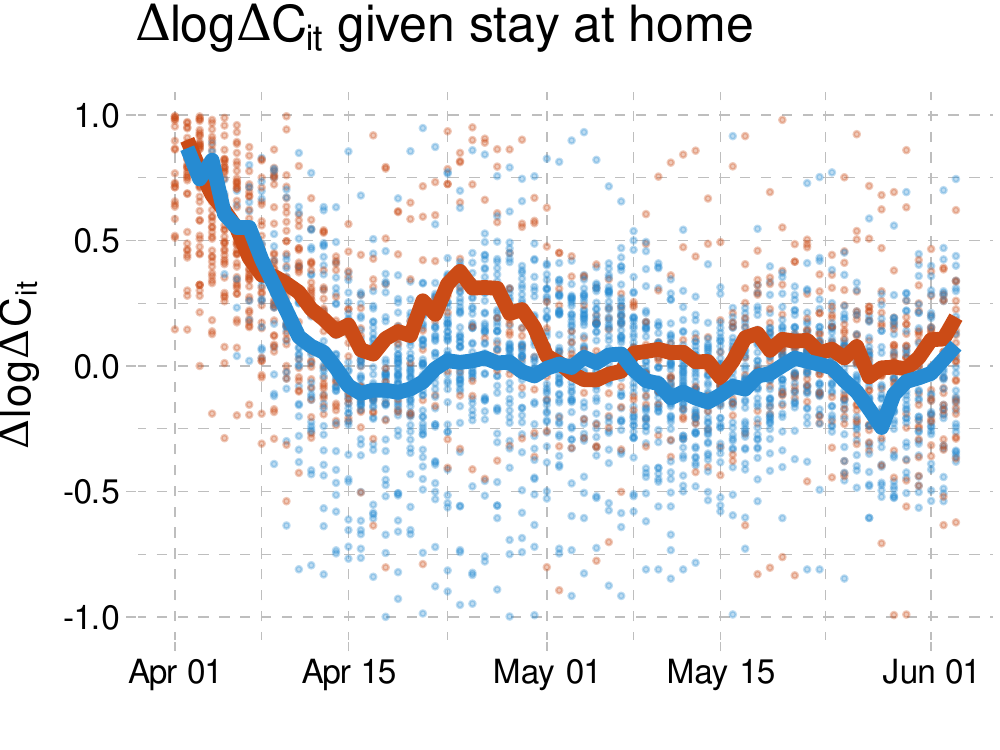}
      &
        \includegraphics[width=0.483\textwidth]{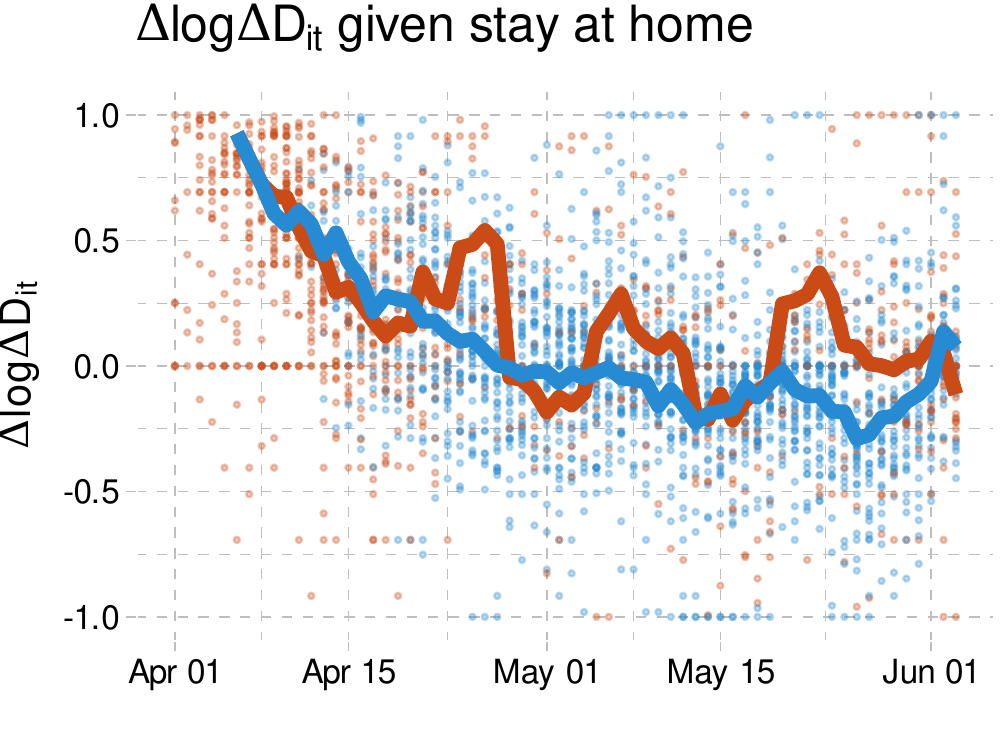}
      \\
    \end{tabular}
  \end{minipage}
     \begin{flushleft}
      \footnotesize In these figures, red points are the case or death
      growth rate in states without each policy 14 (or 21 for deaths)
      days earlier. Blue points are states with each policy 14 (or 21
      for deaths) days earlier. The red line is the average across
      states without each policy. The blue line is the average across
      states with each policy.    \end{flushleft}\end{figure}

\begin{figure}
  \caption{Case and death growth conditional on policies \label{fig:growthpolicies2}}
  \begin{minipage}{\linewidth}
    \centering
    \begin{tabular}{cc}
      \includegraphics[width=0.483\textwidth]{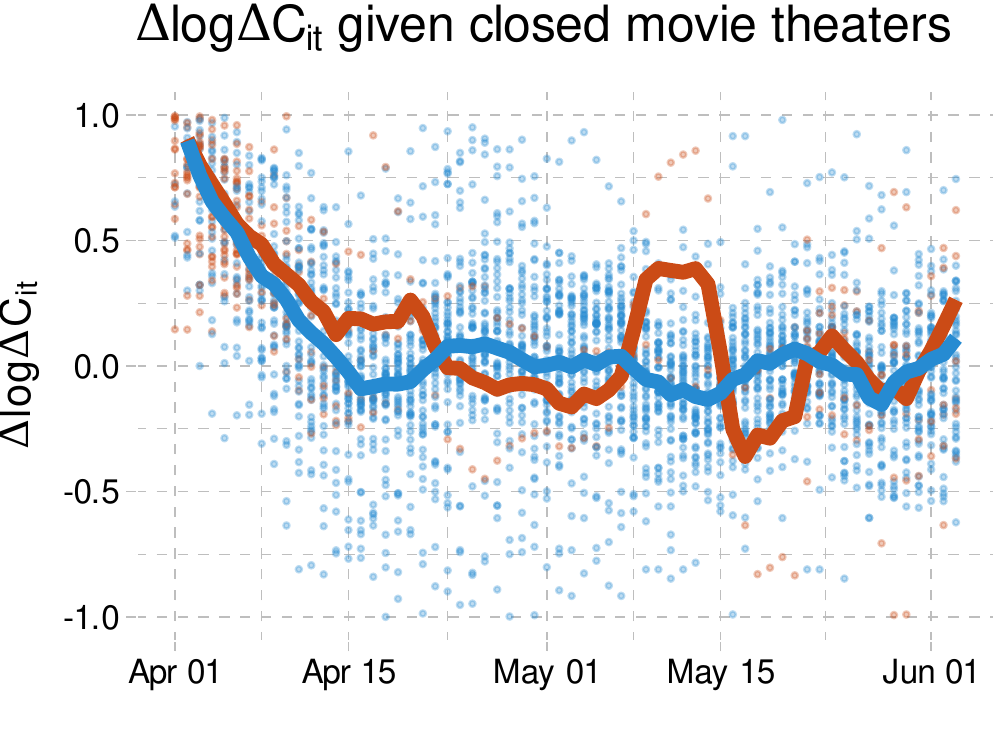}
      &
        \includegraphics[width=0.483\textwidth]{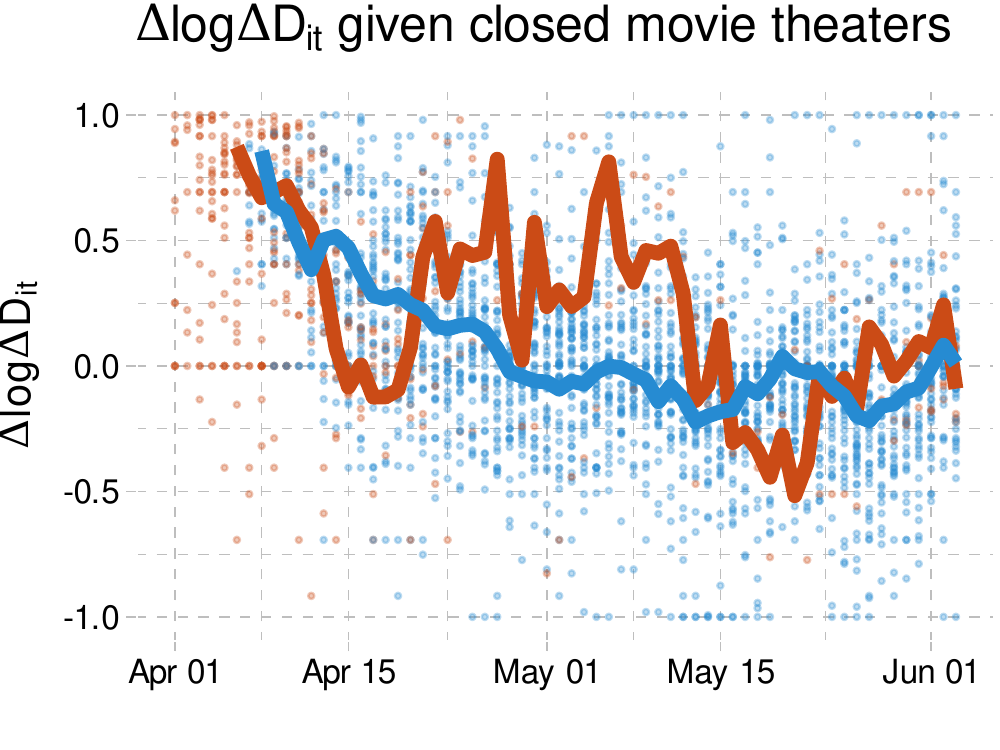}
      \\
      \includegraphics[width=0.483\textwidth]{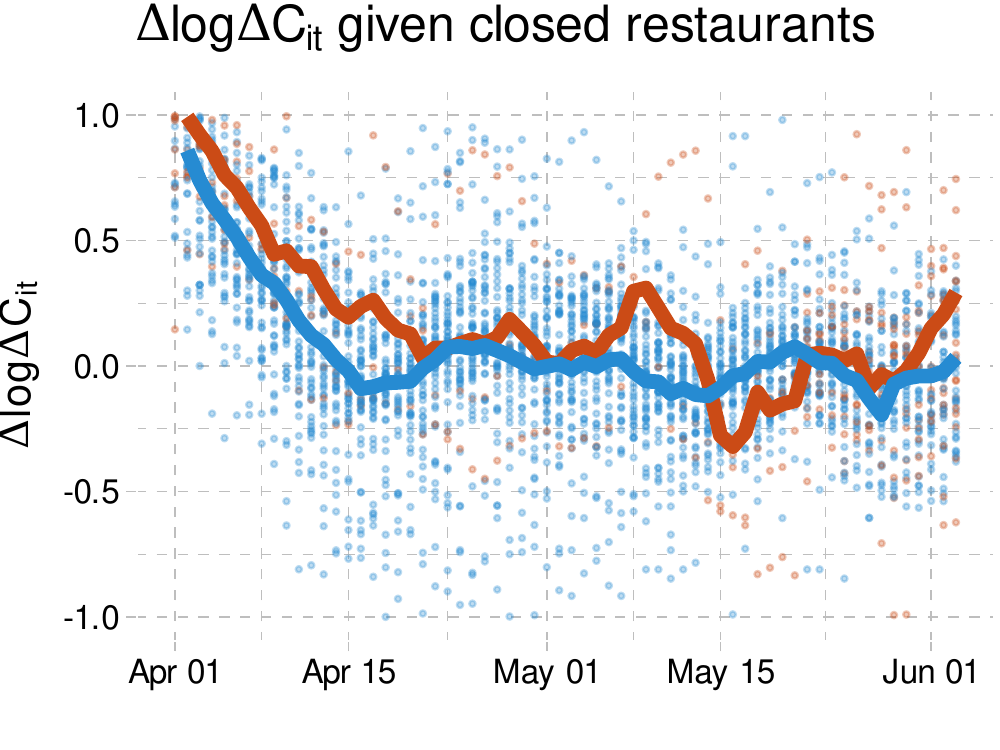}
      &
        \includegraphics[width=0.483\textwidth]{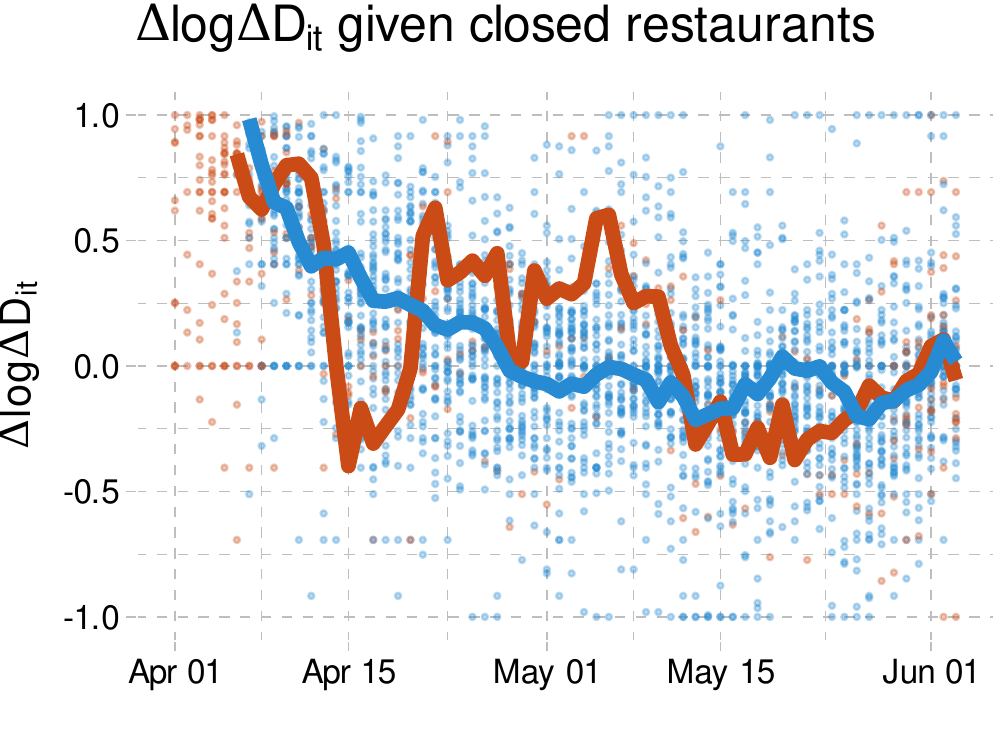}
      \\
      \includegraphics[width=0.483\textwidth]{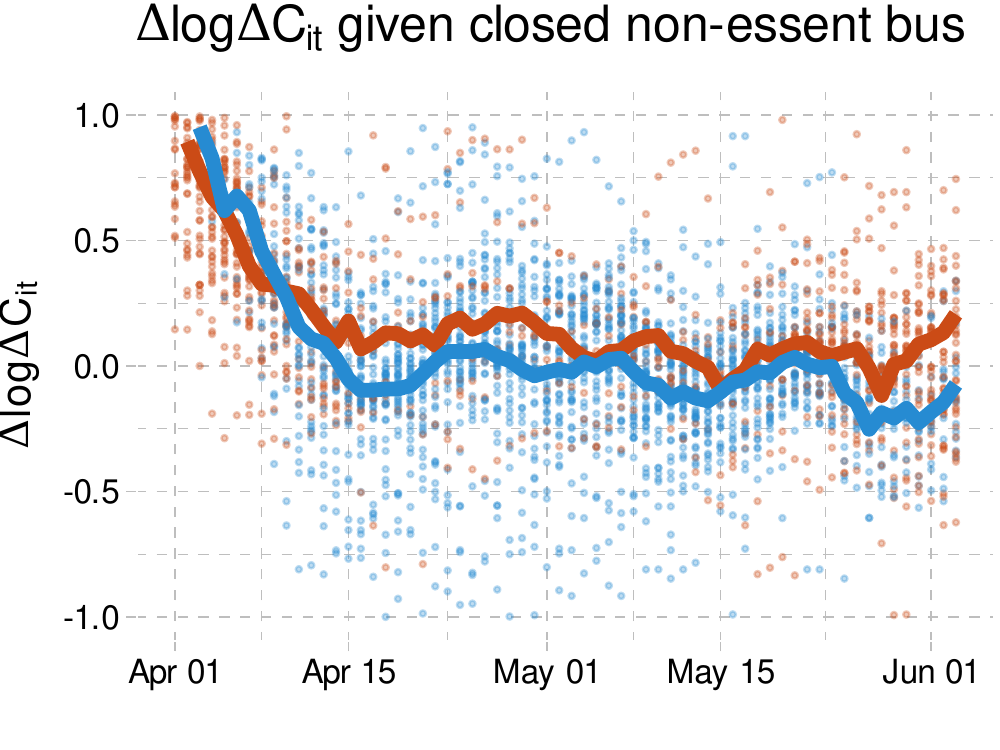}
      &
        \includegraphics[width=0.483\textwidth]{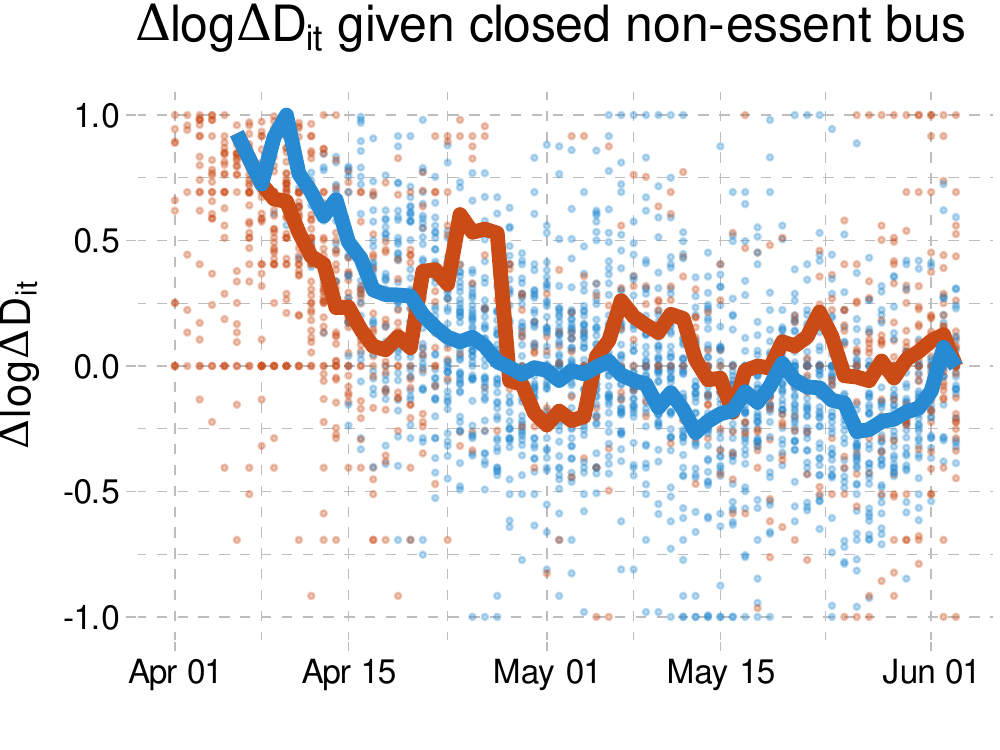}
    \end{tabular}
  \end{minipage}
     \begin{flushleft}
      \footnotesize  In these figures, red points are the case or death
      growth rate in states without each policy 14 (or 21 for deaths)
      days earlier.  Blue points are states with each policy 14 (or 21
      for deaths) days earlier.  The red line is the average across
      states without each policy. The blue line is the average across
      states with each policy.    \end{flushleft}
\end{figure}

Figures \ref{fig:growthpolicies1}-\ref{fig:growthpolicies2} show the  average case and death growth conditional on date and whether each of six policies  is implemented or not.

\afterpage{%
  \clearpage
  \thispagestyle{empty}
  \begin{landscape}
    \centering 
\begin{table}[ht]
\centering
\begin{tabular}{rllll}
  \hline
 & N & Min & Median & Max \\ 
  \hline
State of emergency & 51 & 2020-02-29 & 2020-03-11 & 2020-03-16 \\ 
  Date closed K 12 schools & 51 & 2020-03-13 & 2020-03-17 & 2020-04-03 \\ 
  Closed day cares & 15 & 2020-03-16 & 2020-03-23 & 2020-04-06 \\ 
  Date banned visitors to nursing homes & 30 & 2020-03-09 & 2020-03-16 & 2020-04-06 \\ 
  Closed non essential businesses & 43 & 2020-03-19 & 2020-03-25 & 2020-04-06 \\ 
  Closed restaurants except take out & 48 & 2020-03-15 & 2020-03-17 & 2020-04-03 \\ 
  Closed gyms & 49 & 2020-03-16 & 2020-03-20 & 2020-04-03 \\ 
  Closed movie theaters & 49 & 2020-03-16 & 2020-03-21 & 2020-04-06 \\ 
  Stay at home  shelter in place & 42 & 2020-03-19 & 2020-03-28 & 2020-04-07 \\ 
  End relax stay at home shelter in place & 39 & 2020-04-24 & 2020-05-18 & 2020-06-19 \\ 
  Began to reopen businesses statewide & 50 & 2020-04-20 & 2020-05-07 & 2020-06-05 \\ 
  Reopen restaurants & 48 & 2020-04-24 & 2020-05-18 & 2020-06-22 \\ 
  Reopened gyms & 44 & 2020-04-24 & 2020-05-18 & 2020-07-13 \\ 
  Reopened movie theaters & 37 & 2020-04-27 & 2020-06-01 & 2020-07-13 \\ 
  Resumed elective medical procedures & 35 & 2020-04-20 & 2020-04-30 & 2020-05-29 \\ 
  Mandate face mask use by all individuals in public spaces & 35 & 2020-04-08 & 2020-06-26 & 2020-08-05 \\ 
  Mandate face mask use by employees in public facing businesses & 44 & 2020-04-03 & 2020-05-01 & 2020-08-03 \\ 
  Stop Initiation of Evictions overall or due to COVID related issues & 30 & 2020-03-16 & 2020-03-24 & 2020-05-30 \\ 
  Stop enforcement of evictions overall or due to COVID related issues & 29 & 2020-03-15 & 2020-03-24 & 2020-06-08 \\ 
  Renter grace period or use of security deposit to pay rent & 3 & 2020-04-10 & 2020-04-24 & 2020-05-20 \\ 
  Order freezing utility shut offs & 34 & 2020-03-12 & 2020-03-19 & 2020-04-13 \\ 
  Froze mortgage payments & 1 & 2020-03-21 & 2020-03-21 & 2020-03-21 \\ 
  Waived one week waiting period for unemployment insurance & 36 & 2020-03-08 & 2020-03-18 & 2020-04-06 \\ 
   \hline
\end{tabular}
\caption{State Policies \label{tab:policies}} 
\end{table}

  \end{landscape}
  \clearpage
}

\subsection{Timing\label{sec:timing}}

There is a delay between infection and when a person is tested and
appears in our case data. \cite{midas2020} maintain a list of
estimates of the duration of various stages of Covid-19
infections. The incubation period, the time from infection to symptom
onset, is widely believed to be 5 days. For example, using data from
Wuhan, \cite{li2020} estimate a mean incubation period of 5.2 days.
\cite{siorda2020} reviews the literature and concludes the mean
incubation period is 3-9 days.

Estimates of the time between symptom onset and case reporting or death
are less common. Using Italian data, \cite{cereda2020} estimate an
average of 7.3 days between symptom onset and
reporting. \cite{zhang2020} find an average of 7.4 days using Chinese data
from December to early February, but they find this period declined
from 8.9 days in January to 5.4 days in the first week of
February. Both of these papers on time from symptom onset to reporting
have large confidence intervals covering approximately 1 to 20 days.

Studying publicly available data on infected persons diagnosed outside of Wuhan,
\cite{linton2020}  estimate an average of 15 days  from  onset to death. Similarly, using publicly available reports of 140 confirmed Covid-19 cases in China, mostly outside Hubei Province, \cite{sanche2020} estimate the time from onset to death to be  16.1 days.

Based on the above, we expect a delay of roughly two weeks between
changes in behavior or policies, and changes in reported
cases while a corresponding delay of roughly three weeks for deaths.


\FloatBarrier

\subsection{Direct and Indirect Policy Effects with national case/death variables}
Tables \ref{tab:dieff} and \ref{tab:dieff-con}  present  the estimates of direct and indirect effects of policies for the specification with past national case/death variables. The effects of school closures and  the sum of policies are estimated substantially smaller in Table \ref{tab:dieff} when national case/death variables are included than in Table \ref{tab:dieff-si}. This sensitivity reflects the difficulty in identifying the aggregate time effect---which is largely captured by national cases/deaths---given little cross-sectional variation in the timing of school closures across states. On the other hand, the estimated effects of policies other than school closures are similar between Table \ref{tab:dieff-si} and Table \ref{tab:dieff}; the effect of other policies are well-identified from cross-sectional variations.

\subsection{Double Machine Learning}

To estimate the coefficient of the $j$-th policy variable, $P_{it}^j$, using double machine learning (\cite{chernozhukov18}), we consider a version of (\ref{eq:R4}) as follows:
\begin{align*}
Y_{i,t+\ell} & = P_{it}^j \theta_0 + \phi' X_{it} + g_0(W_{it}) + \xi_{it}, && E[\xi_{it}|P_{it}^j,X_{it},W_{it}]=0,\\
P_{it}^j &= \psi' X_{it} +m_0(W_{it}) +V_{it},&& E[V_{it}|X_{it},W_{it}]=0,
\end{align*}
where $X_{it}=((P_{it}^{-j})',B_{it}',I_{it}')'$ collects  policy variables except for the $j$-th policy variables $P_{it}^{-j}$, behavior variables $B_{it}$, and information variables $I_{it}$. The confounding factors $W_{it}$ affect the policy variable $P_{it}^j$ via the function $m_0(W)$ and the outcome variable $Y_{i,t+\ell}$ via the function $g_0(W)$. We apply Lasso  or  Random Forests to estimate $g_0(W)$  for dimension reductions or for capturing non-linearity while the coefficients of $X_{it}$ are estimated under linearity without imposing any dimension reductions.  We omit the details of estimation procedure here. Please see the discussion in Example 1.1 of  \cite{chernozhukov18} for reference.

\subsection{Debiased Fixed Effects Estimator}

We apply cross-over Jackknife bias correction as discussed in \cite{chen2020} in details. Here, we briefly describe our debiased fixed effects estimator.

Given our panel data with $N=51$ states and $T=75$ days  for case growth equation (\ref{eq:R4}), consider two subpanels as follows:
$$
{\bf S}_1 =  \{(i,t) :  i \leq \lceil N/2 \rceil,  t \leq \lceil T/2 \rceil \} \cup  \{(i,t) : i  \geq \lfloor N/2 + 1 \rfloor, t  \geq \lfloor T/2 + 1\rfloor \}
$$
and
$$
{\bf S}_2 =  \{(i,t) :  i \leq \lceil N/2 \rceil, t  \geq \lfloor T/2 + 1\rfloor \} \cup  \{(i,t) : i  \geq \lfloor N/2 + 1 \rfloor, t  \leq \lceil  T/2 \rceil \}
$$
where $\lceil . \rceil$  and  $\lfloor . \rfloor$ are the ceiling and floor functions.  Each of these two subpanels includes observations for all cross-sectional units and time periods.

We form the debiased fixed effects estimator as
$$
\widehat \beta_{\rm BC}  =  2 \widehat \beta -  \widetilde \beta_{{\bf S}_1\cup{\bf S}_2},
$$
where $\widehat \beta$ is the standard fixed effects estimator while $\widetilde \beta_{{\bf S}_1\cup{\bf S}_2}$ denotes the fixed effects estimator using the data set  ${\bf S}_1\cup{\bf S}_2$ but treats the states in  ${\bf S}_1$ differently from those in  ${\bf S}_2$ to form the fixed effects estimator; namely, we include approximately twice more state fixed effects to compute $\widetilde \beta_{{\bf S}_1\cup{\bf S}_2}$.\footnote{Alternatively, we may form the cross-over jackknife corrected estimator as $\widehat \beta_{\rm CBC}  =  2 \widehat \beta -  (\widetilde \beta_{{\bf S_1}}+\widetilde \beta_{{\bf S_2}})/2$, where $\widetilde \beta_{{\bf S}_j}$ denotes the fixed effect estimator using the subpanel  ${\bf S}_j$ for $j=1,2$. In our empirical analysis, these two cross-over jackknife bias corrected estimators give  similar result.} We obtain bootstrap standard errors by using multipler bootstrap with state-level clustering.

\subsection{Details for Computing Counterfactuals \label{app:counterfactuals}}

We compute counterfactuals from the ``total effect" version of the
model, with behavior concentrated out.
$$
Y_{i,t+\ell}
= \mathsf{a}'
P_{it} + \mathsf{b}' I_{it} +  {\tilde {\delta}}' W_{it}   + {\bar \varepsilon}_{it}
$$
We consider a counterfactual change of $P_{it}$ to $P_{it}^\star$, while
$W_{it}$ and $\bar \varepsilon_{it}$ are held constant. In response to
the policy change, $Y_{i,t+\ell}$ and the part of $I_{it}$ that
contains $Y_{it}$, change to $Y_{i,t+\ell}^\star$ and $I_{it}^\star$. To be
specific, let $C_{it}$ denote cumulative cases in state $i$ on day
$t$. Our outcome is:
$$
Y_{i,t+\ell} \equiv \log
\left(C_{i, t+\ell} - C_{i, t + \ell - 7} \right) - \log
\left(C_{i, t+\ell-7} - C_{i, t + \ell - 14} \right) \equiv \Delta
\log \Delta C_{i,t+\ell}
$$
where $\Delta$ is a 7 day difference operator. Writing the model in
terms of $\Delta \log \Delta C$ we have:
$$
\Delta \log \Delta C_{i,t+\ell} = \mathsf{a}' P_{it} + \mathsf{b}_D
\Delta \log \Delta C_{i,t} + \mathsf{b}_L \log \Delta C_{i,t} +
{\tilde {\delta}}' W_{it}   + {\bar \varepsilon}_{it}
$$
To simplify computation, we rewrite this model in terms of $\log
\Delta C$:
$$
\log \Delta C_{i,t+\ell} = \mathsf{a}' P_{it} + \log \Delta
C_{i,t+\ell-7} + (\mathsf{b}_D + \mathsf{b}_L) \log \Delta C_{i,t} -
\mathsf{b}_D \log \Delta C_{i,t-7} + {\tilde {\delta}}' W_{it}   + {\bar \varepsilon}_{it}
$$
This equation is used to iteratively compute
$\log \Delta C_{i,t+\ell}^\star$
conditional on initial $\{\log \Delta C_{i,s}\}_{s=-\ell-7}^{0}$,
and the entire sequence of $\{W_{i,t}, P_{i,t}^\star, \bar \varepsilon_{it} \}_{t=0}^T$.

Weekly cases instead of log weekly cases are given simply by
$$ \Delta C_{i,t+\ell}^\star = \exp(\log \Delta C_{i,t+\ell}^\star)$$

Cumulative cases can be recursively computed as:
$$
  C_{i,t} = C_{i,t-7} + \Delta C_{i,t} = C_{i,t-7} + \exp(\log \Delta C_{i,t})
$$
given initial $\{C_{i,s}\}_{s=-7}^0$.

Note that $\log \Delta C_{i,t+\ell}^\star$ depends linearly on
$\bar \varepsilon$, so the residuals (and our decision to condition on
versus integrate them out) do not matter when considering linear
constasts of $\log \Delta C_{it}$ or $\Delta \log \Delta C_{it}$, or
when considering relative contrasts of $\Delta C_{it}$.


\subsubsection*{Inference}

Let $S(\theta, \mathbf{\varepsilon})$ denote some counterfactual
quantity or contrast of interest, where $\theta = (\mathsf{a},
\mathsf{b}, \tilde{\delta})$ are the parameters, and
$\mathbf{\varepsilon}$ is the vector of residuals. Examples of $S$
that we compute include:

\begin{itemize}
\item Contrasts of growth rates: $S(\theta, \mathbf{\varepsilon}) =
  \Delta \log \Delta C_{it}^\star - \Delta \log \Delta C_{it}$

\item Relative contrasts of weekly cases: $S(\theta, \mathbf{\varepsilon}) =
  \Delta C_{it}^\star/\Delta C_{it}$

\item Relative contrasts of cumulative cases: $S(\theta, \mathbf{\varepsilon}) =
  \frac{C_{it}^\star - C_{it}}{C_{it}}$
\end{itemize}

The first two examples do not actually depend on
$\mathbf{\varepsilon}$, but the third one does. Inference is by
simulation. Let $\hat{\theta}$ denote our point estimates and
$\hat{\mathbf{\varepsilon}}$ the associated residuals. We draw
$\tilde{\theta}_j$ from the asymptotic distribution of
$\hat{\theta}$. Let $\tilde{\varepsilon}_j$ denote the
residuals associated with $\tilde{\theta}_j$. We then compute
$ \tilde{s}_j = S(\tilde{\theta}_j, \tilde{\varepsilon}_j) $
for $j=1,..., 200$ and plots the mean across $j$ as a point estimate
and quantiles across $j$ for confidence intervals.

\end{document}